# DECISIONING 2023

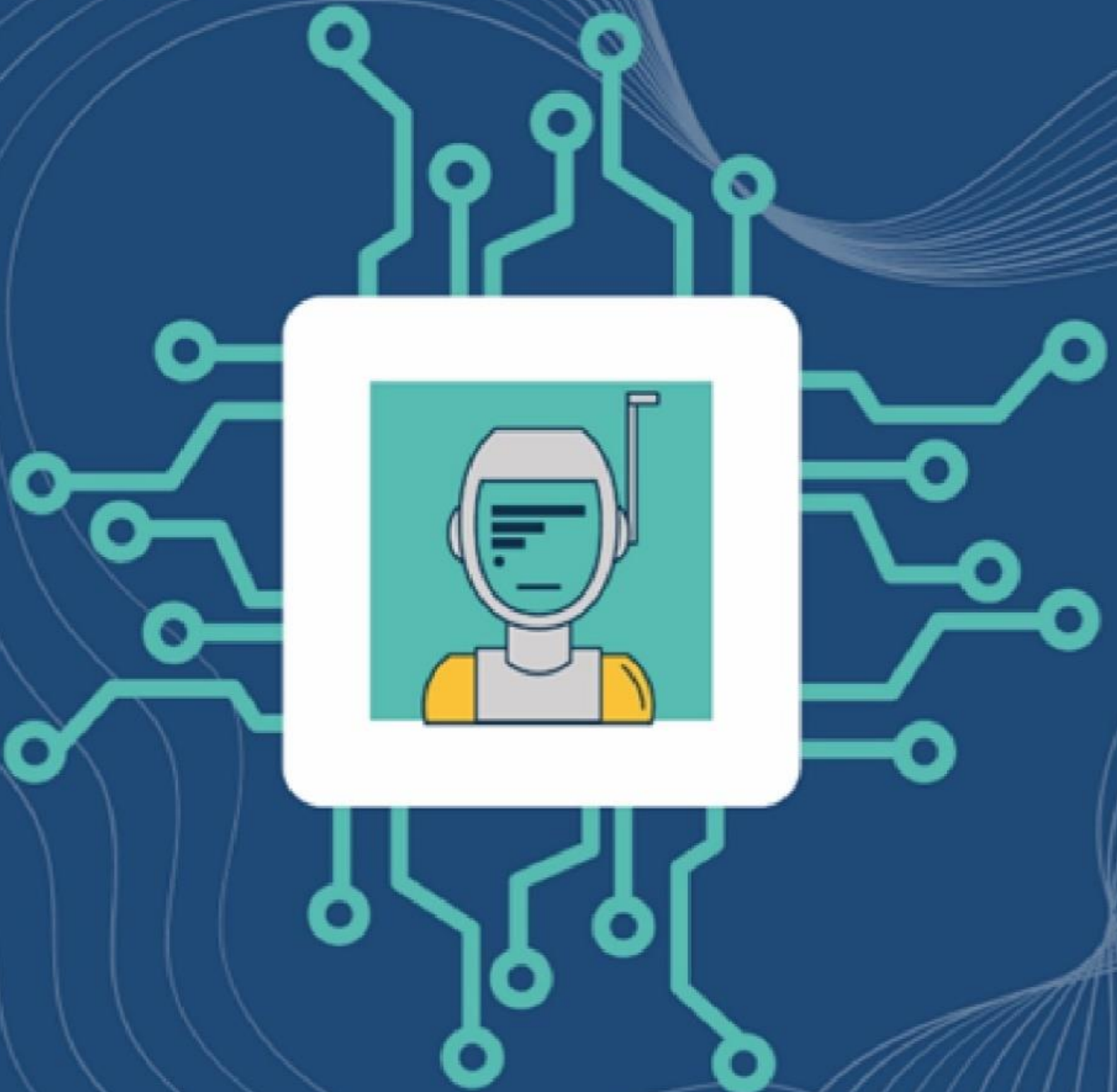

## THE SECOND WORKSHOP ON "COLLABORATION IN KNOWLEDGE DISCOVERY AND DECISIONING MAKING"

UNIVERSITÉ DE LORRAINE

Lifia
Facultad de Informática - Universidad Nacional de La Plata

Universidad Tecnológica de Pereira

Universidad del Cauca

FUNDACIÓN UNIVERSITARIA DE POPAYÁN

Corporación Universitaria Comfacauca
Unicomfacauca

# DECISIONING 2023

The second workshop on
“Collaboration in knowledge
discovery and decision making.”

*Freddy Muñoz Sanabria · Cesar A Collazos ·
Diego Torres · Mario Lezoche ·
Vanessa Agredo · Pablo H. Ruiz
Julio A Hurtado*

***Editors***

Daniel Andres Feriz Garcia
Luis Freddy Muñoz Sanabria
Cesar A. Collazos

***Compilers***

Luis Freddy Muñoz Sanabria
Fundación Universitaria de Popayán
*Popayán, Colombia*

Cesar A. Collazos
Universidad del Cauca
*Popayán, Colombia*

Diego Torres
Universidad Nacional de La Plata
*Argentina*

Mario Lezoche
University of Lorraine
*Francia*

Vanessa Agredo
Corporación Universitaria Comfacauca
*Popayán, Colombia*

Pablo H Ruiz
Corporación Universitaria Comfacauca
*Popayán, Colombia*

Julio A Hurtado
Universidad del Cauca
*Popayán, Colombia*

FUNDACIÓN UNIVERSITARIA DE POPAYÁN

# DECISIONING 2023:

## The second workshop on “Collaboration in knowledge discovery and decision making.”

In a knowledge society, the term knowledge must be considered a core resource for organizations. So, beyond being a medium to progress and to innovate, knowledge is one of our most important resources: something necessary to decide. Organizations that are embracing knowledge retention activities are gaining a competitive advantage. Organizational rearrangements from companies, notably outsourcing, increase a possible loss of knowledge, making knowledge retention an essential need for them. When Knowledge is less shared, collaborative decision- making seems harder to obtain insofar as a “communication breakdown” characterizes participants’ discourse. At best, stakeholders have to find a consensus according to their knowledge. Sharing knowledge ensures its retention and catalyzes the construction of this consensus.

Our vision of collaborative decision-making aims not only at increasing the quality ofthe first parts of the decision-making process: intelligence and design, but also at increasing the acceptance of the choice. Intelligence and design will be done by more than one individual and constructed together; the decision is more easily accepted. The decided choice will then be shared. Thereby where decision-makingcould be seen as a constructed model, collaborative decision-making, for us, is seenas the use of socio-technical media to improve decision-making performance and acceptability. The shared decision making is a core activity in a lot of human activities. For example, the sustainable decision-making is the job of not only governments and institutions but also broader society. Recognizing the urgent needfor sustainability, we can argue that to realize sustainable development, it must be considered as a decision-making strategy. The location of knowledge in the realization of collaborative decision-making has to be regarded insofar as knowledgesharing leads to improve collaborative decision-making: a “static view” has to be structured and constitutes the “collaborative knowledge.” Knowledge has an important role in individual decision-making, and we consider that for collaborative decision-making, knowledge has to be shared. What is required is a better understanding of the nature of group work”. Knowledge has to be shared, but how do we share knowledge?

Decisioning 2023 is the second workshop on Collaboration in knowledge discoveryand decision-making: It has been organized by six research teams from France, Argentina, and Colombia to explore the current frontier of knowledge and applications in different areas related to knowledge discovery and decision-making.The format of this workshop is hybrid (face-to-face or virtual), aims at the discussionand knowledge exchange between the academy and industry members. This workshop has been partially supported by STIC-AmSud projects: AGROFAIR (Agro-Knowledge Integration: Developing a FAIR data science approach for adding valueto the agricultural supply chain) and AICODA: Artificial Intelligence COllaborative Decision making Agriculture.

This workshop was organized in the city of Popayan (Colombia) during June 21 to 23 – 2023, and brought together researchers from different countries and regions of Colombia. Seven long articles, five short articles, and two posters were presented inthis edition.

## Chairs and Committee

### General Chairs

**Dr. Cesar Collazos,** *Universidad del Cauca, Colombia*.
<ccollazo@unicauca.edu.co>
**Dr. Diego Torres,** *Universidad Nacional de La Plata-Universidad Nacional deQuilmes, Argentina.* <diego.torres@lifia.info.unlp.edu.ar>
**Dr. Mario Lezoche**, *University of Lorraine, France*
<mario. lezoche@univ-lorraine.fr>

### Organizing Committee

**Dr. Julio Hurtado**, *Universidad del Cauca, Colombia.*
<ahurtado@unicauca.edu.co>
**Dr. Freddy Muñoz**. *Fundación Universitaria de Popayán,*
<freddy.munoz@docente.fup.edu.co>
**Dr. Julio César Chavarro**, *Universidad Tecnológica de Pereira, Colombia,*
<jchavar@utp.edu.co>
**Mag. Clara Eugenia Satizábal Serna**. *Fundación Universitaria de Popayán,*
<sistemas@fup.edu.co>
**Mag. Daniela Iboth Gutiérrez Idrobo**. *Fundación Universitaria de Popayán,*<daniela.gutierrez@docente.fup.edu.co>
**Mag. Luis Alfonso Vejarano**. *Fundación Universitaria de Popayán*
**Mag. Daniel Andrés Feriz García**. *Fundación Universitaria de Popayán*
**Mag. Sonia Del Consuelo Gaviria Armero.** *Fundación Universitaria de Popayán*

### Doctoral Symposium Chair

**Dr. Cesar Collazos**, *Universidad del Cauca, Colombia.*
<ccollazo@unicauca.edu.co>
**Dr. Alejandro Fernández,** *Universidad Nacional de La Plata-Universidad Nacional de Quilmes, Argentina* <alejandro.fernandez@lifia.info.unlp.edu.ar>

**Publicity Chairs**

**Dr. Vanessa Agredo Delgado,** *Corporación Universitaria Comfacauca-Unicomfacauca, Colombia.* <vagredo@unicomfacauca.edu.co>
**Dr. Pablo H. Ruiz,** *Corporación Universitaria Comfacauca-Unicomfacauca,Colombia.* <pruiz@unicomfacauca.edu.co>
**Dr. Leandro Antonelli,** *Universidad Nacional de La Plata, Argentina.* <leandro.antonelli@lifia.info.unlp.edu.ar>

**Program Committee Members**

| | | |
|---|---|---|
| Alfonso | Infante | *alfonso.infante@decd.uhu.es* |
| Alicia | Mon | *amon@unlam.edu.ar* |
| Ana | Esteso | *aesteso@cigip.upv.es* |
| Ana Isabel | Molina Díaz | *AnaIsabel.Molina@uclm.es* |
| Ana Paula | Costa | *apcabral@cdsid.org.br* |
| Andres | Rodriguez | *asrodriguez@gmail.com* |
| Andres | Rodriguez | *arodrig@lifia.info.unlp.edu.ar* |
| Antonio | Silva Sprock | *asilva.sprock@gmail.com* |
| Carlos | Buil Aranda | *cbuilaranda@gmail.com* |
| Claudia | Pons | *cpons@info.unlp.edu.ar* |
| Cristina | Manresa-Yee | *cristina.manresa@uib.es* |
| Danielle | Morais | *dcmorais@insid.org.br* |
| Diego | Firmenich | *diego.firmenich@gmail.com* |
| Fernando | Moreira | *fmoreira@upt.pt* |
| Francesca | Pozzi | *pozzi@itd.cnr.it* |
| Francisco Luis | Gutiérrez Vela | *fgutierr@ugr.es* |
| Gabriela | Arevalo | *gabriela.b.arevalo@gmail.com* |
| Germán Ezequiel | Lescano | *gelescano@unse.edu.ar* |
| Gineth Magaly | Cerón Rios | *gceron@unicomfacauca.edu.co* |
| Gustavo | Constain Moreno | *gustavo.constain@gmail.com* |
| Guy | Camilleri | *Guy.Camilleri@irit.fr* |
| Horacio | Del Giorgio | *horacio.delgiorgio@gmail.com* |
| Horacio | Del Giorgio | *hdelgiorgio@ing.unlam.edu.ar* |
| Huizilopoztli | Luna-García | *hlugar@uaz.edu.mx* |
| Isabelle | Linden | *isabelle.linden@unamur.be* |
| Jaime | Díaz | *jaime.diaz87@gmail.com* |
| Jaime | Muñoz-Arteaga | *jmauaa@gmail.com* |
| Jason | Papathanasiou | *jasonp@uom.gr* |
| Javier Alejandro | Jiménez Toledo | *javierjx@gmail.com* |
| Javier | Berrocal | *jberolm@unex.es* |
| Jesús | Gallardo Casero | *jesus.gallardo@unizar.es* |
| Jose | García-Alonso | *jgaralo@unex.es* |
| Jose | Pow | *japowsang@pucp.pe* |
| Juan Enrique | Garrido Navarro | *juanenrique.garrido@udl.cat* |
| Juan Manuel | Murillo Rodríguez | *juanmamu@unex.es* |
| Julián | Muñoz | *julianfer87@gmail.com* |
| Julio | Hurtado | *jhurtado@dcc.uchile.cl* |
| Julio | Ponce | *julk_cpg@hotmail.com* |
| Klinge Orlando | Villalba-Condori | *kvillalbac@unsa.edu.pe* |
| Leandro | Antonelli | *leandro.antonelli@lifia.info.unlp.edu.ar* |
| Leandro | Fernandes | *leandro.fernandes@mackenzie.br* |
| Liliana | Vizzuett | *lilianavizz@hotmail.com* |
| Luis Freddy | Muñoz Sanabria | *lfreddyms@hotmail.com* |
| Luis Mariano | Bibbo | *lmbibbo@lifia.info.unlp.edu.ar* |
| Manuel | Ibarra | *manuelibarra@gmail.com* |
| Marco Javier | Suarez Baron | *marcojaviersuarezbaron@gmail.com* |
| María Del Mar | Alemany-Díaz | *mareva@omp.upv.es* |
| Maria Dolores | Lozano | *maria.lozano@uclm.es* |
| Miguel | Redondo | *Miguel.Redondo@uclm.es* |
| Natalia | Padilla | *natalia.padilla@unir.net* |
| Nathalie | Aussenac-Gilles | *aussenac@irit.fr* |
| Oscar | Revelo Sánchez | *orevelo@udenar.edu.co* |
| Oscar | Franco | *oscarhf2002@hotmail.com* |
| Pablo | Torres-Carrion | *pvtorres@utpl.edu.ec* |
| Pablo | Santana Mansilla | *psantana@unse.edu.ar* |
| Pablo E. | Martínez López | *fidel@unq.edu.ar* |
| Pablo Hernando | Ruiz Melenje | *pruiz@unicomfacauca.edu.co* |
| Patricia | Paderewski | *patricia@ugr.es* |
| Pascale | Zárate | *pascale.zarate@ut-capitole.fr* |
| Patricio | Galdames | *pgaldames@ubiobio.cl* |
| Pavlos | Delias | *pdelias@af.ihu.gr* |
| Philippe | Palanque | *palanque@irit.fr* |
| Pierre-mmanuel | Arduin | *pierre-emmanuel.arduin@dauphine.psl.eu* |
| Ricardo | Tesoriero | *ricardo.tesoriero@uclm.es* |
| Rosanna | Costaguta | *rcostaguta@hotmail.com* |
| Sergio | Ochoa | *sochoa@dcc.uchile.cl* |
| Silvana | Aciar | *silvanav.aciar@gmail.com* |
| Susana | Bautista | *susana.bautista@ufv.es* |
| Wilson Javier | Sarmiento | *wilson.sarmiento@unimilitar.edu.co* |
| Yasamin | Eslami | *yasamin.eslami@ec-nantes.fr* |
| Yosly Caridad | Hernandez | *yoslyhernandez@gmail.com* |

# *Contents*

# SEMIoTICA - Security Scenarios Modeling forIoT-based Agriculture Solutions

**Julio Ariel Hurtado[1], Leandro Antonelli[2], Santiago López[1], Adriana Gómez[3], Juliana Delle Ville[2], Frey Giovanny Zambrano[1], Andrés Solis[1,4], Marta CeciliaCamacho[5], Miguel Solinas[6], Gladys Kaplan[7], and Freddy Muñoz[8]**

[1] IDIS, Universidad del Cauca, Colombia
{ahurtado, santiagolopez94, freyzambrano, afsolis}@unicauca.edu.co.edu.co
[2] Lifia - Facultad de Informática, Universidad Nacional de La Plata, Argentina
{lanto, jdelleville}@lifia.info.unlp.edu.ar
[3] universidad Tecnologica de Pereira, Colombia
adrianagomezr@utp.edu.co
[4] Corporacion Universitaria Comfacauca, Colombia
[5] Institución Universitaria Colegio Mayor del Cauca, Colombia
cecamacho@unimayor.edu.co
[6] LARYC, FCEFyN, Universidad Nacional de Cordoba, Argentina
miguel.solinas@unc.edu.ar
[7] Universidad Nacional de La Matanza, Argentina
gladyskaplan@gmail.com
[8] Fundacion Universitaria de Popayan, Colombia
lfreddyms@gmail.com

**Abstract.** Agriculture is a vital human activity contributing to sustain- able development. A few decades ago, the agricultural sector introduced the Internet of Things (IoT), playing a relevant role in precision and smart farming. IoT developments in farms require a lot of connected devices working cooperatively. It increases the vulnerability of IoT de- vices mostly because it lacks the necessary built-in security due to their constrained context and computational capacity. Additionally, storage and data processing connecting with edge or cloud servers are the rea- son for many security threats. To ensure that IoT-based solutions meet functional and non-functional requirements, particularly security, soft- ware companies should adopt a security-focused approach to their spec- ification. This paper proposes a method for specifying security scenarios integrating requirements and architecture viewpoints in the context of the IoT in agriculture solutions. The method comprises four activities. First, the description of scenarios for the intended software. After that, scenarios with incorrect system use should be described. Then, these are translated into security scenarios using a set of rules. Finally, the securityscenarios are improved. Additionally, this paper describes a preliminary validation of the approach, which software engineers in Argentina and Colombia performed. The results show that the approach proposed al- lows software engineers to define and analyze security scenarios in the IoT and agriculture contexts with good results.

## 1 Introduction

The International Telecommunication Union (ITU) defines IoT (Internet of the Things) as a "global infrastructure for the information society that provides ad- vanced services through the connection of objects (physical and virtual) relying on the interoperability of current and future knowledge and communication tech- nologies" [31]. According to [17] as an interconnected network, IoT contributes to making decisions based on the information collected, and its interaction does not need human intervention. The definition includes the concept of a Cyber- Physical system, which is a complex abstraction that requires a conceptual map [5] rather than a simple definition to state its concept.

Regarding software development, requirements analysis is a critical activ- ity for defining software functionalities, attributes, and quality properties. This process is particularly different for software construction by using emergent tech- nologies like IoT. Traditional software development practices must adapt to these new technologies and business contexts [17]. Requirements engineering involves collaboration between clients and development teams for incorporating the right features into the finished product [27]. Inconsistencies between initial require- ments and the final product can lead to re-engineering processes, increasing the project scope and cost [26]. Requirements engineering works with both types of knowledge, explicit and tacit [1]. Tacit knowledge is difficult to communicate because experts and development teams often have different backgrounds and use distinct terminologies [20], making it challenging to elicit information from stakeholders.

Software products are defined by a set of functional and non-functional re- quirements. The latter is responsible for the software product's quality and is most frequently considered when developing an IoT system according to its spe- cific application domain [17]. A way to specify software requirements is to de- scribe use scenarios through storytelling techniques. This approach is effective because it is a way to incorporate details that are essential to provide a rich con- solidation of knowledge. Scenarios employ natural language, allowing experts to use them without complex formalisms. This makes them highly effective in promoting communication and collaboration among diverse groups of experts [3].

The main challenges associated with these requirements in developing theseproducts are limited processing and storage capacity, performance reliability, availability, accessibility, interoperability, security, privacy, scalability flexibility,and context awareness [17, 19]. Following this reasoning, security is an aspect that is highly relevant in IoT-based software because it protects resources suchas modules, code, and others from unauthorized access [19]. With scenarios, experts can describe various situations and work together to improve them, learning from one another in the process. This can be especially valuable when dealing with complex problems that require inputs from multiple perspectives. Overall, scenarios can be a powerful tool for fostering cooperation and achieving better outcomes in a wide range of domains.

The software architect must consider designing the whole system when stake- holders identify security concerns rather than adding security technologies in an ad-hoc manner [15]. As Bruce Schneier points out [25], security is a process and a chain that is only as strong as its weakest link. Therefore, software providers must adopt a security-centric approach to designing and developing IoT-based solutions that conform to functional and non-functional requirements like secu- rity [14].

The agricultural sector now requires data collection and advanced technolo- gies to improve production while using limited resources. Sustainable agriculture can help preserve nature without compromising the needs of future enerations [6, 13]. The Food and Agriculture Organization (FAO) has identified population growth, resource scarcity, and degradation as key challenges. There is a need to increase efficiency, productivity, and quality in agrifood systems while protectingthe environment [24]. To achieve this, new developments and technologies mustbe introduced to automate traditional farming methods and make farm labormore efficient. The Internet of Things (IoT) appears as technology to transform conventional processes [6, 13].

IoT systems encounter distinct security challenges when compared to tradi- tional IT systems primarily due to the presence of resource-constrained devices. These limitations make IoT systems more susceptible to a wide range of at- tack vectors, posing potential threats to their security [7]. Therefore, accurately identifying and understanding the specific security requirements is crucial when developing such systems.

This paper proposes a scenario-based method for specifying the architec-ture's security aspects. The method is composed of four essential activities. The first activity consists in describing scenarios of the intended software applica-tion. The second activity consists in describing scenarios related to the previous ones but referring to incorrect usage of the application. The third activity con- sists in applying a set of rules to map attributes from the previous scenariosto the architecture scenarios. Finally, the four activity consists in describingthe architectural scenarios in more detail. Moreover, this document describes a preliminary evaluation of the proposed approach. Given the security challenges facing IoT agriculture, the research question in his paper is: how adequately elicit security requirements in IoT- based smart agriculture solutions?

The paper is organized in the following way. Section 2 describes some back- ground about the scenarios. Then, section 3 describes some related work. Section 4 details our contribution, which is the proposed approach. Section 5 describesthe tool to support the proposed method. Section 6 presents the preliminary evaluation. Finally, Section 7 discusses some conclusions.

## 2 Background

This section describes two types of scenarios. First, it describes scenarios that focus on the functionality of a software application. Afterward, the section describes scenarios that focus on architectural security concerns.

### 2.1 Scenarios for describing functionality

A scenario [3] is an artifact that describes situations (in the application or the software domain) using natural language. It describes a specific situation that arises in a certain context to achieve some goal. There is a set of steps (the episodes) to reach that goal. In the episodes, active agents as actors use materials, tools, and data as resources to perform some specific action. Although there are many templates to describe scenarios, this paper will use the scenarios proposedby Leite et al. [21]. Figure 1 summarizes the template.

| Attribute | Scenario title | Goal | Context | Actors | Resources | Episodes |
|---|---|---|---|---|---|---|
| Description | Id | Objective | Starting point (time, place, activities previously achieved) | Active agents | Passive elements (tools, materials, data) | List of actions, simple breakdown with no conditions, no iterations |

***Fig. 1.*** *Template for describing scenarios that focus on functionality.*

Let's consider the following example that describes a scenario about how the irrigation system is activated. This task can be done in different ways regard- ing the technological infrastructure that the farm has. For example, an operator can manually start the irrigation by physically accessing the machine room withthe pumps. In this situation, there is no IoT software application. This paperis going to focus on another scenario where a software application performs theactivation of the pumps. For example, an agriculture expert evaluates the fieldconditions to determine whether irrigating is necessary and provides the infor-mation to the farm supervisor. Then, the supervisor activates the irrigation pipethrough an IoT-based web application. Table 2 summarizes the situation.

| Attribute | Scenario title | Goal | Context | Actors | Resources | Episodes |
|---|---|---|---|---|---|---|
| Description | Attempt to access the water irrigation infrastructu re by an authorized person. | Protect access to the water irrigation system to ensure responsible use of the water. | The field counts with an irrigation infrastructure (pipes, tanks, pumps, and valves) to water (irrigate) the field. | Expert Supervisor | The checklist to determine whether it is necessary to irrigate the field. The security protocol to access and operate the pump and valves. | An expert evaluates the conditions of the field to determine whether it is necessary to irrigate. The expert writes a report to the supervisor with the recommendation to irrigate. The supervisor logs in to the IoT web application The supervisor starts the pump and opens the valves. |

***Fig. 2.*** *Authorized attempt to start the irrigation system.*

The previous scenario describes an authorized person's legitimate use of the software application to activate the irrigation system. This scenario could be similar to Use Cases or User Stories [20]. Nevertheless, the software system can be vulnerable to hack attacks, where a malicious user desires to break into theweb software application to start the irrigation system just for fun or to destroythe crop. This incorrect and harmful description of the software application is related to misuse cases [12].

### 2.2 Scenarios for describing architectural security concerns

Software architecture is the designing process of the system's fundamental struc- ture and organization to achieve specific quality attributes, which are the criti-cal non-functional characteristics determining the system's overall effectiveness. Quality attributes are specified through quality scenarios, which define how thesystem should behave under various conditions. A Quality Attribute (QA) Sce-nario is a specific, testable scenario that demonstrates how a quality attribute requirement is satisfied. A QA scenario is typically structured with an id, a stimulus that triggers the interaction with the software application, the environ-ment where the interaction occurs, the artifact affected, the response, and somequantitative description of the response. Table 3 summarizes the template.

| Attribute | Scenario Id | Source of the stimulus | Stimulus | Environment | Artifact | Response | Response measure |
|---|---|---|---|---|---|---|---|
| Description | Unique identifica tion of the scenario | Some human, system, or any other actor generates a stimulus to the system. | It is an input condition that generates a response from the system | The stimulus occurs under a certain context. The system may have an overload context, normal operation, or some other relevant state. For many systems, "normal" operation can refer to one of a number of modes. For these kinds of systems, the environment should specify in which mode the system is executing | The stimulated artifact. This may be an ecosystem, a whole system, a component, or some set of components. | It is the response as the result of the arrival of the stimulus. | The response should be measurabl e so that the requireme nt can be tested. |

***Fig. 3.*** *Security scenario template.*

Security refers to the system’s capability to defend against danger, ensure itssafety, and protect system data from unauthorized disclosure, modification, or destruction. Security involves protecting computer systems themselves throughtechnical and administrative safeguards. Additionally, security can refer to thedegree to which a particular security policy is enforced with some level of as- surance. The three fundamental types of security concerns are confidentiality, integrity, and availability. Confidentiality refers to the protection of data and processes from unauthorized disclosure or access by individuals or entities thatare not authorized to access it. Integrity refers to protecting data and processesfrom unauthorized modification, intentional or accidental. It includes ensuringthat data is not tampered with or corrupted during storage, processing, or transmission. And availability refers to the protection of data and processes from de-nial of service attacks or other forms of disruption that can prevent authorized users from accessing or using them. This includes ensuring that systems are avail- able and responsive when needed and that they can handle high levels of trafficor activity without becoming overloaded or crashing. Figure 4 describes an ex-ample of a security scenario that refers to the same situation of the requirementscenario described in Figure 2.

| Attribute | Identification | Source of the stimulus | Stimulus | Environment | Artifact | Response | Response measure |
|---|---|---|---|---|---|---|---|
| Description | S01 | An unauthorized individual attempts to access the water irrigation system through an IoT connected device. | The individual attempts to gain access to sensitive data (sensor measurements) or to manipulate the system's functionality (change valve behavior). | Normal execution. The system counts with IoT connected devices used to access the functionality of the solution, such as sensors, actuators, processors. | The security protocols and access control subsystem. | The security protocols detect the unauthorized access attempt and block the individual from accessing the access control subsystem. | Ensure the security of the system, it's important that attacks are detected quickly, ideally within 0.5 seconds. Additionally, the system must have a high rate of success in resisting attack attempts, with a target of greater than 95%. |

***Fig. 4.*** *Security scenario example.*

## 3 Related work

The complexity of IoT software applications is a concern identified several re- searchers. Thus, there are some proposals to deal with this complexity. Nguyenet al. [16] propose FRASAD, a model-driven software development frameworkto manage the complexity of Internet of Things (IoT) applications. Karadu- man et al. [11] is another proposal to deal with the complexity. Their approach includes activities such as requirements development, domain-specific design, verification, simulation, analysis, calibration, deployment, code generation, and execution. Nevertheless, these proposals do not consider security, which is our main concern.

Some other approaches consider the security issue, but it is considered in terms of implementation, while our proposal considers the security in terms of the specification of requirements. Cardenas et al. [2] propose a process and a tool to apply formal methods in Internet of Things (IoT) applications using theUnified Modeling Language (UML). They have developed a plug-in tool to vali- date UML software models regarding the design of a secure software application. Slovenec et al. [28] present a taxonomy of security requirements to consider themwhen designing and implementing the software application. El-Gendy et al. [7]propose a security architecture to provide security enabled IoT services, and provide a baseline for security deployment. The architecture solution outlined in this context plays a crucial role in addressing the security requirements of IoT systems. These security requirements are useful components of our security scenarios proposal. By focusing these requirements, we can effectively establisha robust security framework at requirements level. Sotoudeh et al. [29] studyaims to establish security requirements for IoT systems, with a focus on en- hancing the security of smart home applications. The identified requirements complement our proposal as they introduce a significant vocabulary for express- ing security scenarios in the IoT and smart farm context. By incorporating these elicited requirements, we can effectively address the specific security challenges and considerations associated with IoT and smart farm environments.

There are some approaches that consider security in requirements, but theydo not emphasize the way to specify security requirements precisely. Iqbal etal. [9] present a literature review about an In-Depth Analysis of IoT Security Requirements, but the work they present does not refer about how to specifythem. Özkaya et al. [18] proposal deal with different non-functional requirements: security, scalability, and performance. And they try to balance the different requirements or decide which one to satisfy when there is a conflict. Carvalhoet al, [4] also deals with conflicts, but their approach deals with non-functional Requirements.

Finally, Kammuller et al. [10] present an approach to specify security re- quirements for IoT applications. They combine a framework for requirement elicitation with automated reasoning to provide secure IoT for vulnerable usersin healthcare scenarios. They map technical system requirements using high- level logical modeling. Then they perform an attack tree analysis. And finally, asecurity protocol analysis. Their work pays more attention to the tree analysis to identify the situation, while our approach pays attention to how to describe security requirements precisely.

## 4 Our approach

This section is organized in the following manner: firstly, we provide a descriptionof the general approach, followed by a detailed explanation of each step.

### 4.1 Our approach in a nutshell

Our proposed approach consists of several steps. Firstly, we describe scenariosthat outline the intended usage of the software. Next, we create scenarios that describe incorrect usage of the application in an attempt to exploit any vul- nerabilities. We then establish rules for converting these scenarios into securityscenarios. Lastly, we refine and improve the security scenarios. Figure 5 provides a summary of our approach.

### 4.2 Description of the scenarios with the correct use of the indentedsoftware application

This step describes the scenarios that focus on the correct use of the software application regarding security concerns. This step should be executed by a re- quirements engineer or analyst (or a group of them) that must interact with the

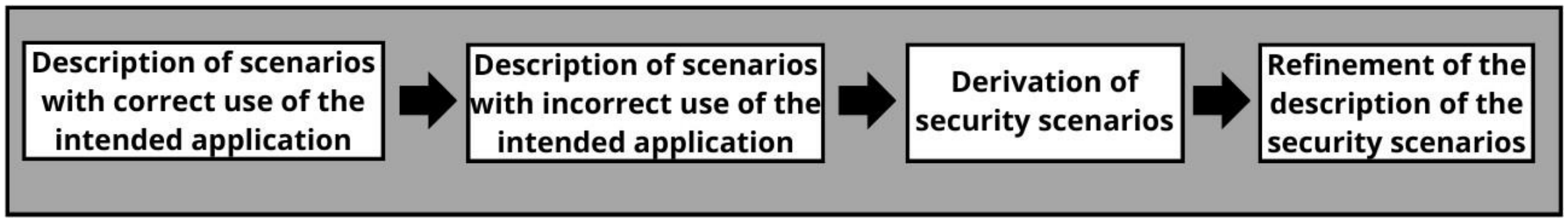

***Fig. 5.*** *Our approach in a nutshell.*

Experts of the domain (clients, users, and stakeholders in general) to capture the software application’s requirements and specify scenarios. Those should describe the functionality of the intended software, and they should also consider secu-rity concerns. That is why the analyst eliciting and defining scenarios should have some background in security non-functional requirements to consider this concern in the specification. The result of this step is a set of scenarios that describes the functionality like the one described in Figure 2.

### 4.3 Description of the scenarios with the incorrect use of theindented software application

This step consists in analyzing the scenarios described in the previous step to find security issues and describing the ones that explode the problems and compro- mise the security of the software application. Ideally, this step should be done by the same requirement engineer (or group of them) that participated in the pre- vious tasks. They should analyze every scenario in detail, and considering guides like the ones proposed by Gupta et al. [8] and Yazdinejad [32], they must describescenarios of incorrect use of the software application. Basically, they should de- scribe scenarios that explode possible vulnerabilities.

For example, consideringthe scenario that describes the correct use of the software application to activate the irrigation system (Figure 2), the requirements engineer can determine that the access to the system (and therefore the access to the activation of the pumps) determine a security breach. Hence, the analyst describes a scenario where an unauthorized person gets access to the software application and, consequentlyto the irrigation infrastructure). Figure 6 describes the complete scenario.

| Attribute | Scenario title | Goal | Context | Actors | Resources | Episodes |
|---|---|---|---|---|---|---|
| Description | Attempt to access the water irrigation infrastructure by an unauthorized person. | Protect access to the water irrigation infrastructure to ensure responsible use of the water. | The field counts with an irrigation infrastructure (pipes, tanks, pumps, and valves) to water (irrigate) the field. An unauthorized person attempts to manipulate the pump and the valve to irrigate the field. | An unauthorized person | The checklist to determine whether it is necessary or not to irrigate the field. The security protocol to access and operate the pump and valves. | An unauthorized gains access to the IoT web application.<br>An unauthorized person starts the pump and opens the valves. |

***Fig. 6.*** *Unauthorized attempt to start the irrigation system.*

### 4.4 Derivation of security scenarios

This step describes a set of rules to map the information contained in a scenario that describes the incorrect use of the intended system to obtain a first draftof a scenario to describe security concerns. It is essential to mention that the scenario with incorrect usage will not provide complete information about the security scenario. The rules proposed will use only four attributes (title, context, actors, and resources), and this information will be used to fill four attributes to the security scenario (stimulus, environment, source of the stimulus, and arti- fact). Therefore, with this information, the following step can refine the security scenario. Figure 7 summarizes the mapping between attributes of both types of scenarios. Following the example of the scenario that describes the incorrect use of the software application described in Figure 6, the scenario obtained applying the proposed rules is the one described in Figure 8. It is important to mention that this scenario (the resulting from the mapping rules) needs to be refined inthe following step, which is why this simple mapped security scenario is still far from the security scenario (like the one described in Figure 4).

| The attribute of the incorrect usage scenario | Title | Context | Actors | Resources |
|---|---|---|---|---|
| Attribute of the security scenario | Stimulus | Environment + Source of the stimulus | Sources of the stimulus | Artifact |

*Fig. 7.* *Mapping rules between attributes of the incorrect and security scenarios.*

| Attribute | Stimulus | Environment + Source of stimulus | Source of stimulus | Artifact |
|---|---|---|---|---|
| Description | Attempt to access the water irrigation infrastructure by an unauthorized person. | The field counts with an irrigation infrastructure (pipes, tanks, pumps) to water (irrigate) the field. An unauthorized person attempts to manipulate the pump and the valve to irrigate the field. | Unauthorized person. | The checklist to determine whether it is necessary or not to irrigate the field. The security protocol to access and operate the pump and valves. |

*Fig. 8.* *Mapping rules between attributes of the incorrect and security scenarios.*

## 4.5 Refinement of the security scenarios

Some adjustments and improvements should be made to the scenarios derived from the mapping rules in the previous step. Some new information should be added, and some information should be rephrased. Consequently, the require- ments engineering should use his experience and knowledge to provide further information and paraphrase some others based on the elicitation meeting andhis expertise in the field. For example, the identification of the security scenario should be provided. But this is a minor issue since the indication must be an idto identify the scenario inside the software development process, and it is more related to documentation definitions. Afterward, the attributes of stimulus, envi-ronment, source of stimulus, and artifact should be rephrased considering the in-formation obtained in the previous step. The attributes environment and sourceof stimulus mainly captured data from the same attribute in the incorrect usagescenario, so in the security scenario, the information should be split and dividedinto two attributes. Finally, the attributes response and response measure shouldbe completed. Although the mapping rules do not provide information to meet these attributes, the information provided in the rest of the scenario provides the context necessary so the requirement engineers can describe these two attributes. It is important to mention that the response measure attribute, in particular, should be described with quantitative measures. Therefore, engineering require- ments should pay attention to that, although the tool described in the following section provides some support. Figure 9 summarizes the refinements. Then, the scenario described at the beginning of this paper in Figure 4 is an example of the scenario that this approach pursues to obtain.

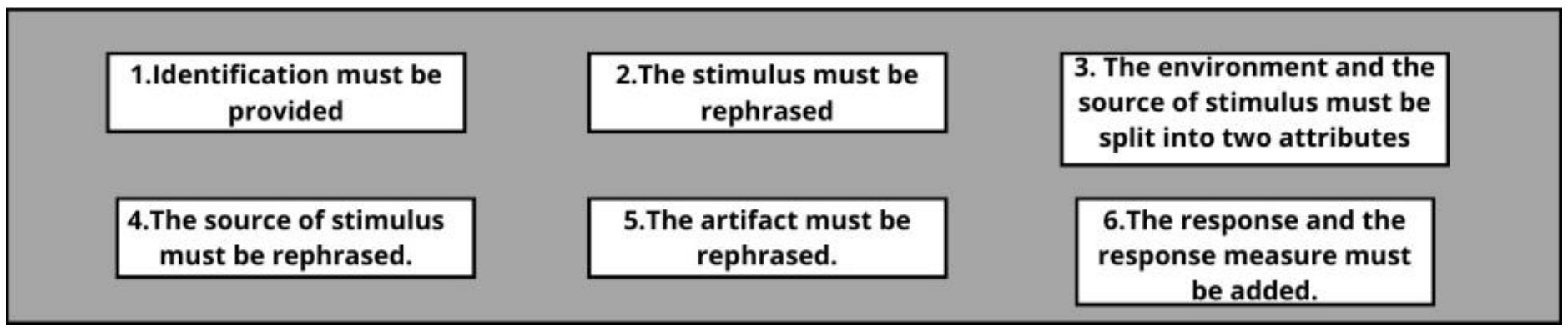

*Fig. 9.* *Refinement to the security scenarios.*

Security scenarios in smart farms and IoT require a specific vocabulary for ac-curately expressing them. There are several concerns taken into account as partof these scenarios as propose [29]. One such concern is technology-dependent se- curity for IoT devices (artifact), which refers to the security measures required in the IoT context (environment). Another important aspect is the authentication of IoT objects and individuals (sources of stimulus) using various mechanismsto prevent or detect attacks (responses). These responses to potential security threats have several limits (response measurement). Requirements engineerscould use this vocabulary as a lexicon and semiotic tool.

# 5 Assessment of the approach

## 5.1 Assesment Desing

Our aim is to assess the acceptance of our approach by security experts in the context of IoT-based smart agriculture, using the Technology Acceptance Model (TAM) to guide our evaluation. Specifically, we are interested in understandinghow much our approach is accepted by this target audience. To evaluate the usefulness and ease of use of our approach, we have adopted the well-known and widely used metrics of Perceived Usefulness and Perceived Ease of Use asdefined in Fred D. Davis's work. To this end, we have designed and administered a survey to a group of expert security professionals who are representative of our target audience and possess experience in eliciting security requirements. By administering the survey to a group of expert security professionals who arerepresentative of our target audience and possess experience in eliciting securityrequirements, we will be able to gather valuable feedback and insights. These insights will help us identify areas of strength and weakness in our approach, and ultimately guide improvements that enhance its overall acceptance.

## 5.2 Survey Application and Data Collection

We conducted a survey with a group of five experts in the software and net- working security field. Prior to the survey, we presented our methodology to thegroup and spent approximately 40 minutes discussing and addressing any ques-tions they had. Once we presented our approach, we administered a survey thatincluded 17 closed-ended questions and three open-ended questions. The survey aimed to gather insights from the experts on the easy to use and usefulness per-ception of our method. Most of the experts found the proposed security scenariomethod to be a useful tool for specifying the requirements of agricultural IoT solutions. Half of the experts surveyed agreed that the proposed method simpli-fies the process of specifying security requirements, resulting in better quality and control of the specification. The experts noted that the proposed method iswell-defined, easy to understand, and flexible, making it ideal for defining sce-narios. Additionally, the evaluation revealed that the majority (over 60 percent)found it to be clear, well-structured, and interactive in its development.

## 5.3 Results and Analysis

While the method was generally perceived as useful and easy to use for devel-oping security scenarios, it was suggested that it needs to be more specific to determine its usefulness in practice. The experts suggested that the method canbe enhanced to include specific aspects of cybersecurity, as well as developmentand implementation elements that are essential to ensuring the security of agri-cultural IoT systems. This would enable a complete specification of the securityrequirements of these systems. Furthermore, it was noted that users need to in-teract with the method to remember its steps. During the evaluation, experts identified some areas for improvement such as incorporating vulnerabilities andrisks commonly found in IoT systems, considering different types of users andadversaries, and taking into account various attack vectors.

By doing so, the proposed method can be further refined to better meet the needs of users and enhance the security of agricultural IoT systems, particularlyadding this information in the lexicon associated.

## 6 Prototype of the tool support

A software tool was prototyped in order to provide support to the proposed approach. The tool was implemented in Python [22] using libraries such as Spacy [30], an nlp processing library and textblob [23]. This tool consists of a web application, that can be used on desktop computers as well as mobile phones. The application manages different projects and different kinds of artifacts usingnatural language. Scenarios are one kind of artifact, but the application canbe extended easily to manage other artifacts (User Stories, Use Cases, etc.). The prototype provides support for the different activities of the approach. Theprototype provides some edition form to allow users to write scenarios of correct and incorrect use of the software application.

The prototype also provides a form to list all the scenarios describing the functionality of the intended software, and by selecting one or more scenarios, theprototype performs the derivation of security scenarios by applying the mapping rules proposed. Then, the security scenarios can be edited in order to improve their description.

The prototype includes some natural language processing tools that make itpossible to provide support to assist the requirements engineering to describe the security scenarios. For example, the prototype can verify the use of terms that belong to a glossary. Thus, using Lemmatization and Stemming techniques, theprototype can verify whether certain expressions are used. This is very important in the attributes of response to ensure the use of the correct technique to cope with the issue the scenario is describing. Another feature is the identification ofquantitative descriptions. Natural language processing tools make it possible toassess whether this type of expression is present (for example, in the response measure attribute) to be sure that the scenario is correctly written.

## 7 Conclusion

This paper proposed an approach to describing security scenarios to design a robust software architecture considering IoT technology in the agricultural do- main. Developers of IoT applications should be concerned about security (andsome other non-functional requirements) since the risk of exposing physical ar-tifacts to intruders is considerable. Moreover, it is difficult to identify the threatand design a countermeasure. Generally, these issues are identified when it is too late when some intruder explodes the vulnerability. Therefore, this paper presents a lightweight approach that begins with a description of the functionalrequirements. The misuse of the application is identified in order to design coun- termeasures to deal with it. The paper also described a prototype tool to help apply the proposed approach. Finally, a preliminary assessment was also pro- vided.

The survey applied to five security experts found that the proposed security scenario method is generally useful for specifying agricultural IoT solutions, but needs improvement in certain areas. Experts suggested incorporating specificcybersecurity aspects, vulnerabilities and risks commonly found in IoT systems, and different types of users and adversaries. They also noted the method needsto be more specific and interactive for users to remember its steps. The results provide valuable insights for refining and improving the method to meet user needs and enhance security.

Currently, the most widely used development process is agile development,but we propose a complementary and lightweight technique specifically for IoTapplications the smart farm field. As future work, we aim to enrich the proposalwith additional guidelines for writing scenarios for each stage. Additionally, fur- ther experimentation is necessary before we make the approach more complex.However, we firmly believe that the approach should be strengthened and mademore robust.

**Acknowledgements** This paper is partially supported by funding provided bythe STIC AmSud program, Project 22STIC-01.

## References

1. Ahmed, U.: A review on knowledge management in requirement engineering. IEEE **5** (2018)
2. Cardenas, H., Zimmerman, R., Viesca, A.R., Al Lail, M., Perez, A.J.: Formal uml-based modeling and analysis for securing location-based iot applications. In: 2022 IEEE 19th International Conference on Mobile Ad Hoc and Smart Systems (MASS). pp. 722–723 (2022). https://doi.org/10.1109/MASS56207.2022.00109
3. Carrol, J.: Five reasons for scenario-based design. In: Proceedings of the 32nd Annual Hawaii International Conference on Systems Sciences. 1999. HICSS-
4. Abstracts and CD-ROM of Full Papers. vol. Track3, pp. 11 pp.– (1999). https://doi.org/10.1109/HICSS.1999.772890
5. Carvalho, R.M.: Dealing with conflicts between non-functional requirements of ubi-comp and iot applications. In: 2017 IEEE 25th International Requirements Engi- neering Conference (RE). pp. 544–549 (2017). https://doi.org/10.1109/RE.2017.51
6. CPS2023: Cyber-physical systems. https://ptolemy.berkeley.edu/projects/cps/(2023)
7. Dhanaraju, M., Chenniappan, P., Ramalingam, K., Pazhanivelan, S., Kaliaperu- mal, R.: Smart farming: Internet of things (iot)-based sustainable agriculture. Agri-culture **12**(10), 1745 (2022)
8. El-Gendy, S., Azer, M.A.: Security framework for internet of things (iot). In: 2020 15th International Conference on Computer Engineering and Systems (ICCES). pp. 1–6 (2020). https://doi.org/10.1109/ICCES51560.2020.9334589
9. Gupta, M., Abdelsalam, M., Khorsandroo, S., Mittal, S.: Security and privacy in smart farming: Challenges and opportunities. IEEE Access **8**, 34564–34584 (2020). https://doi.org/10.1109/ACCESS.2020.2975142
10. Iqbal, W., Abbas, H., Daneshmand, M., Rauf, B., Bangash, Y.A.: An in-depth analysis of iot security requirements, challenges, and their countermeasures via software-defined security. IEEE Internet of Things Journal **7**(10), 10250–10276 (2020). https://doi.org/10.1109/JIOT.2020.2997651
11. Kammüller, F., Augusto, J.C., Jones, S.: Security and privacy require- ments engineering for human centric iot systems using efriend and is- abelle. In: 2017 IEEE 15th International Conference on Software Engineer-ing Research, Management and Applications (SERA). pp. 401–406 (2017). https://doi.org/10.1109/SERA.2017.7965758
12. Karaduman, B., Mustafiz, S., Challenger, M.: Ftg+pm for the model-driven devel- opment of wireless sensor network based iot systems. In: 2021 ACM/IEEE Inter- national Conference on Model Driven Engineering Languages and Systems Com- panion (MODELS-C). pp. 306–316 (2021). https://doi.org/10.1109/MODELS- C53483.2021.00052
13. Khamaiseh, S., Xu, D.: Software security testing via misuse case modeling.In: 2017 IEEE 15th Intl Conf on Dependable, Autonomic and Secure Com- puting, 15th Intl Conf on Pervasive Intelligence and Computing, 3rd IntlConf on Big Data Intelligence and Computing and Cyber Science and Tech- nology Congress (DASC/PiCom/DataCom/CyberSciTech). pp. 534–541 (2017). https://doi.org/10.1109/DASC-PICom-DataCom-CyberSciTec.2017.98

14. Khan, N., Ray, R.L., Sargani, G.R., Ihtisham, M., Khayyam, M., Ismail, S.: Current progress and future prospects of agriculture technology: Gateway to sustainable agriculture. Sustainability **13**(9), 4883 (2021)
15. Martin, T., Geneiatakis, D., Kounelis, I., Kerckhof, S., Nai Fovino,I.: Towards a formal iot security model. Symmetry **12**(8) (2020). https://doi.org/10.3390/sym12081305, https://www.mdpi.com/2073-8994/12/8/1305
16. Myagmar, S., Lee, A.J., Yurcik, W.: Threat modeling as a basis for se- curity requirements. In: Symposium on Requirements Engineering for Infor- mation Security (SREIS). University of Pittsburgh (August 2005), http://d- scholarship.pitt.edu/16516/
17. Nguyen, X.T., Tran, H.T., Baraki, H., Geihs, K.: Frasad: A framework for model- driven iot application development. In: 2015 IEEE 2nd World Forum on In- ternet of Things (WF-IoT). pp. 387–392 (2015). https://doi.org/10.1109/WF- IoT.2015.7389085
18. Ojo-Gonzalez, K., Bonilla-Morales, B.: Requerimientos no funcionales para sis- temas basados en el internet de las cosas (iot): Una revisión. I+ D Tecnológico **17**(2), 30–40 (2021)
19. Ozkaya, O., Ors, B.: Model based node design methodology for secure iot applica- tions. In: 2018 26th Signal Processing and Communications Applications Confer- ence (SIU). pp. 1–4 (2018). https://doi.org/10.1109/SIU.2018.8404490
20. Pal, S., Hitchens, M., Rabehaja, T., Mukhopadhyay, S.: Security requirements for the internet of things: A systematic approach. Sensors **20**(20), 5897 (2020)
21. Potts, C.: Using schematic scenarios to understand user needs. In: Proceed- ings of the 1st Conference on Designing Interactive Systems: Processes, Prac-tices, Methods, and Techniques. p. 247–256. DIS ’95, Association for Computing Machinery, New York, NY, USA (1995). https://doi.org/10.1145/225434.225462, https://doi.org/10.1145/225434.225462
22. do Prado Leite, J.C.S., Hadad, G.D.S., Doorn, J.H., Kaplan, G.N.: A scenario construction process. Requirements Engineering **5**, 38–61 (2000)
23. Python: Python framework. https://www.python.org/ (2023), accessed: 2023-03- 11
24. Python: Textblob library. https://pypi.org/project/textblob/ (2023), accessed: 2023-03-11
25. Rose, D.C., Wheeler, R., Winter, M., Lobley, M., Chivers, C.A.: Agriculture 4.0: Making it work for people, production, and the planet. Land use policy **100**, 104933(2021)
26. Schneier, B.: Cryptography is harder than it looks. IEEE Security & Privacy **14**(1), 87–88 (2016). https://doi.org/10.1109/MSP.2016.7
27. Serna M, E., Serna A, A.: Proceso y progreso de la formalización de requisitosen ingeniería del software. Ingeniare. Revista chilena de ingeniería **28**(3), 411–423(2020)
28. Shankar, P., Morkos, B., Yadav, D., Summers, J.D.: Towards the formalization of non-functional requirements in conceptual design. Research in engineering design**31**, 449–469 (2020)
29. Slovenec, K., Vuković, M., Salopek, D., Mikuc, M.: Securing iot services basedon security requirement categories. In: 2022 International Conference on Soft- ware, Telecommunications and Computer Networks (SoftCOM). pp. 1–6 (2022). https://doi.org/10.23919/SoftCOM55329.2022.9911319
30. Sotoudeh, S., Hashemi, S., Garakani, H.G.: Security framework of iot-based smart home. In: 2020 10th International Symposium onTelecommunications (IST). pp. 251–256 (2020). https://doi.org/10.1109/IST50524.2020.9345886
31. Spacy: Spacy library. https://spacy.io/ (2023), accessed: 2023-03-11
31. Y.2060, I.T.: Itu-t y.2060. https://handle.itu.int/11.1002/1000/11559 (2012)
32. Yazdinejad, A., Zolfaghari, B., Azmoodeh, A., Dehghantanha, A., Karimipour, H., Fraser, E.D.G., Green, A.G., Russell, C., Duncan, E.: A review on securityof smart farming and precision agriculture: Security aspects, attacks, threats and countermeasures. Applied Sciences (2021)

# Application of the Methodological Framework forDesigning Collaborative Systems with Awareness Mechanisms. Case study of the Miski bakery

**Arboleda. J.D.** [1] **César A. Collazos**[2]

[1] Universidad del Cauca, Facultad de Ingeniería Electrónica y Telecomunicaciones. Maestría en Computación. Popayán, Cauca. juanarboleda@unicauca.edu.co

[2] Grupo IDIS, Universidad del Cauca, Facultad de Ingeniería Electrónica y Telecomunicaciones. Popayán, Cauca, ccollazo@unicauca.edu.co

**Abstract.** In the commercial processes of companies specialized in the elabora- tion of customized cakes, a negotiation activity is performed with the client to collaboratively create the design of the decoration of the product. Currently, thisactivity is performed through different digital communication channels, mainly WhatsApp chat and social network chats such as Facebook and Instagram, as these allow the exchange of images, facilitate synchronous or asynchronous com- munication, and a record of the conversations remains in the history. The currentdynamics of e-commerce mark a trend oriented to the digital transformation of business processes, which is why the need to design a software tool that facilitatescomputer-assisted cooperative work has been identified. The objective of this re-search paper is to apply a methodological framework to incorporate Awareness mechanisms to the design of a software that will provide strategic support in thecollaborative negotiation process with the purpose of improving communicationand shared understanding among the actors of the pastry shop. To achieve the objective, the methodological framework proposed by the Awareness theory is applied for the development of the groupware, which is composed of five stages:Identify the objective of the Awareness information in the process to be inter- vened. Identify the components of the Awareness information in the communi- cation context. Modeling of the software process to achieve an integration of Awareness mechanisms in the business process to be intervened. Design of the information distribution strategy among the actors involved in the process. Cre- ation of the user interface with Awareness mechanisms.

## 1 Introduction.

### 1.1 Context.

In Colombia in 2022 it was determined [1] that 80% of Micro, Small and Medium En-terprises (MSMEs) are in a state of digital transformation, which consists mainly in thetactical appropriation of technologies that allow them to be more competitive, impro their productivity levels and better adapt to the dynamics, trends and challenges im- posed by e-commerce.

For the case study, a microenterprise in the city of Popayán called Miski Pastelería was selected. The bakery is currently conducting a diagnosis of the digital status of the business [2] in order to start the process of formulating the digital transformation path, which will allow it to implement technological solutions relevant to its cultural context [3]. An analysis of all the processes involved in the elaboration of a customized cake in the Miski bakery was carried out. To facilitate the interpretation of its context, a modeling of the processes is made in Business Process Modeling Notation (BPMN)[4], which allows to identify graphically and categorically, the beginning (Customer Need), the three macro-processes: Strategic (Commercial), Productive, Support, and the end of the process (customer satisfaction).

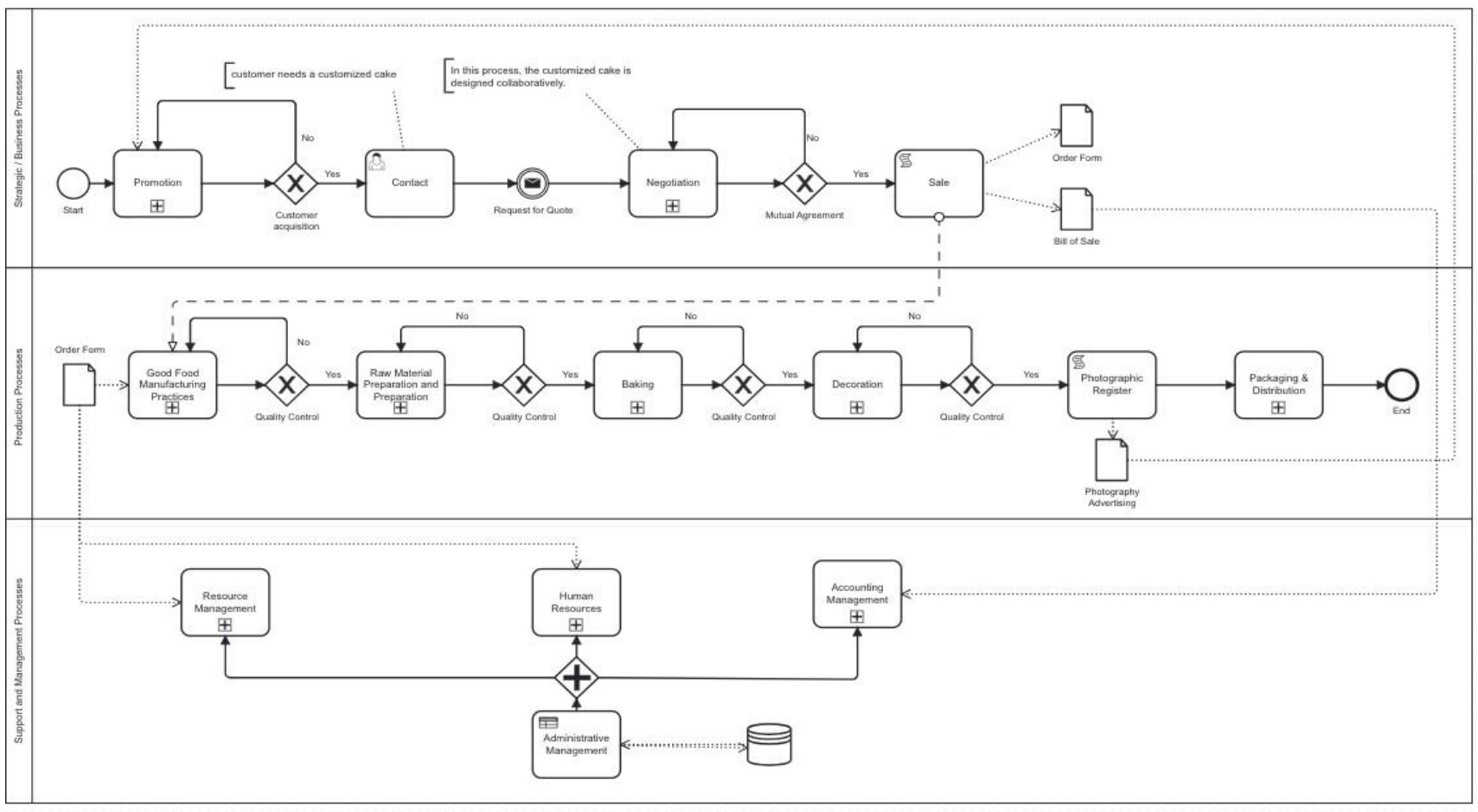

***Fig. 1.*** *Model of the pastry processes.*

### 1.2 Brief description of the pastry processes.

The process starts with the customer's need for the bakery. Regarding the needs of the pastry shop's customers, it was identified that 81% of the customer archetype cor- responds to women, 70% of whom are mothers, and 57% are between 25 and 34 yearsof age. These customers need 60% of the time a birthday cake for children between 0 and 10 years old. The design of a customized cake is the result of a collaborative processbetween the client and different members of the bakery, which is why each product is unique and unrepeatable.

Strategic Processes (Business Processes). This macro-process is composed of four processes: Promotion, Contact, Negotiation and Sale. The Promotion process corre- sponds to all advertising activities, positioning and visibility in social networks and Internet search engines. In addition, marketing strategy activities (Get, Keep, Grow) are carried out to capture the interest, loyalty and recommendation of the customer ar-chetype described. The contact process is developed through various communication channels, the customer can go directly to the store and talk to the seller or can use digitalmedia when they are geographically dispersed. In the Miski pastry shop 60% of con- tacts are made through the chat of (WhatsApp 40%, Instagram 35%, Facebook 25%), 10% by e-mail, 20% by phone calls, and 10% through the web. The activity carried out by the company in this process is to apply the customer service protocol; which, mainly consists of giving timely response to the quotations made and helping the cus- tomer to make decisions in front of the product. The customer frequently uses images of reference cakes that he/she has previously searched for on the Internet, or that he/shehas seen in the bakery's catalog. The image is a support tool that allows the customer to express his desires, expectations and requirements for the customized product, and satisfies the limitations of natural language representation [5].

In the Negotiation process, a mediation is made between the customer's requirements(imagined cake object of desire), the company's production capacity, and the customer'spurchasing power (what he is willing to pay) this activity is known as quotation. This negotiation is a collaborative process [6], as the customer, the vendor and the pastry chef have a common goal, to design a customized cake decoration for a special event.

The negotiation process ends when a mutual agreement is reached regarding the for- mal characteristics of the product and its price. It is the duty of the bakery's salespersonto initiate the activities that correspond to the formal survey of the requirements of thecustomized cake. The requirements gathering is documented through the use of a paperform called an order form. The order form is in charge of storing the customer infor- mation (Id, contact...), the product delivery and distribution information (date, time...),and the information of the formal characteristics of the cake (Size, Flavor, Color, Typeof Decoration).

The productive macro-processes correspond to the materialization of the design of the cake product of the negotiation. To carry out this task, seven processes are carried out, categorized as follows: Production Process 1: Good Manufacturing Practices. Pro- ductive Process 2: Preparation of raw materials. Productive Process 3: Baking. Produc-tive Process 4: Decoration. Productive Process 5: Photographic Registration. Produc- tive Process 6: Packaging and storage. Productive Process 7: Distribution. At the end of the productive macro-processes, the project of elaboration of a personalized cake isfinished, thus achieving the satisfaction of the needs of the company's client. To guar- antee the success of the productive macro-processes, there are the support macro-pro- cesses, whose core is composed of administrative processes, accounting management processes, human management processes, and resource management processes.

### 1.3 Description of the observed problem.

Particularly, ambiguities have been identified in the use of natural language to de- scribe the chromatic characteristics of the cake, for example: "I need it to be blue", theproblem is that, with the colorants used in food, more than 50 different shades of blue can be achieved, therefore, there is a high probability that the blue color imagined by the baker is different from the color imagined by the customer. The decoration of a personalized cake obeys the theme of the event, and is loaded with a high symbolic [7]and emotional value [8], the emotional aspects of cake decoration make the client have very high design expectations, which generally imply a high degree of technical diffi- culty for the decorator, which translates into a high selling price of the product, and sometimes customers are not willing to pay such a high value, which is why the cus- tomer with the help of the seller and the pastry chef begin to make more realistic ex- pectations of the product and initiate a process of co-creation to design the right cake for the customer's needs and budget, and that is within the production capabilities of the company.

The paper format plays a fundamental role in the production processes; in addition to storing the customer's requirements, it is a communication object [9] that interacts with the different actors, processes and production units of the bakery. It was observedthat copies of the order form are generated and sent to the accounting processes, wherethe information is recorded in the cash flow, daily balance, and income forms. Anothercopy goes to the client and is delivered as an invoice/ticket to receive the cake. The lastcopy goes to the production processes to be interpreted by the pastry chef/decorator [10].

When the pastry chef must read and interpret [11], [12] the paper format, problems of understanding arise due to different factors such as: the order form must be filled outin calligraphy by the salesperson, the salesperson must fill in more than 90 boxes wheredifferent data are requested to create a written description of the customer's wishes, thelayout of the visual elements that make up the order form are subject to the size of the paper, This results in illegible texts due to the size of the font, or tight spaces, which, when filling out manually, force the overlapping of lines when writing, or that the writ-ten words overflow the margins of the form. These factors make it difficult for the baker to interpret the requirements, since the elicitation of requirements through paper formsis constantly susceptible to human error.

The requirements documentation process, which is carried out in a traditional (hand-made) way, causes frequent spelling errors, legibility problems and crossings out, prob-lems for easy reading and interpretation due to the high rate of grammatical, syntax, semantic or pragmatic errors made by the person in charge, which directly affects the production processes, hinders mental processes related to hermeneutics [13] and there-fore reduces the understanding of the tasks to be performed by the pastry chef/decora-tor. This causes that the customized product is not elaborated under the requested re-quirements, and, therefore, generates a high degree of dissatisfaction, frustration and sadness [11] in the customer at the moment of receiving the product. For traditional bakeries, this situation manifests itself as a critical problem of communication and un-derstanding between production units, since the main processes depend on paper for- mats [10]. Therefore, it is presumed that the design of these formats does not allow thegeneration of a representation of an efficient mental model of the user [10], thus causingconfusion, reprocessing and errors in repetitive tasks.

To approach this problem, Section 2 of the article establishes a conceptual frame- work, which allows us to analyze the state of the art of the existing literature on the situation of the pastry industry. Section 3 defines a methodological approach to identify and apply awareness mechanisms in collaborative processes. In section 4, the method-ological framework defined is applied in the Miski bakery. Section 5 discusses the re- sults, and section 6 discusses future work.

## 2 Conceptual Framework.

It is necessary to investigate the existing literature to identify trends in technologies and methodologies that can contribute to the closing of technological gaps identified inthe communication processes where customers and micro enterprises of the niche cor-responding to Bakeries / Pastry Shops participate collaboratively.

The aim is to provide a possible solution approached from a scientific (Computing) and methodological (Design) approach, to the communication problems that affect thedifferent actors when working as a team, due to cultural aspects [14] representation andunderstanding, which arise in business processes, and hinder understanding when cap-turing data related to the requirements for the development of customized cakes (FoodIndustry).

*Nucleus 1: Computer Science.* The research in this area of knowledge focuses on software engineering. Emphasis is placed on the search of scientific literature related to the processes of elicitation and requirements analysis. In addition, it is important toinquire about modeling languages. It is desired to know the application of models, methods and tools provided by collaboration engineering; therefore, it is important to investigate about the concepts related to Groupware, Awareness, Thinklets and spe- cially to shared understanding in virtual environments.

*Nucleus 2: Communication and graphic design.* It is very useful to learn about meth-odologies and processes related to user-centered design (UCD) and culture-centered design (CCD), since these methodologies allow a detailed and empathetic understand-ing of the context of the actors involved in the process to be intervened. In addition, these methodologies favor the creation of software models with better levels of com- munication and understanding between users, thus generating a better interaction.

*Nucleus 3: Business Administration / Food Industry.* In this area of knowledge, it isof interest for the research to analyze the trends, dynamics and processes of companiesin the gastronomic sector, particularly in the market niche corresponding to pastry shops specialized in the elaboration of customized cakes, for this reason, it is necessaryto investigate the existing literature on the food industry and the technological innova-tions that exist in this field such as 3D printing of food, biodegradable packaging, and environmentally friendly manufacturing. It is necessary to know case studies where optimization or automation of production systems is performed through software, to analyze the impact of digital transformation in supply chains, in the manufacture of customized artifacts, and in customer service processes, and thus, to obtain references of technologies applied to the food industry.

**Systematic literature mapping.** For the selection of the subject of the systematic map-ping a research protocol [15] consisting of three phases was designed: First, the prepar-atory phase, whose activities consist of: Identifying the thematic nuclei related to the situation or problem to be addressed. Perform a critical and strategic analysis of each thematic nucleus. Analyze the key words, categorize them and make a list of selected words. Second, execution phase: Create and apply inclusion and exclusion criteria. Construct the final search string. Execute the search string. Select the database. Compile found studies. Narrow down articles. Perform snowball sampling. Third, documen-tation phase: Categorize articles by level of relevance and contribution. Conduct a lit- erature review. Analysis of main findings and trends. Draw conclusions.

The key words of each core are listed and categorized from the most general to the most specific thematic areas according to their level of contribution to the project. A superficial search is performed in the meta-search engine Scopus, using the filter: (Search within: Article title, Abstrack, Keywords). This work was carried out for eachterm to estimate its potential in terms of number of publications. For the construction of the search string, the keywords considered to be of greatest relevance and contribu- tion to the research should be selected, and it should also be verified that there is a considerable number of articles.

**Search chain.** (( ( "Awareness" AND "groupware" AND "CSCW" AND "Work- space" ) ) OR (("Cake design" OR (“Gastronomy” AND “Supply Chain”)) OR ("Foodindustry" AND "Bakery" AND "software") OR (“collaborative cooking” OR “cake Decoration”)) OR ((“Semiotic Engineering” AND “Usability”) OR (“User Experience”AND “Web Design” AND “User Interfaces” AND "User Centered Design") OR (“Se-miotic Engineering” AND “collaborative design”))) ANDNOT (("health" OR "Farm- ing" OR "farm" OR "Agriculture" OR "Chemical" OR "tourism" OR “Video” OR “Medical” OR “chemistry” OR “Healthcare” OR “School” OR “Medical images” OR“fruits”))

A total of 114 related documents were obtained as a result of applying the search string. An analysis was carried out and those articles considered to contribute to the interests of the project were selected. Twenty-three articles were identified that are aligned with the research interests of the project. This is equivalent to 20.17% of the total number of documents resulting from the search chain. The articles were evaluatedand categorized according to their level of contribution in order to narrow down the research. At the time of narrowing down the 23 resulting articles, the following resultswere obtained: 6 articles categorized as being of greater relevance and contribution (26.08%). 7 articles were identified as highly relevant (30.43%). 5 articles defined as medium relevance (21.73%). 5 articles identified as low relevance (21.73%). It can beconcluded that 56.51% of the articles delimited are considered for in-depth literature review. In the process of analysis and documentation of the findings, researches such as CakeVr [16] validate the relevance of the communication problem identified in the collaborative activities of cake shops. A language pattern for collaborative cooking [17], and interactive cake decorating models [18] were identified. The most important theoretical contribution is related to the methodological framework for designing col- laborative systems with awarenes mechanisms [19].

On the other hand, in the area of software process engineering [20], the main contri-bution is in the understanding and identification [21] of critical processes that can be supported through software. In the area of Semiotic Engineering the contributions are related to the construction of graphical interfaces oriented to communicability and un- derstanding [22] based on elements of Culture-Centered Design [22] & User-CenteredDesign [23] and tools were discovered to perform an emotional evaluation of the usa- bility [24], Interaction [25] and Accessibility of the software model.

## 3 Methodological Framework.

Collaborative work involves the effort of two or more people to achieve a commongoal [26], collaborative work increases group efficiency, fosters social relationships and improves individual well-being [27] Understanding the characteristics of group work enables the design of appropriate information technology to support collaborative work processes. CSCW studies the impact of technology on group interaction within shared virtual spaces [19] For efficient interaction, it is important that members feel part of the group and obtain sufficient information to perceive, understand and adapt tothe shared work environment. This contextual information, which group members needto reduce metacognitive effort, is known as awareness [19] Awareness mechanisms require the identification of the type of information that is needed for the group mem- bers to understand and adapt to the shared work environment.

Awareness mechanisms require identifying the type of information to be provided, information distribution strategies, and a method of representing contextual infor- mation. Group members interact through artifacts or direct communication channels and must be aware of other group members (activities, changes, states...). In groupware,people can communicate information either explicitly, or implicitly through interaction with shared artifacts, which involve non-verbal, more symbolic communication.

The methodological framework for designing collaborative systems with awarenessmechanisms [19] is based on five stages: Awareness goals and support: consists of identifying the type of information that the members of the work group should perceive, and the context in which it is produced. Awareness Information identification: allowsto obtain a semantic interpretation of the information model and to give it meaning. Modeling, it is necessary to define and model the information to be provided and its content once the awareness objectives have been established. Distribution allows to have an explicit knowledge of the information and its mechanisms for socialization. awareness User Interfaces: define how the awareness information is going to be repre-sented to be used.

## 4 Application of the Methodological Framework to design collaborative systems with awareness mechanisms for theMiski bakery.

The methodology is applied to the case study of the pastry shop Miski in the city ofPopayán.

### PHASE 1: Awareness Information Goals.

It is necessary to determine the most im-portant type of Awareness to successfully perform the activity of negotiating the re- quirements of the customized product. To design the decoration of a cake in a collabo-rative manner requires synchronous or asynchronous interaction between the three mainactors identified in the process, which are geographically dispersed: the customer, the seller and the baker. At the beginning of this process, it is necessary for the client to obtain information regarding the status of the other members of the group, i.e. the client wants to know if there is another human being accompanying him/her in the process, who that person is, what expertise he/she has, what position he/she holds, will he/she understand my tastes and desires for the cake, etc. Another relevant factor is the level of urgency with which the customer needs the product, if the level is low, the commu-nication is generally asynchronous, but if the level of urgency is high, the communica-tion is synchronous.

The cultural context in which the customer's special event takes place directly influ-ences the customer's imaginary construct of the cake. Due to the limitations to describethe cake through natural language, the client uses images of reference, which visually allow him to make explicit his requirements regarding the product, for this reason, it isimportant that the client is informed about the shared work space, because that is wherethe decoration design is going to be made, therefore, first, he must be able to visualizethe cake (work area), second, the client must know the different tools and decoration possibilities that the software allows (context), as well as its restrictions and third, the client must have a practical understanding of each of the steps (Activities) that are fol-lowed for the elaboration of a personalized cake, thus, through the information providedby the software, the client is guided or immersed in the decoration process. Each ele- ment used in the workspace is a visual metaphor that emulates the elements and utensilsused in the decoration station, therefore, it must provide a context and a background meaning, so that the client can use them and thus achieve the desired design, also gen-erating a good user experience.

Every time a client generates a design, this information is available so that the pastrychef, from his perspective and expertise, can make recommendations to improve the aesthetic quality of the product, to achieve this activity, it is necessary to have clear information about the preferences and interests shared by the client.

***Table 1.*** *Different types of awareness in pastry business processes.*

| Awareness | Importance | Explanation |
|---|---|---|
| Group | High | The customer wants to know about her cake design partners. |
| Workspace | Very High | The customer must be able to visualize the cake and interact with the decoration elements in the shared workspace. |
| Context | Very High | The customer must be familiar with the tools and elements that make up the shared workspace. |
| Peripherial | Low | For the time being, they are not considered important. |
| Activity | Very High | For the pastry chef, it is important to know and visualize the customer's requirements and the latest modifications to the product. |
| Availability | High | The customer wants to know when the salesperson or pastry chef is available to provide support if needed. |
| Perspective | Very High | The salesperson needs to capture and understand the customer's expectations and desires for the cake. |
| Community | High | The customer expresses his ideas in front of the cake, the baker helps to make aesthetic decisions and the salesperson calculates and informsthe costs of the product. |
| Presence | Medium | the salesperson must be informed each time a customer enters the shared workspace. |
| Rhythm | Medium | The customer wants to track the processing status of his product. |

## PHASE 2: Awareness Information Identification.

Component 1: People.

*2.1.1. People's Structure: Social norms, Conventions, and roles.* In the negotiation process, the three main actors of the process (Client, Pastry Chef, Salesperson) interactdirectly. The salesperson must act according to customer service protocol, must respondas quickly as possible, and must be very courteous and polite following social norms. The salesperson is the mediator between the natural language used by the customer andthe technical language used by the pastry chef and must also capture all the customer'srequirements for the product in order to calculate the costs. The pastry chef is the expertin decoration, his role is to guide the customer in the decoration decisions, to achieve aesthetic quality levels, and to capture all the requirements, expectations and desires ofthe imagined product. The client is the actor who initiates the process, needs a cake, gets in touch, shares information about how he imagines the product object of desire, and expects his expectations to be fully satisfied. In short, the customer communicateshis ideas, the pastry chef cooperates with his expertise in the design of the cake deco- ration, and the seller coordinates the requirements, costs, and available resources.

*2.1.2 People's State: Availability, activity and emotions.* The salesperson needs to beinformed about the customer's emotions during the interaction with the shared work- space artifacts, e.g., is everything going well, is the customer confused, does the cus- tomer need help? The customer must know if he/she has the company of a human in the process. The pastry chef must understand the customer's need for support and knowthe degree of satisfaction with the product decoration. All stakeholders must visualize the current status of the customized cake.

*2.1.3. People's Location: Presence, distance, visibility, space-place, and metaphors.* The negotiation process is done through digital media because they are geographicallydispersed, the shared workspace to be designed for the software is a metaphor of the processes, it is based on the production logic followed by the company, it simulates inthe virtual environment the elements of decoration, the utensils that are used in the company, and the decoration space, which gives context to the artifacts with which thecustomer interacts when designing the cake.

*2.1.4. People's actions: events and distortion (aggregation, animation, etc.).* The pastry chef must be informed and updated regarding changes made by the client to thedecoration of the product. Likewise, he/she must be informed about the raw material inventories that are necessary for the elaboration of the product, he/she must know theavailability of human resources, time and raw materials during the negotiation processwith the client in order to determine the production capacity of the company. The Clientmust be aware that each action performed in the cake decoration activity in the sharedspace work has a cost, for example; each figure or decoration motif that is added to thecake has a price, therefore, the more figures, the higher the monetary value of the cake.The client with his actions, and if he is well informed, can adjust the price of the cake to his purchasing power or established budget. The salesperson in his role as coordina-tor is in charge of generating the order form according to customer requirements and decoration elements used in the shared space, and his duty is to guide the customer's actions so that the process is satisfactory.

*2.1.5. People's activity: activity and goals*. The software design is based on the pro-duction logic where first the size of the cake is chosen from different options ranging from ¼ of a pound (8 portions for the client) to 4 pounds and can be from 1 floor to 3,also its shape is defined (round or square). Second, the flavor is chosen, where the cus-tomer must select their favorite flavors, and even make combinations, or choose any type of filling available. Third, the customer must choose the base color, and finally thetype of decoration is chosen, each activity is sequential, and the change of activity, tasksto be performed, and the tasks performed, or achievements obtained are explicitly re- ported.

## 2.2. Component 2: Task or project.

*2.2.1. Task structure: process planning.* The salesperson should guide the customerand let him/her know how the shared workspace behavior works. The client in the shared workspace must be able to easily understand the dynamics and mechanics of decorating a cake virtually. The pastry chef must know the complexity of the decorationto be able to organize his work team in production (number of people in the team, ex- pertise of the work team).

*2.2.2. Task or project state: state-based workflow.* The customer must obtain infor-mation regarding each phase of decoration performed, and must be able to evaluate theinteraction experience, and ask for help if necessary. The salesperson must follow up and must know when all the tasks have been completed in order to generate the order form, and the Pastry Chef needs to be updated to know in which task the customer needssupport, or if he/she can start the productive macro-processes.

## 2.3. Component 3: Resources.

*2.3.1. Resource structure: spatial and semantic networks.* In the shared virtual work- space, interaction artifacts that emulate furnishing elements and utensils are in charge of providing information regarding their use, value and formal characteristics. It is thetask of the vendor to familiarize the customer with the concepts assigned to each inter-active artifact, creating a network of meanings associated with each phase of cake dec-oration. Being able to visualize the cake in the shared work area allows to generate a shared understanding among the actors involved in the process.

*2.3.2. Resources state: availability.* Make informed decisions and plan contingent actions by knowing the state of resources, e.g., knowing how many decoration elementsare being used and their cost.

*2.3.3. Resources location: availability and resource discovery.* In the shared virtual space, artifacts are virtual representations of physical objects that are available for the customer to decorate his cake.

## PHASE 3: Modeling.

To model a collaborative process in UML, a collaboration dia- gram can be used. This diagram shows how the elements interact with each other to achieve a common goal. First, the key elements must be identified: start by identifyingthe key elements involved in the collaborative process, such as users, systems and re- sources. Then a collaboration diagram is created. Add awareness information elements to the diagram. The interactions between the elements are established, showing the communications and data exchanges that occur during the collaborative process. Then labels are added to the communications and data exchanges to describe the type of mes-sage being sent. implement constraints. If there are constraints on the process, such as response time or resource availability, the diagram is analyzed for constraints. The di-agram is reviewed and validated.

It is important to keep in mind that collaborative processes are often complex and may require several collaboration diagrams to fully model all interactions. It is also important to document the collaborative process in a specification or requirements doc-ument so that all stakeholders can clearly understand how it works.

## PHASE 4: Distribution.

The following are used for the distribution of awareness in- formation:

*4.1 Notification Mechanisms.* Notification mechanisms are very important to guide the client during the decoration process through the graphical interface.

*4.2. User-driven discovery.* The information design allows the user to discover within the interface the Logical Model to decorate a cake, the client is guided through a series of steps.

## PHASE 5: Awareness User Interfaces.

A prototype of a graphical interface is de- signed where the awareness mechanisms are applied. The interface is a metaphor that simulates in a virtual way the work space that exists in a pastry shop, simulates the toolsand utensils necessary for the decoration, and invites the client to follow a collaborativedesign process, where he has the possibility to interact and visualize 3D graphics withthe rendering of the cake decoration. The interface must be composed of the followingawareness mechanisms.

*People structure:* It is necessary to identify the role of each actor in the cake designprocess. For the client it is important to differentiate the seller from the baker. The graphic interface will make use of avatars to identify each actor, and the tasks in whicheach actor can actively collaborate, and their hierarchy will be presented in a kind of organization chart.

*People's status:* Different people's statuses can be represented, the first one is the availability status, which determines if the communication is going to be synchronousor not. When the actor is available, a green circle will light up around the avatar image,blue when the actor is busy or slow to respond, and gray when the actor is not available.It is important to evaluate the emotional states of the actors, mainly that of the custom-ers, since they are not as familiar with the interface as the bakery workers. In this casethe avatar will have the ability to represent emotions through its facial gestures, so it can represent emotions that can be generated at the time of design as: Confusion if youdo not understand the task to be developed, or can represent sadness if the customer isfrustrated because it fails to successfully complete a task, or can represent happiness, ifeverything is going in order. If a group member notices a negative emotion in the cli- ent's avatar, it is his or her duty to go immediately. To generate support for the task being performed in order to generate a good user experience.

*Placement of people*: The idea is to generate a visual metaphor in the virtual envi- ronment that simulates the cake decorating station, so that the customer can interact with the design artifacts and can immersivity decorate the cake.

*People actions:* The customer can logically interact with the artifacts that are used in the cake decorating process, each interaction with a design element. Each time an object is interacted with, it responds by enlarging its size to make it easy to identify andmanipulate. In the interface a diagram is designed that represents the activities and tasksand indicates in certain tasks what actions have been performed by the actors in the process.

*People's activity:* An activity log is generated that allows the actors to track changes and modifications made in each activity and its corresponding tasks, this allows groupmembers to be informed of developments against the product design.

*Task structure:* A structure of activities is created whose workflow is similar to the decoration process in the real environment, through four basic steps such as choosing the size, choosing the flavor, choosing the topping and its color, and choosing the dec-oration design. Each of the activities is supported by a series of tasks that allow at the end of the process to have an accurate visual representation of the requirements that thecustomer has in front of the cake.

*Resource structure:* Each task is related to a particular type of resource, for example,when choosing the flavor of the cake, the available flavors must be shown in the inter-face, and so each structured activity is in charge of its own resources.

*Status of resources*: For the customer it is important to know the availability of dec-oration elements that he has to achieve his goal, he must also know the value of each element, for this task, an element counter is designed that adds up the price of each object used exactly, so that the customer can know when the type of decoration chosenexceeds his budget, and so he can have control over his expenses. In the same way, the pastry chef must know what resources he has available for the elaboration of the cake.

## 5 Discussion of results.

By applying the methodological framework, it was possible to identify the actors involved, their roles within the process, their specific activities and tasks, and to cate- gorize the activities into three types: individual, cooperative, or collaborative.

By identifying the roles and tasks of each actor (Table 2), the awareness mechanismsnecessary to support the collaborative process can be established.

Table 3 shows the type of awareness mechanisms assigned to each activity of the negotiation process. The process of defining the elements of decoration is used to ex- emplify the type of awareness elements that were used at the time of designing the prototype of the client interface for this activity (Figure 2).

To satisfy the client's needs for awareness information during the process of definingdecoration elements, a low-resolution prototype of the client's interface was designed, where the semiotic representation of the awareness mechanisms applied and the type of information, they transmit can be graphically visualized (Table 4). When validatingthe graphical user interfaces designed with awareness mechanisms with the actors in- volved, a high degree of acceptance was identified by the users because the design of the metaphor that simulates the virtual environment of decorating a cake is based on business logic, and takes part of the symbolic language of the environment to generate a shared vocabulary, which allows generating a better understanding compared to the paper format, since the groupware tool to be designed allows visualizing in real time how the personalized cake looks like.

***Table 2.*** *Table of participation in the negotiation process of a cake.*

| **Activity / Roles** | **Customer** | **Salesperson** | **Pastry chef** | **type** |
|---|---|---|---|---|
| Request for quote | x | | | Individual |
| Respond to request | | x | | Individual |
| Define delivery conditions | x | x | | Cooperative |
| Define type of celebration | x | x | | Cooperative |
| Define cake flavor | x | x | x | Collaborative |
| Define cake size | x | x | x | Collaborative |
| Define decoration elements | x | x | x | Collaborative |
| Define price of the cake. | x | x | x | Cooperative |

***Table 3.*** *Table of Awareness mechanisms used in each activity within the negotiation process of a customized cake.*

| **Activity / Roles** | **Awareness Mechanisims** | | |
|---|---|---|---|
| | **Customer** | **Salesperson** | **Pastry chef** |
| Request for quote | - People´s State: Availability | | |
| Respond to request | | - People´s State: Availability<br>- People's action: Event alert<br>- People´s Activity: goals | |
| Define delivery conditions | People's action: Event alert<br>- Norms: Time constraint | - People's action: Event alert<br>- Resources state: Availability | |
| Define type of celebration | - People´s State: Availability<br>- Task Structure: workflow<br>- Task State: started / in process / done | - People´s State: Availability<br>- Task State: started / in process / done | - People´s State: Availability<br>-Task State<br>- People´s State: Emotions |
| Define cake flavor | - Task Structure: Workflow.<br>- Resources State: Availability<br>- People`s Actions:decision tracking.<br>- Task State: started / in process / done | - Task Structure: Workflow.<br>- Resources State: Availability<br>- People`s Actions:decision tracking.<br>- Task State: started / in process / done | - Task Structure: Workflow.<br>- Resources State: Availability<br>- People`s Actions:decision tracking.<br>- Task State: started / in process / done |
| Define cake size | - Task Structure: Workflow.<br>- Resources state: Availability<br>- People`s Actions:decision tracking. | - Task Structure: Workflow.<br>- Resources state: Availability<br>- People`s Actions:decision tracking. | - Task Structure: Workflow.<br>- Resources state: Availability<br>- People`s Actions:decision tracking. |
| Define Decoration elements<br><br>Define price ofthe cake. | - People´s State: Availability<br>- Task Structure: Workflow.<br>- Resources state: Availability<br>- People`s Actions: Event Modify (add, move, quit)<br>- Task State: started / in process / done<br>- People´s Action: Perspective. | - People`s Actions:decision tracking.<br>- Task Sate: Done | - People's action: Event alert<br>- People´s Action: Perspective.<br>- People´s State: Availability<br>- Task Structure: Workflow.<br>- Resources state: Availability<br>- People`s Actions: Event (modify).<br>- Task State: started / in process / done<br>- People´s State: Emotions<br>- People`s |
| | - Task Structure: Workflow.<br>- People`s Actions:price tracking.<br>- Task Sate: Done | - People`s Actions:decision tracking.<br>- Task Sate: Done | - Actions:decision tracking.<br>- People´s State: Emotions<br>- Task Sate: Done |

***Table 4.*** *Association of the information needs of the client, the assigned awareness mechanism,and its graphical representation in the interface.*

| Activity Define decoration elements | | |
|---|---|---|
| **Customer needs** | **Awarenes Mechanisims** | **Graphical representation on the interface** |
| The customer needs to know if the pastry chef is available. | - People´s State: Availability | 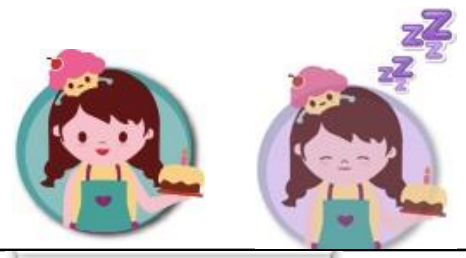 |
| The customer needs to know where he is in the process and what the tasks are. | - Task Structure: Workflow. | 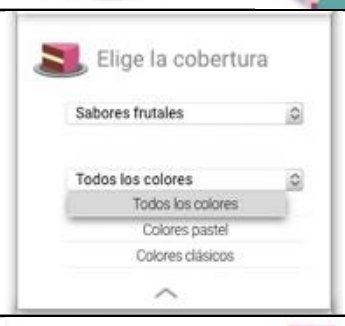  |
| The customer needs to know what resources are available to perform the task. | - Resources state: Availability | 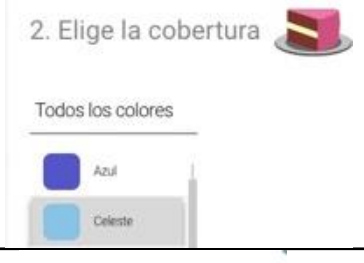  |
| The customer needs to visualize the effect of his actions. | - People`s Actions: Event Modify (add, move, quit) | 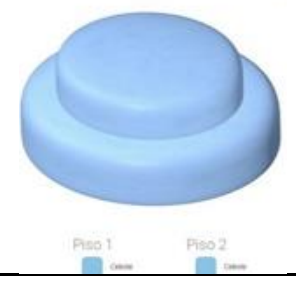 |
| The client should be aware of the cost of his choices before finalizing the process. | - Task State: started / in process / done | 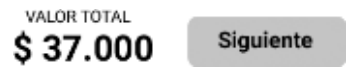  |

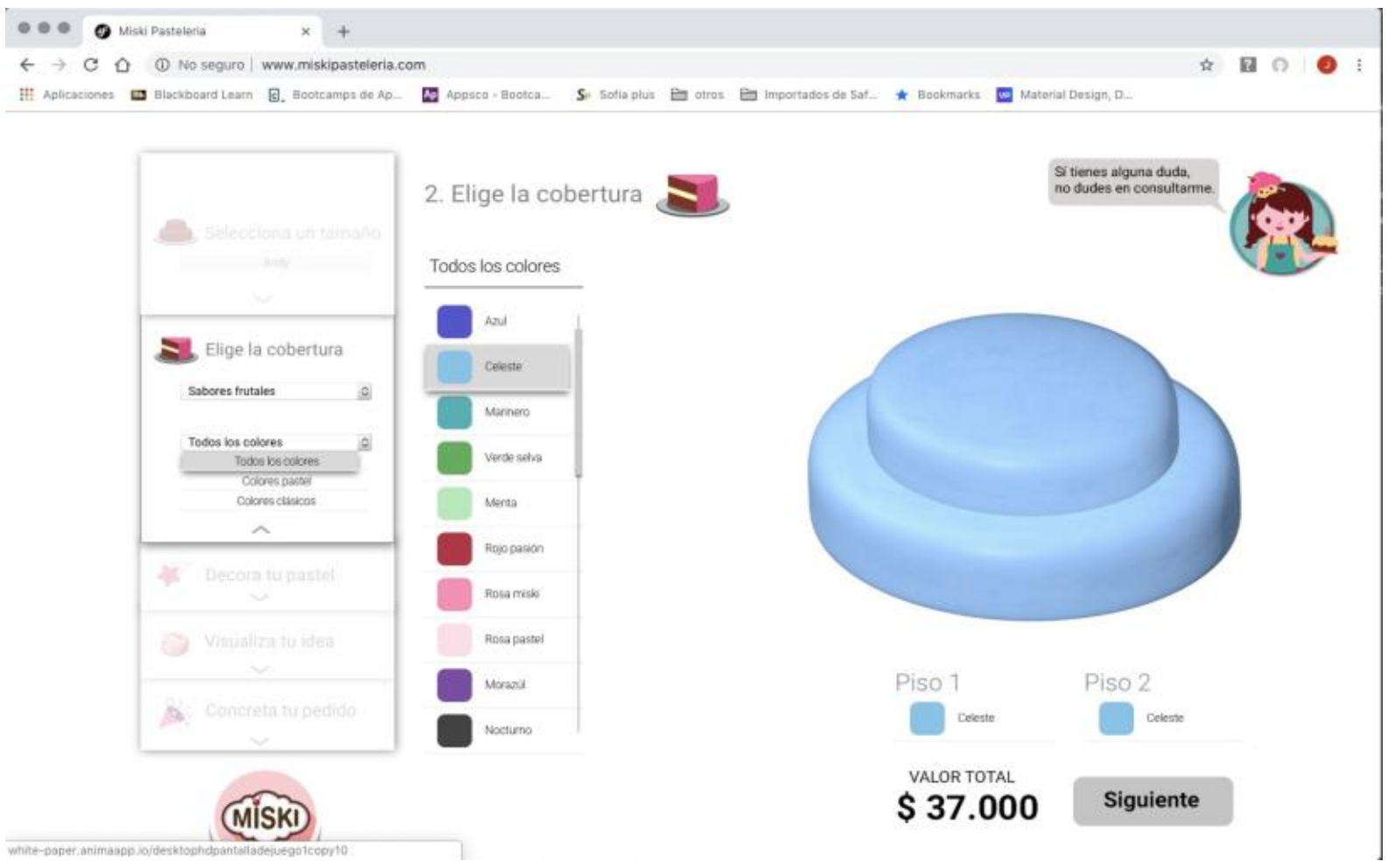

***Fig. 2.*** *prototype image of graphical user interface with awareness mechanisms.*

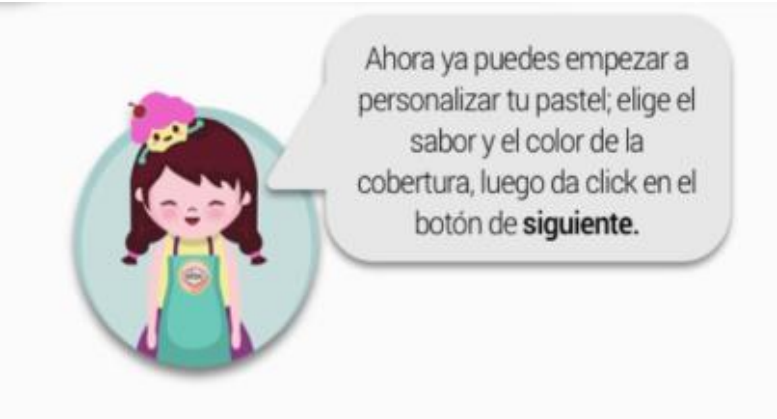

***Fig. 3.*** *Avatar representing the pastry chef during the decoration process, this awareness infor-mation mechanism allows to guide the customer during the activities.*

## 6 Conclusions and future work.

The prototype was validated with the actors of the bakery, and feedback was receivedto improve the interface and the workflow. The importance of the articulation of the natural and technical language within the processes and their graphic representation was emphasized.

Using awareness mechanisms in the negotiation process of the bakery improves theunderstanding and comprehension of the requirements of the customized cake. It allowsthe customer to make informed decisions regarding the cost of his requirements, increases motivation since he can visualize the imagined product, strengthens relationships between people since the customer does not feel alone, on the contrary, he knows that he can count on the support of experts whenever necessary and generates an improvement in satisfaction levels due to the reduction of misunderstandings causedby communication problems.

A clear example is the awareness mechanism that allows the client to be informed about the cost of his decoration decisions. The placement of this element before the "next" button (end activity) allows the client to reflect and mediate between his expec-tations and his budget before moving on to the next activity.

In the future, it is expected to be more formal in the notation and to be able to applymore precise methodologies [28], when describing collaborative processes that need awareness mechanisms within a groupware system.

## 7 References

1. MinTIC and Innpulsa, “Centros de transformación digital empresarial,” *https://www.centrosdetransfor- maciondigital.gov.co/695/w3-channel.html*, 2023.
2. MinTIC and Innpulsa, “MODELO DE MADUREZ PARA LA TRANSFORMACIÓN DIGITAL INNPULSA COLOMBIA-INNPULSA DIGITAL MINTIC-DIRECCION DE TRANSFORMACIÓN DIGITAL,” 2019.
3. C. A. Collazos and R. Gil, “Using cross-cultural features in web design patterns,” in *Proceedings - 20118th International Conference on Information Technology: New Generations, ITNG 2011*, IEEE Computer Society, 2011, pp. 514–519. doi: 10.1109/ITNG.2011.95.
4. T. Djatna, M. F. Koswara, and D. K. R. Kuncoro, “A system analysis and design for mobile digital busi-ness traceability at a food manufacturing,” in *IOP Conference Series: Earth and Environmental Science,*Institute of Physics, 2022. doi: 10.1088/1755-1315/1063/1/012051.
5. P. Bourdieu, *Capital Cultural, Escuela y Espacio Social ISBN. 968-23-2054-2,* Primera. 1997.
6. D. Marc Kilgour Colin Eden, “Second Edition Handbook of Group Decision and Negotiation.”
7. E. Johansson and D. Berthelsen, “The birthday cake: Social relations and professional practices around mealtimes with toddlers in child care,” in *Lived Spaces of Infant-Toddler Education and Care: Exploring Diverse Perspectives on Theory, Research and Practice*, Springer Netherlands, 2014, pp. 75–88. doi: 10.1007/978-94-017-8838-0_6.
8. S. R. Charsley, “Wedding cakes and cultural history ISBN 0-203-40414-9 Master e-book ISBN,” 1992.

9. H. Fuks, A. Raposo, M. A. Gerosa, M. Pimental, and C. J. P. Lucena, "The 3C collaboration model," in *Encyclopedia of E-Collaboration*, IGI Global, 2007, pp. 637–644. doi: 10.4018/978-1-59904-000- 4.ch097.
10. J. Iván and T. Arbelaez, "Generación de la interfaz de usuario de negocio a partir de patrones de negocios basadas en los fundamentos metodológicos de TD-MBUID," 2016.
11. L. Correa, "Revista Faz / N° 2 - Julio 2008."
12. S. Valtolina, B. R. Barricelli, and Y. Dittrich, "Participatory knowledge-management design: A semiotic approach," *J Vis Lang Comput*, vol. 23, no. 2, pp. 103–115, Apr. 2012, doi: 10.1016/J.JVLC.2011.11.007.
13. G. Stahl, "Group Cognition Computer Support for Building Collaborative Knowledge Acting with Tech- nology Series MIT Press 2006 Contents Group Cognition: Computer Support for Building Collaborative Knowledge."
14. F. Dong, S. Sterling, D. Schaefer, and H. Forbes, "BUILDING the HISTORY of the FUTURE: A TOOL for CULTURE-CENTRED DESIGN for the SPECULATIVE FUTURE," in *Proceedings of the Design Society: DESIGN Conference*, Cambridge University Press, 2020, pp. 1883–1890. doi: 10.1017/dsd.2020.63.
15. K. Petersen, S. Vakkalanka, and L. Kuzniarz, "Guidelines for conducting systematic mapping studies in software engineering: An update," *Inf Softw Technol*, vol. 64, pp. 1–18, Aug. 2015, doi: 10.1016/j.infsof.2015.03.007.
16. Y. Mei, J. Li, H. De Ridder, and P. Cesar, "Cakevr: A social virtual reality (vr) tool for co-designing cakes," *Conference on Human Factors in Computing Systems - Proceedings*, May 2021, doi: 10.1145/3411764.3445503.
17. T. Isaku and T. Iba, "Creative cocooking patterns: A pattern language for creative collaborative cooking," in *ACM International Conference Proceeding Series*, Association for Computing Machinery, Jul. 2015. doi: 10.1145/2855321.2855366.
18. M. Miyatake, A. Watanabe, and Y. Kawahara, "Interactive Cake Decoration with Whipped Cream," vol. 5, no. 20, 2020, doi: 10.1145/3379175.
19. C. A. Collazos, F. L. Gutiérrez, J. Gallardo, M. Ortega, H. M. Fardoun, and A. I. Molina, "Descriptive theory of awareness for groupware development," *J Ambient Intell Humaniz Comput*, vol. 10, no. 12, pp. 4789–4818, Dec. 2019, doi: 10.1007/s12652-018-1165-9.
20. J. Münch, O. Armbrust, M. Soto, and M. Kowalczyk, "Software Engineering for Embedded Systems Software Process Definition and Improvement."
21. L. García-Borgoñón, M. A. Barcelona, J. A. García-García, M. Alba, and M. J. Escalona, "Software pro- cess modeling languages: A systematic literature review," *Information and Software Technology*, vol. 56, no. 2. pp. 103–116, Feb. 2014. doi: 10.1016/j.infsof.2013.10.001.
22. C. A. Collazos, M. Paula González, A. Neyem, and C. Sturm, "LNCS 4560 - Guidelines to Develop Emotional Awareness Devices from a Cultural-Perspective: A Latin American Example," 2007. [Online]. Available: http://www.we-make-money-not-art.com/archives/007274.php
23. A. Moquillaza *et al.*, "Developing an ATM Interface Using User-Centered Design Techniques," 2017, doi: 10.1007/978-3-319-58640-3_49.
24. T. Clemmensen, Q. Shi, J. Kumar, H. Li, X. Sun, and P. Yammiyavar, "LNCS 4559 - Cultural Usability Tests – How Usability Tests Are Not the Same All over the World," 2007.
25. F. Nake and S. Grabowski, "Human–computer interaction viewed as pseudo-communication," *Knowl Based Syst*, vol. 14, no. 8, pp. 441–447, Dec. 2001, doi: 10.1016/S0950-7051(01)00140-X.
26. A. Herrera, D. Rodríguez, and R. García-Martínez, "Taxonomía de Mecanismos de Awareness."
27. C. Collazos, A. Solano, and H. M. Fardoun, "Collaboration engineering: Supporting the collaborative processes design for the accessible and usable interactive systems design," in *ICSOFT 2018 - Proceedings of the 13th International Conference on Software Technologies*, SciTePress, 2019, pp. 786–793. doi: 10.5220/0006941407860793.
28. J. Gallardo, C. Bravo, and A. I. Molina, "A framework for the descriptive specification of awareness support in multimodal user interfaces for collaborative activities," *Journal on Multimodal User Interfaces*, vol. 12, no. 2, pp. 145–159, Jun. 2018, doi: 10.1007/s12193-017-0255-x.

# A Method to obtain a Knowledge Representation from a Natural Language Specification of the Domain using the Glossary LEL

**Leandro Antonelli[1], Mario Lezoche[2], Juliana Delle Ville[1]**

[1]Lifia, Fac. de Informática, UNLP, La Plata, Bs As, Argentina
[2] Université de Lorraine, Nancy, Lorraine, France

leandro. antonelli, juliana.delleville}@lifia.info.unlp.edu.armario.lezoche@univ-lorraine.fr

**Abstract**— Good requirements (correct, consistent, unambiguous, etc.) are crucial to software development success. Errors made in the requirements stage can cost up to 200 times if they are discovered once the software is delivered to the client. Natural language artifacts are the most used tool to write requirements, since they are understandable by the both parties that participate in the software development: the stakeholders and the development team. Nevertheless, natural languagecan introduce many defects (ambiguity, vagueness, generality, etc.). Formal reasoning is a good strategy to check whether requirements satisfy the attributes of good requirements or not, but formal reasoning cannot be applied to natural language specification with defects. Thus, this paper proposes an approach to write a good specification and obtain knowledge from it. The approach uses a particular lexicon, the glossary LEL, and it suggest guidelines to write good specification, and it also suggest rules to obtain knowledge (concepts and relations) from the glossary LEL. The paper also presents a prototype to assist to this approach, and a preliminary evaluation of the approach.

## 1. Introduction

Good requirements are crucial to software development success. If requirements are not good, the software application developed from them will not satisfy the need, wishes and expectations of the stakeholders. Moreover, if requirements are not good, a lot of work will need to be done to repair the software application to satisfy their needs. It is estimated that errors made in requirements stage can cost up to 200 times if they are discovered when the software is delivered to the client [7].

The expression "good requirements" has a broad meaning. There are many characteristics that are desirable that a software specification has. Does not matter if the development process is agile with minimum of documentation or if it classic with an orthodox software requirements specification of hundreds of pages. It is desirable that the specification would be correct, consistent, unambiguous, etc. These are called quality attributes. Although it is impossibly to satisfy all of them, since there is a huge effort of rework to correct errors, it is necessary to invest the effort in trying to produce the best specification as possible.

Natural language is the most used tool to write requirements [21]. Requirements are specified usually requirements engineering or business analysts, and they should interact with the stakeholder who has the requirements and the knowledge to include in the software application. And the specification also needs to be understandable and precise enough so the software development team can develop the application from it. Natural language specifications are understandable for these both sides [26]. Nevertheless, natural language can introduce many defects: ambiguity, vagueness, generality, etc. [6].

Formal reasoning means infer knowledge from a specification. For example, considering an agriculture specification that states that "plants need water, and tomato is a plant", it can be inferred that "tomatoes need water". It is necessary to have requirements without defects originated by the natural language in order to make possible the forma reasoning [25]. Moreover, with forma reasoning it will be possible to automatize many tasks in order to obtain good requirements [15] [17]. For example, consistency is one of the quality attributes of the specification. And consistency means that there is no contradiction in any given pairs of requirements. Having the specification "plants need water, tomato is a plant, and tomato does not need water" it can inferred that "tomato needs water and tomato does not need water". Since it is a contradiction, it can be concluded that the specification is not consistent.

The glossary LEL is a semi structured artifact with the goal of describing the language of the domain. The glossary LEL is easy to learn, it is easy to use, and it has good expressiveness. We have usedthe glossary LEL in many domains, some of them very complex, and we had good results. Cysneiros etal. [11] report the use of LEL in a complex domain as the health domain. The glossary LEL uses natural language to describe the language, but it also categorize the vocabulary in 4 categories: subjects, objects, verbs and states. The description of each term is done through two attributes: notion and behavioral responses. Thus, the glossary LEL is more than a simple glossary, it represents the knowledge of a domain (captured through its language) in natural language.

The concept of the kernel sentence was introduced in 1957 by linguist Z.S. Harris [16] and featured in the early work of linguist Noam Chomsky [10]. Kernel sentences are also known as basic sentences. They are declarative constructions, in active voice, always affirmative with only one verb. We argue that the use of kernel sentences in the description of the glossary LEL, as well as some other guidelines to make the natural language near a controlled language, allow us to obtain a good specification.

Having a good specification, it is feasible to perform a formal reasoning to infer new knowledge. Nevertheless, before performing formal reasoning, it is necessary to synthetize the knowledge captured in the language through the glossary LEL. Thus, this paper proposes an approach to obtain the knowledge (concepts and relationships) from a glossary LEL. The inference of the new knowledge is outside the scope of this proposal. This paper also presents a prototype, that is, a tool to assist during the application of the proposed approach. Moreover, a preliminary evaluation of the approach is also described.

The rest of the paper is organized in the following way. Section 2 describes some background about glossary LEL and kernel sentences. Section 3 reviews some related work. Section 4 details our contribution namely, the proposed approach. Section 5 describes the tool to support the process. Section 6presents the preliminary evaluation. Finally, Section 7 discusses some conclusions.

## 2. Background

This section describes basic concepts of the glossary LEL and the kernel sentences.

### 2.1. Glossary LEL

The Language Extended Lexicon (LEL) is a glossary that describes the language of an application domain, where not necessarily there is a definition of a software application. The glossary LEL is tied to asimple idea: "understand the language of a problem without worrying about the problem" [19]. The language is captured through symbols that can be terms or short expressions. They are defined through two attributes: notion and behavioral responses. Notion describes the denotation, that is, the intrinsic and substantial characteristics of the symbol, while behavioral responses describe symbol connotation, that is, the relationship between the term being described and other terms (Fig. 1). Each symbol of the LEL belongs to one of four categories: subject, object, verb or state. This categorization guides and assists the requirements engineer during the description of the attributes. Table 1 shows each category with its characteristics and guidelines to describe them.

**Category:** symbol
**Notion:** description
**Behavioral responses:**
Behavioral response 1
Behavioral response 2

**Fig. 1.** Template to describe a LEL symbol

***Table 1.*** *Template to describe LEL symbols according to its category*

| Category | Notion | Behavioral Responses |
|---|---|---|
| Subject | Who is he? | What does he do? |
| Object | What is it? | What actions does it receive? |
| Verb | What goal does it pursue? | How is the goal achieved? |
| State | What situation does it represent? | How this situation is checked? |

Let's consider a bank domain where there are clients that open account in order to save money. The client can deposit money into his account, and he can also withdraw money if the account has a positive balance. In this simple example there are many terms that should be defined into the glossary LEL to capture the language. For example, "client" should be a symbol of subject category. Then, "account" should be a symbol of object category. “Deposit” and “withdraw” should be symbols of verb category. Finally, “positive balance” should be a symbol of state category. According to the template described in Table 1, the symbols “client”, “account”, “withdraw” and “positive balance” are described in Table 2.

The behavioral responses of symbol “client” describes just two responsibilities: to open and to withdraw money, but there are other for example to deposit that is not mentioned. Then, the behavioral responses of the account only describes the responsibility of the bank to calculate interest, but the responsibilities of the client: open, deposit and withdraw should also be mentioned if the glossary LEL is complete. Then, the behavioral responses, of the verb “withdraw” describe how the task if performed by the bank, who first reduces the balance, then check if the resulting balance is positive, and in that case provide the money to the client. Otherwise, the balance is increased to its original value and no money is provided to the client. Finally, the behavioral responses of the state “positive balance” describe in detail what “positive” means. In this cases, it means greater or equal to zero.

***Table 2.*** *Glossay LEL symbols for the banking example*

| Category | Notion | Behavioral Responses |
|---|---|---|
| Subject: Client | Person that saves money in the bank. | The client opens an account.<br>The client withdraws money from the account. |
| Object: Account | Instrument used by the client to save money. | The bank calculates interest for the account. |
| Verb: Withdraw | Act of extracting money from an account. | The bank reduces the balance of the account.<br>If the account has a positive balance the bank provides the money to the client.<br>If the account has not a positive balance the bank increases the balance of the account. |
| State: Positive balance | Situation where the account has money. | The balance is greater or equal to zero. |

### 2.2. Kernel Sentences

A kernel sentence is a simple construction with only one verb. It is also active, positive and declarative. This basic sentence does not contain any mood. It is termed as “kernel” since it is the basis upon which other more complex sentences are formed.

For example, Fig. 2 describes two kernel sentences. The first one states that a subject (“client”) performs an action (“opens”) on a certain object (“account”). The second sentence has the same structure while it describes a different action (“deposits”) and the object is a little bit more complex because it states that the object “money” is deposited into the object “account”. Fig. 3 shows two sentences that are not kernel. The first sentence (“The client deposits money and calculates interest into the account”) has two verbs, that is why it is not a kernel sentences. Moreover, the sentence is partially correct, since it is true that the client deposits money into the account. But, it is not true that the client calculates the interests. It is the bank who performs this calculation. That is why simple (kernel sentences) are suggested to describe specification. A simple sentence is more probable to be completely true or false. The second sentence (“The account is opened”) is written in passive voice, while kernel sentences should be written in active voice. Since it is written in passive voice, the subject of the sentence (“the account”), is not the subject who performs the action. In fact, the real subject who performs the action is not mentioned. According to the example provided, the real subject should be “the client”, but it could also be the company that the client work for and the company is the one who decided to open an account to pay his salary. These are two simple examples about how can kernel sentences writing style help to improve specifications.

The client opens an account.
The client deposits money into the account.

**Fig. 2** Kernel Sentences

The client deposits money and calculates interest into the account.
The account is opened.

***Fig. 3*** *No Kernel Sentences*

## 3. Related works

Vu et al. [34] propose a method to transform specification to source code. Their approach systematically extracts the formal semantics of ARM instructions from their natural language specifications. Although ARM is based on RISC architecture and the number of instructions is relatively small, there are various series including Cortex-A, Cortex-M, and Cortex-R. Their approach applies translation rules enriched by the sentences similarity analysis.

Some other proposal deal with more abstract representation of the knowledge, for example data base table, schemas or object oriented designs. Geetha et al. [14] identify the schema and relationship of a data base from the natural language requirements specification. Their work starts with identifying the table schema and their properties. Then the Primary Key attribute is identified based on adjectives. It is done with rules and also with a machine learning component trained with statistical data. The relationships are identified using the primary key to identify the Foreign Key.

Bargui et al. [5] propose and approach to specify requirements through a template that represents the concepts of a decision making process. They also provide a linguistic pattern for the acquisition of analytical queries. This pattern facilitates the automatic analysis of the queries without being too restrictive to the writing styles.

Kuchta et al. [18] propose a method and a tool to extract concepts based on a grammatical analysis of requirements written in English without the need to refer to specialized ontology. These concepts can be further expressed in the class model, which then can be the basis for the object-oriented analysis of the problem. The method uses natural language processing techniques to recognize parts of speech and to divide sentences into phrases and also the WordNet dictionary to search for known concepts and recognize relationships between them.

Rigou et al. [28] are concerned about a formal or a semi-formal model description of functional requirements in the context of a model-driven engineering approach, where a platform-independent model, is used to derive automatically or semi-automatically the source code of a system. They claim that generally, the approaches use a predefined set of rules which impose several restrictions on the way a specification is written. Thus, they propose a deep learning approach.

Then, there are proposal that deals with conceptual model earlier in the software development life cycle. For example, domain models, concepts, and entities, some approaches are based on syntactic rules and some other are enriched with semantic. Shen et al. [31] propose an approach to assist the phases of requirements elicitation and analysis using masked language models, which have been used to learn contextual information from natural language sentences and transfer this learning to natural language processing (NLP) tasks. They claim that the masked language models can be used to predict the most likely missing word in a sentence, and thus be used to explore domain concepts encoded in a word embedding. Their approach extracts domain knowledge from user-authored scenarios using typed dependency parsing techniques. They also explore the efficacy of a complementary approach of using a BERT-based masked language models to identify entities and associated qualities to build a domain model from a single-word seed term.

Szwed et al. [33] propose an approach to extract concepts from unstructured Polish texts with special focus the morphological forms. Since Polish is a highly inflected language, detected names needto be transformed following Polish grammar rules. They propose a method for specification of transformation patterns, which is based on a simple annotations language. Annotations prepared by a user are compiled into transformation rules. During the concept extraction process the input document is split into sentences and the rules are applied to sequences of words comprised in sentences. Recognized stringsforming concept names are aggregated at various levels and assigned with scores.

Lit et al. [20] presents an approach that employs LSTM-CRF model for requirement entity extraction. Their proposal consist of four phases: (i) model construction, where it is built a LSTM-CRF model and an isomorphic LSTM language model for transfer learning; (ii) LSTM language model training, where it is captured the general knowledge and adapt it to requirement context; (iii) LSTM-CRF training, where it is trained the LSTM-CRF model with the transferred layers; (iv) requirement entity extraction, where it is applied the trained LSTM-CRF model to a new-coming requirement, and automatically extracts its requirement entities.

Wang et al. [35] propose an approach that selects frequent verbs from software requirement specification documents in the e-commerce domain, and built the semantic frames for those verbs. Then the selected sentences are labeled manually and the result is used as training examples for machine learning. They also correct the parsing result of the Stanford Parser. During the labeling process, they adopt a sequential way in which the previous labeled results will be used to construct dynamic features for the identification of the subsequent semantic roles.

Sadoun et al. [29] presents an approach that model the domain knowledge through an ontology and to formally represent user requirements by its population. Their approach of ontology population focuses on instance property identification from texts. They use extraction rules automatically acquired from a training corpus and a bootstrapping terminology. These rules aim at identifying instance property mentions represented by triples of terms, using lexical, syntactic and semantic levels of analysis. They aregenerated from recurrent syntactic paths between terms denoting instances of concepts and properties.

Finally, some approaches are concerned in obtaining knowledge representation similar to our proposal.

Schlutter et al. [30] propose a knowledge extraction approach based on an explicit knowledge representation of the content of natural language requirements as a semantic relation graph. Their approach is fully automated and includes an NLP pipeline to transform unrestricted natural language requirements into a graph. They split the natural language into different parts and relate them to each other based on their semantic relation.

An et al. [1] explore a set of state-of-the-art pre-trained general-purpose and domain-specific language models to extract knowledge triples for metal-organic frameworks. They created a knowledge graph benchmark with 7 relations for 1248 published metal-organic frameworks synonyms.

# 4. The proposed approach

This section describes the proposed approach. First, it describes the generality of the approach, that consist of three steps: (i) glossary LEL description, (ii) glossary LEL revision, and (iii) knowledge extraction. Then, every of the step is described in a different subsection.

## 4.1. Our approach in a nutshell

The proposed approach has the objective of obtaining a synthetic representation of the knowledgecaptured in the language of the application domain captured through the glossary LEL.

The proposed approach needs the knowledge as input and produces a knowledge representation as output. The knowledge can be obtained from different sources. It can be obtained mainly directly from the stakeholders, though interviews or other type of technic (focus group, questionnaires, etc.). But the knowledge can be obtained from documentation.

Thus, the proposed approach is thought to be applied by a requirements engineer or a business analyst who describe the glossary LEL, review it and finally synthetize the knowledge. Moreover, the approach can be used not only by one persona, many people can participate. That is, many requirements engineering can participate describing the glossary LEL in a collaborative way. Moreover, stakeholder experts of the domain can participate writing symbols of the glossary LEL. That is why, the approach consists of three steps: (i) glossary LEL description, (ii) glossary LEL revision, and (iii) knowledge extraction. When more than one person participate of the construction of the LEL, the step (ii) is vital. Nevertheless, when only one person participate, and the volume of information is huge, the step (ii) is also vital. Moreover, the need of a tool to automatize the revision is crucial.

Summing up, the proposed approach can be applied in different situations: (i) to summarize documents produced by other people, (ii) to consolidate and summarize documents produced by only one analyst eliciting from multiple sources, and (iii) to consolidate and summarize documents produced collaborative by a group of analyst or experts.

The three steps that compose the proposed approach are: (i) glossary LEL description, (ii) glossary LEL revision, and (iii) knowledge extraction. The first step proposes a set of guidelines to help people to describe the glossary LEL. The second step proposes a set of guidelines to review some aspects of the description. Some revision (not all of them) are linked with guidelines of the step (i). Finally, the step (iii) provides rules to obtain knowledge from the glossary LEL specified, which is described in terms of concepts and relations. Fig. 4 summarizes the approach.

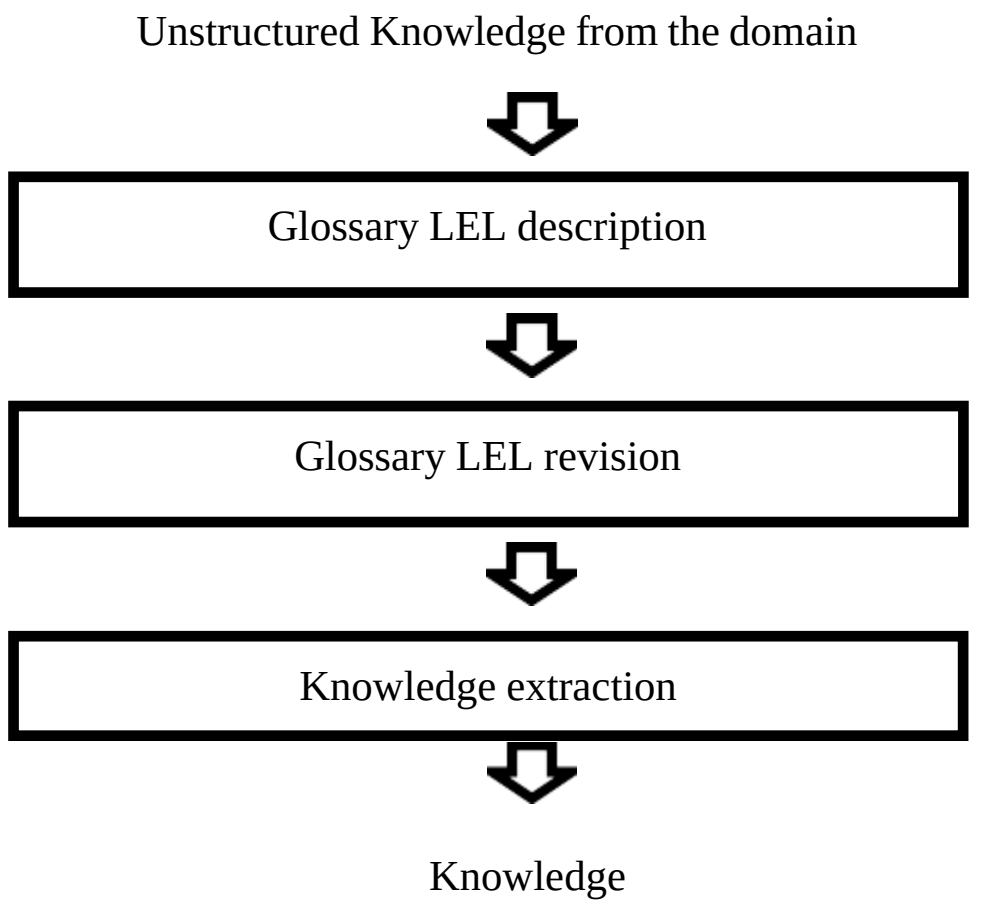

***Fig. 4.*** *Our approach in a nutshell*

Let’s consider the glossary LEL described in Table 2. We can assume that is was described in the step 1 of the proposed approach, and reviewed in the step 2. In fact, the example of the Table 2 satisfies all the guidelines suggested. If step 3 is applied to that LEL it will obtained that the concept “client” and the concept “account” are linked by the relation “withdraw”. Then, the whole relation (“client”, “withdraw”, “account”) is restricted by the condition “positive balance”.

## 4.2. Step 1, guidelines to write the glossary LEL

This section describes every one of the six guidelines proposed to describe the glossary LEL. There are some other proposals about how to describe the glossary LEL. Some of them suggest the basic template described in Table 1 and some other suggests specific structures [2] [3] [Antonelli 2023]. These guidelines reinforce the need to use kernel sentences and reference to symbol already defined, and moreover, this guidelines provides suggestion about how to write compound sentences using subordination (if then), negation (no) and coordination (and). Finally, these guidelines also consider the use of state symbols in subordination expressions. The rest of the section describes and provides example for every one of the guidelines.

**Guideline 1. Sentences to describe behavioral responses must suit kernel sentences philosophy.** Behavioral responses of every category (subject, object, verb and state) must be written according to the philosophy of kernel sentence, that is, an explicit subject followed by only one verb in turn followed by an object that receives the action. Fig. 2 shows examples of sentences well written according to this guideline that are used to describe the behavioral responses of the symbol “client” in Table 2.

**Guideline 2. Terms used to describe behavioral responses must be defined in the LEL glossary.** This guideline suggest that the subject, the verb and the object mentioned in the previous guidelines, should also be symbols already defined in the glossary LEL. For example, the symbol “client” defined in Table 2, has the following description in the behavioral responses “The client opens an account. The client withdraws money from the account.” If it is considered that Table 2 has a complete description of a glossary LEL, the symbols “client”, “account” and “withdraws” are defined. But the symbol “opens” is not defined and it should be defined according to this guideline.

**Guideline 3. Use subordination (if then) in behavioral responses of verb symbols to describe conditions.** Let's consider the example of verb "Withdraw" in Table 2. One of the behavioral responses states "If the account has a positive balance then bank provides the money to the client.". That means that the bank only provides the money to the client, when the account has positive money after reducing the balance with the amount that the client want to withdraw. That is, it is checked whether the balance remains positive after the extraction. It is important to mention that the application of this guideline is not a contradiction to the guideline 1, since the use of a subordination requires kernel sentences, although there are two kernel sentences in the same sentence, there are still kernel sentences associated.

**Guideline 4. State symbols should be used in the condition of a subordinated sentence.** Since the state symbols describes situation, this situation should be used as a condition of the subordinated sentence. Moreover, the condition should be included in the kernel sentence. For example, let's consider the verb "Withdraw" in Table 2. One of the behavioral responses states "If the account has a positive balance then ...", "positive balance" is a state symbol describe in Table 2, and the expression "the accounthas a positive balance" suit the rules of kernel sentences.

**Guideline 5. Use negation (not) to invert a condition or an action.** For example, the verb "Withdraw" in Table 2 contains a behavioral response that states "If the account has not a positive balance, then ...". Nevertheless, the negation can be used to deny some behavioral response. That is, behavioral response of subject for example, describe all the activities that subject can perform. The symbol "client" defined in Table 2, has the following description in the behavioral responses "The client opens an account. The client withdraws money from the account." Generally, (and according to guideline 1 that recommends the use of kernel sentences that should be positive by definition), behavioral responses should be positive. Nevertheless, some times, it is very important to make explicitly definition that something cannot be done. In this case, the negation can be used. For example, considering a bank that only works with organization and not with private clients, the responsibility of opening account relies on the organization and not into the client. Thus, a behavioral response of a client could state "The client does not open an account".

**Guideline 6. Use coordination (and) to join two subjects, objects or kernel sentences.** Although multiple kernel sentences can be written, using the coordinator "and" more simple and natural sentences can be written. For example, instead of writing "The client deposits money in the account. The bank deposits money in the account", it could be written "The client and the bank deposits money in the account".

## 4.2.- Step 2, guidelines to review the glossary LEL

This section describes some guidelines to review the glossary LEL described with the guidelines of the previous sections. These guidelines are simply a way of checking whether the previous guidelines were used correctly. Although these guidelines can be used as a checklist after some description was done, the following section describes a prototype that can perform these checks automatically.

**Guideline 1. Check for explicit subject and only one verb written in active voice in a behavioral response.** This guideline has the objective to be sure that kernel sentences are used to describe behavioral responses. Thus, the use of only one verb, written in active voice should be used. Moreover, the sentence should have at least one explicit subject.

**Guideline 2. Check for the definition of every subject, verb and object in a behavioral response.** This guideline complements the guideline number 2 of the previous section. The objective is to check thatevery element (with semantic meaning) used in a sentence is describes within the glossary LEL.

**Guideline 3. Check for the definition of every condition of a subordination as a state.** This guideline complements guideline number 4 of the previous section. The objective is similar to the objective of guideline 2 of this section.

**Guideline 4. Replace modal verbs with a subordination.** The expression “The client withdraws money from the account” states that the client is allowed to do that and he will do it when he need. If a modal verb is used “The client might withdraw money from the account” it means that there is certain condition.Thus, the sentence should be rewritten as “If (some condition related to the client) then the client withdraw money from the account”

**Guideline 5. Rephrase expressions with “no” to be sure of expressing prohibition.** Guideline 5 of the previous section suggest the use of negation to invert a condition or an action. When it is used to invert an action, that is, to state that some subject does not do some action, it is important to be sure that the description states a prohibition to perform that action. For example, the expression “The client does not open an account” can be rephrased as “The client is not allowed to open an account” to be sure that a prohibition is described.

**Guideline 6. Check for unambiguous terms used in notion.** When technical specification are written, the goal to be precise and unambiguous is naturally pursued, that is why it is not added a complementary guideline in the previous section. Nevertheless, it is important to pay attention to this issue and to consider this guideline for the revision. Although this guideline can be applied manually, maybe the tool support using some tool of word-sense disambiguation is more useful.

### 4.3.- Step 3, knowledge extraction

This section describes some rules to extract knowledge form the glossary LEL described and reviewed with the guidelines of the previous sections.

**Rule 1. Concepts are obtained from symbols of subject and objects categories.** Considering the glossary LEL described in Table 2 two concepts are obtained: “client” and “account”, the first one is a subject and the second one is an object, but both of them are concepts.

**Rule 2. Relations are obtained from symbols of verbs categories.** Considering the glossary LEL described in Table 2 one relation is obtained: “withdraw”. According to the behavioral response of the symbol “client”, this action (“withdraw”) relates the concepts “client” and “account”.

**Rule 3. Conditions are obtained from symbols of state categories.** Considering the glossary LEL described in Table 2 one condition is obtained: “positive balance”.

**Rule 4. Relations obtained by rule 2 are restricted by conditions.** Considering the glossary LEL described in Table 2 the state: “positive balance” gives origin to a condition. This state is mentioned in the “withdraw” verb that gives origin to the relation between “client” and “account”. Thus, the whole relation (“client”, “withdraw”, “account”) is restricted by the condition “positive balance”.

## 5. Tool support

A software prototype was implemented that can be used to support the application of the proposed approach. The prototype is a web application implemented following a service-oriented architecture. The core of the application and its services are implemented in Python [27], while the web component is implemented with Django [12], and the APis are implemented with Flask [13]. Python [27] is also used to communicate to the Spacy [32] and NTLK libraries [23] used to deal with natural language processing.

The application is responsive, that means, the interface of the application adapts itself to the device used: a computer (desktop or laptop) or mobile device (phone or tablet). Thus, the application provides a variety of platforms to be used and experts will have a wide range of possibilities to contribute with knowledge acquisition.

The prototype implements two roles of users: (i) administrators of the projects and (ii) experts. The administrator of the tool can create project and the experts contribute with knowledge to the projects. The prototype is in fact a framework, since it can manage different type of artifacts based on natural language (Fig. 5). And the prototype also is easily extended to perform different process to derive more information from the artifacts.

```
Class LelElement():
symbol (length = 255, label "Symbol") category
(length = 10, label "Category")Notion (length = 1000,
label "Notion")
BehavioralResponses (length = 1000, label "Behavioral responses")
```

***Fig. 5.*** *Configuration of artifacts*

The prototype uses Spacy [32] and NLTK [23] libraries to perform natural language processing. The POS (part of speech) tagging is an appropriate tool to determine the revisions suggested by guideline 1 as well as the identification of modal verbs mentioned in guideline 4. Then, word disambiguation is another powerful tool mentioned in guideline 6. Then, guidelines 2 and 3 requires to check the presence of some words as symbols of the glossary LEL. In order to do that, the stemming and lemmatization tools are very useful, since they convert words to their root form in order to make the comparison.  Fig. 6 shows how the tool check kernel sentences.

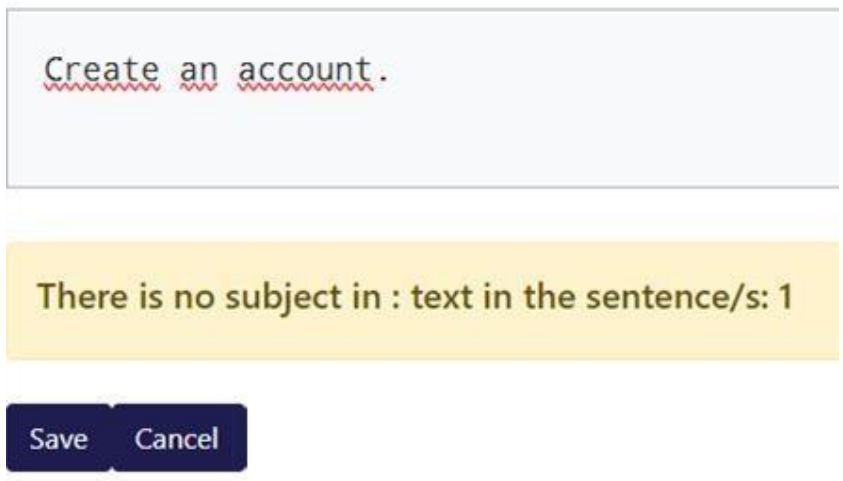

***Fig. 6.*** *Kernel sentence checking*

The framework Plant UML [24] is used to display the model (Figure Fig. 7).

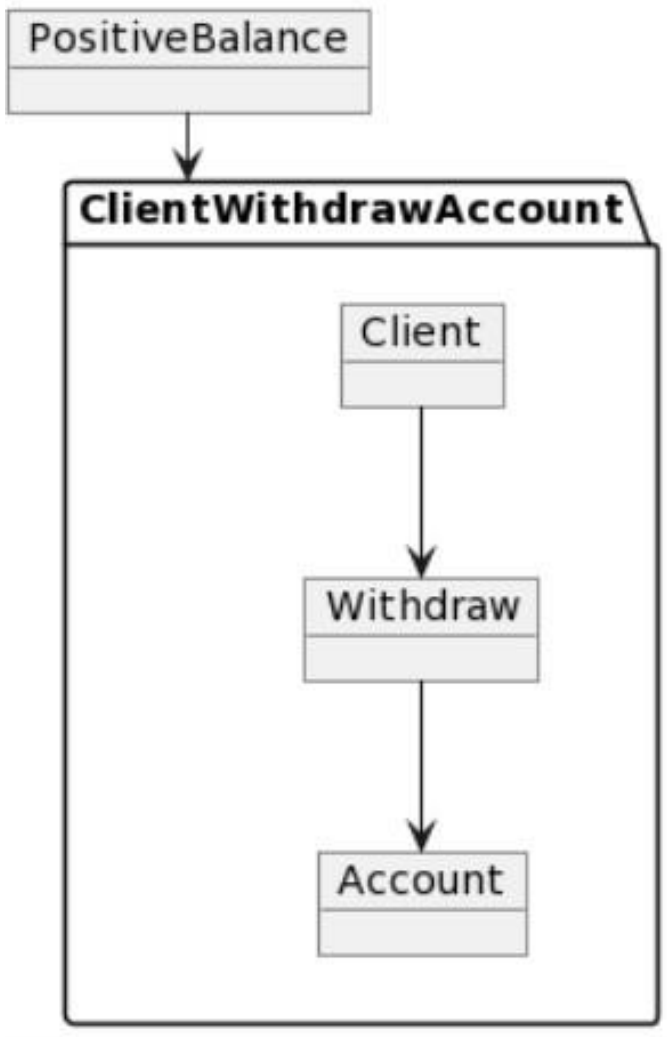

***Fig. 7.*** *Knowledge represented graphically*

## 6. Evaluation

The proposed approach was evaluated, although the prototype was not used in the evaluation because the goal of the evaluation was to assess the perceived usability of the proposed approach instead of evaluating the tool.

The Systems Usability Scale (SUS) was used to perform the evaluation [8] [9] in terms of the usability of the proposed approach. Although SUS is mainly used to assess usability of software systems, it was probe to be effective to assess products and processes [Bangor 2008].

The System Usability Scale (SUS) consists of a 10-item questionnaire; every question must be answered in a five-options scale, ranging from ”1” (” Strongly Disagree”) to ”5” (” Strongly Agree”). Although there are 10 questions, they are related by pairs, asking the same question but in a complementary point of view in order to obtain a result of high confidence. The questions are the following:

1.- I like to use this approach.
2.- I find this approach to be more complicated than it should be.3.- I think the approach is simple and easy to use.
4.- I need technical support to use this approach.
5.- I find the approach functioning smoothly and is well integrated.6.- I think there are a lot of irregularities in the approach.
7.- I think most people can learn this approach quickly.8.- I find this approach to be time-consuming.
9.- I feel confident using this approach.
10.- I think there are a lot of things to learn before I can start using this approach.

The participants of the evaluation were 5 students (with an age average of 24.8 years old) that participate in research project. It is important to mention that all of the students have experience in industry since in Argentina, students generally begin to work in industry in second year of their undergraduate studies. Thus, participants are suitable to this evaluation since they have experience in reading specification and deciding about the understandability of it. So, their opinion about whether the knowledge summarization is good or not to is important to us. They received training on the proposed approach and they were requested to answer the questionnaire.

The calculation of the SUS score is performed in the following way. First, items 1, 3, 5, 7, and 9 are scored considering the value ranked minus 1. Then, items 2, 4, 6, 8 and 10, are scored considering 5 minus the value ranked. After that, every participant’s scores are summed up and then multiplied by 2.5to obtain a new value ranging from 0 to 100. Finally, the average is calculated. The approach can haveone of the following results: “Non acceptable” 0-64, “Acceptable” 65-84, and “Excellent” 85-100 [22]. The score obtained was 88.5. Thus, the approach can be considered as “Excellent”.

## 6. Conclusions

This paper has presented an approach to obtain knowledge specification from a natural language specification. Particularly, we proposed the use of the glossary LEL to capture the language of the domain, hence, its knowledge. The proposed approach consists in a set of guidelines to help to capture and organize the natural language specification, in order to obtain the knowledge. The knowledgeobtained can be used in different way. It synthetizes the glossary LEL, so it can be used to provide a global view. But it also make possible to infer new knowledge, and this is very important to provide complete and consistent specifications, moreover if we consider that they use natural language. This isour main further work. We plan to extend this proposal to infer knew knowledge to be used to add to the glossary LEL so stakeholder can validate and use them. And also to identify inconsistencies and conflict, so they can be solved. Another further work is related to improve the glossary LEL with a categorization of subjects as countable and uncountable. This categorization will make possible to be precise in certain descriptions involving them. We plan to enrich the LEL with magnitudes (numbers with a unit) and categorical elements (a list of possible value). We think that both characteristics are important to write precise specification and obtaining inferred knowledge. Particularly, this will make possible to verify requirements and designing tests from it.

## Acknowledgment

This paper is partially supported by funding provided by the STIC AmSud program, Project 22STIC-01.

## References

1. An, Y. et al.: "Exploring Pre-Trained Language Models to Build Knowledge Graph for Metal- Organic Frameworks (MOFs)" 2022 IEEE International Conference on Big Data (Big Data), Osaka, Japan, pp. 3651-3658 (2022)
2. Antonelli, L., Leite, J.C.S.P., Oliveros, A., Rossi, G.: "Specification Cases: a lightweight approach based on Natural Language", Workshop in Requirements Engineering (WER), Brasilia, Brasil, Agosto 23 – 27 (2021)
3. Antonelli, L., Fernandez, A., Ruffolo, N., Sansone, E., Torres, D.: "A Collaborative Approach to specify Kernel Sentences using Natural Language", Workshop in Requirements Engineering (WER), Natal, Brasil, Agosto 23 – 26 (2022).
4. Antonelli, L., Delle Ville, J., Dioguardi, F., Fernandez, A., Tanevitch, L., Torres, D.: "An Iterative and Collaborative Approach to Specify Scenarios using Natural Language", Workshop in Requirements Engineering (WER), Natal, Brasil, Agosto 23 – 26 (2022).
5. Bargui, F., Ben-Abdallah, H., Feki, J.: "Multidimensional concept extraction and validation fromOLAP requirements in NL," 2009 International Conference on Natural Language Processing and Knowledge Engineering, Dalian, China, pp. 1-8 (2009)
6. Berry, D., Kamsties, E. and M. Krieger. From Contract Drafting to Software Specification:
7. Linguistic Sources of Ambiguity. University of Waterloo. 2003
8. Boehm, B.W.: Software Engineering, Computer society Press, IEEE, 1997.
9. Brooke, J.: "SUS-A quick and dirty usability scale" Usability evaluation in industry, 189(194),pp. 4-7, 1996.
10. Brooke, J: "SUS: a retrospective", Journal of usability studies 8.2, pp.29-40, 2013.
11. Chomsky, N.: The Logical Structure of Linguistic Theory. Plenum Press, New York, 1975.
12. McLellan, S., Muddimer, A., Peres, S. C.: "The effect of experience on System Usability Scale ratings." Journal of usability studies 7.2, pp. 56-67, 2012.
13. NLTK, https://www.nltk.org/, accessed: 2023-03-05
14. PlantUML, available at: https://plantuml.com/, accessed on 27th February 2023
15. K. Pohl, The Three Dimensions of Requirements Engineering, in: Proc. 5th Int. Conf. of Advanced Information Systems Engineering 1993, Paris, Springer, Berlin, pp. 275-292, 1993
16. Potts, C.: "Using schematic scenarios to understand user needs," in Proceedings of the 1st conference on Designing interactive systems: processes, practices, methods, & techniques, 1995
17. Python, https://www.python.org/, accessed: 2023-03-05
18. Rigou, Y., Lamontagne, D., Khriss, I.: "A Sketch of a Deep Learning Approach for Discovering UML Class Diagrams from System's Textual Specification," 2020 1st International Conference on Innovative Research in Applied Science, Engineering and Technology (IRASET), Meknes, Morocco, pp. 1-6 (2020).
19. Sadoun, D., Dubois, C., Ghamri-Doudane, Y., Grau, B.: "From Natural Language Requirements to Formal Specification Using an Ontology," 2013 IEEE 25th International Conference on Tools with Artificial Intelligence, Herndon, VA, USA, pp. 755-760 (2013).
20. Schlutter, A., Vogelsang, A.: "Knowledge Extraction from Natural Language Requirements into a Semantic Relation Graph". In Proceedings of the IEEE/ACM 42nd International Conference on Software Engineering Workshops (ICSEW'20). Association for Computing Machinery, New York, NY, USA, pp 373–379 (2020)
21. Shen, Y., Breaux, T.: "Domain Model Extraction from User-authored Scenarios and Word Embeddings" 2022 IEEE 30th International Requirements Engineering Conference Workshops (REW), Melbourne, Australia, pp. 143-151 (2022)

22. Spacy https://spacy.io/, accessed 2023-03-05
23. Szwed, P.: "Concepts extraction from unstructured Polish texts: A rule based approach," 2015 Federated Conference on Computer Science and Information Systems (FedCSIS), Lodz, Poland, pp. 355- 364 (2015.
24. Vu, A.V., Ogawa, M.: “Formal Semantics Extraction from Natural Language Specifications forARM”. In: ter Beek, M., McIver, A., Oliveira, J. (eds) Formal Methods – The Next 30 Years. FM 2019. Lecture Notes in Computer Science (), vol 11800. Springer, Cham (2019)
25. Wang, Y.: "Semantic information extraction for software requirements using semantic role labeling," 2015 IEEE International Conference on Progress in Informatics and Computing (PIC), Nanjing,China, pp. 332-337 (2015)

# Quantum Computing meets Software Engineering: a Road Map for CollaborativeResearch

**Jose Garcia-Alonso[1], Julio Ariel Hurtado Alegria[2], Juan M. Murillo[3], JuanBraña[4], and Alejandro Fernandez[5]**

[1] Universidad de Extremadura, España
[2] Universidad del Cauca, Colombia
[3] CénitS - COMPUTAEX, España
[4] I-314, Argentina
[5] Universidad Nacional de La Plata, Argentina

**Abstract.** Software quantum developers need to implement efficient so- lutions taking advantage of the quantum computing properties. However, quantum computing is still in its early stages, and software developmentthat exploits the advantages of quantum computing remains a challenge,as a lot of methodologies and tools are rooted in the essential conceptsof classical computation. In this paper, we introduce the Iberoamerican Seedbed for Quantum Software Engineering initiative, which aims to pro- mote research, innovation, and teaching at the intersection of softwareengineering and quantum computing. The initiative will facilitate joint research and innovation projects, staff and student exchange programs, and seminars and workshops. It will also establish a network of quantumsimulators, create a learning program for quantum software engineering, and disseminate research findings to the scientific community and the public.

## 1 Introduction

The principles of quantum mechanics were established by Max Planck and Niels Bohr. Since then, advances in both quantum hardware and quantum information theory have made Quantum Computing an undeniable reality.

Quantum computers rely on the use of quantum bits or qubits, the basicunit of information in a quantum computer, instead of to the classical bits in traditional computers. These qubits use the quantum concepts of superposition and entanglement to open the gates to a new type of computing. The use of superposition allow qubits to be on multiple states at the same time, until they are measured. While entaglement allow qubits to be related to other qubits in a way that their state cannot be described independently of the state of the others.

The use of these properties allow quantum computers to solve bounded- error quantum polynomial time problems (BQP). A class of decision problems solvable by a quantum computer in polynomial time within a margin of error.It is suspected that BQP contains P and that there are some problems in theBPQ class which are outside the P class.

To be able to address BQP problems, quantum developers need to implementefficient algorithms that take advantage of the quantum properties. There arealready some well-known algorithms of this type like Grover algorithm [4] thatsearches in an unordered database faster than any classical algorithm or Shoralgorithm [7] that factorizes an integer faster than any known classical algorithm.

However, as classical software engineering has demonstrated, is not enough to develop efficient algorithms. Modern software solutions are complex systems that rely on multiple layer of infrastructure, tools and techniques that allow software engineers to create high quality solutions. Most of these elements are still in a very early stage in quantum computing [2].

The field of quantum software engineering is an emerging topic that focuson bringing the advantages of software engineering to the development of quan-tum software. Discussion forums like the International Workshop on Quantum Software Engineering at ICSE, the IEEE International Conference on QuantumSoftware at IEEE Services or the Quantum Software Engineering and Technol-ogy Workshop at IEEE Quantum Week are starting to appear.

Nevertheless, additional efforts are needed to advance quantum software engi- neering [6]. There is still a low critical mass of researchers and, therefore, there isa need for training researchers and students in the specifics of quantum softwareengineering. The amount of resources needed to develop and test quantum solu-tions is still important, therefore, there is a need to join efforts between groups and institutions interested in this field. Also, due to the intrinsic interdisciplinary nature of quantum development, with the necessary input from mathematiciansand physicist, collaborative approaches are necessary.

Specifically, we propose a road map for the creation of a collaborative research initiative in the field of quantum software engineering. To do that we present a research road map. Then, we propose the creation of a Seedbed for Quantum Software Engineering and provide some initial planning and expected outcomes for this initiative.

## 2 A research road map

As mentioned above, quantum software engineering is still in an early stage of development. Therefore, additional research is needed in almost every aspect ofthis new discipline [5]. In this section we focus on those topics that we consider the most relevant given the current state of the area.

### 2.1 Software tools to support quantum computing

As a classical software engineering has evolved, so have done the tools that support it. Nowadays, software engineers are used to have at their disposal a wide range of advanced tools, frameworks, and SDKs that help them in theirdaily task.

To simplify the transitions from classical software engineering to quantum software engineering and to train new quantum software engineers similar toolsare needed. Not only because they make life easier to software engineers, butbecause they help develop software with better quality attributes [3].

Current quantum computing systems rely on initial versions of these tools. Most vendors provide environments and SDKs to help developers. For example, IBM offers developers a set of real hardware quantum computers up to 5 qubits and several simulators up to 5000 qubits for free through its platform” IBM QUANTUM EXPERIENCE”. This offering is very useful for developing proof of concept projects, writing academic papers and learning to program quan-tum computers. Other companies like Google, Microsoft, Amazon and startups around the world provide different SKDs and tools for these tasks.

These SDKs are based on some of the existing quantum programming lan- guages like Qiskit (IBM) and Cirq (Google), both written in Python or “Q- Sharp” (Microsoft). Amazon offers an environment named “Amazon Braket”to get started in the quantum computing world, for academic or commercial projects, and this company also offers one hour of its suite for free to try this technology. Other language that is often used to program Amazon Braket com- puters is Julia, through its library “Braket.jl”. These languages, although usually based on a high-abstraction level programming language like python provide a very low-abstraction level language for the programming of quantum hardware, based on the use of quantum gates and quantum annealing approaches.

Another example, is the intersection between quantum computing and ma- chine learning. Several libraries have been developed for this purpose, one of themost well known framework is" quantum-tensorflow", published by Google.

Other tools in this domain are the transpilers that allow developers to trans- late their quantum program so it can be run on a different quantum computer using a different set of gates or with a different architecture. Also, some frame- works are starting to appear that focus on the application of quantum computingto specific domains like machine learning or chemistry.

All these tools, although they provide important advantages for the develop- ment of quantum software, are still very limited in terms of software engineering.The low abstraction levels at which quantum hardware is operated and the fo- cus on addressing some of the hardware limitations, like noise or decoherence time, make the concerns about software quality attributes seem very far away. However, as quantum hardware evolve and more advanced quantum software is created, quality attributes of software will rise in relevances, as happended in classsical software engineering, and quantum software engineers will need tool support.

### 2.2 Hybrid quantum-classical systems

Additionally, although quantum computers can solve some problems quicker that classical computers, modern software almost never work in an isolated way.

Current systems usually are created as global, interconnected set of software components. It seems natural, that quantum software would follow a similar pattern in which hybrid system will be the standard way of development [8].

In fact, the initial steps of quantum computing are taking that hybrid shape. Several relevant quantum algorithms already need a classical part. Also, most quantum computers are still managed through classical systems. So efforts are needed to define hybrid architectures for quantum software engineering.

Moreover, most of the existing quantum computers are offered through the cloud. This kind of access is well known for classical software engineers and provide important advantages to software that justify its omnipresence in mod-ern software. Starting from this point it seems reasonable to address quantum software enginerring from a Service Oriented Compunting point of view.

All of these, make clear the relevance of working on hybrid quantum-classicalsystems for a quantum software engineering.

### 2.3 Collaboration and quantum computing

Finally, advances in quantum software engineering need a collaborative ap- proach. Not only between software engineers, which is always a requirementfor complex systems, but also with professional of other disciplines [1]

Nowadays, advances in quantum computing are mostly created by profes- sional with strong backgrounds in mathematics or physics. The low abstraction level at which quantum software is developed make these professionals irreplace-able. However, they lack the skills and the background needed to create softwarewith the appropriate quality attributes. On the other side, software engineers with the abilities to create quality software usually lack the background andskills needed to create quantum software in its current state. Therefore, collab- oration is needed to produce contributions in the quantum software engineeringdomain.

Additionally, other kind of collaborations are needed to develop useful soft- ware. Collaboration with domain experts are needed to understand the domain requirement and co-creation processes have been proven as one of the best ap- proach for that. Also, collaboration with super-computing centers where quan- tum computers or powerfull simulators are hosted are also required in order to obtains access to the resources needed for development.

Again, all of these make clear the need of a collaborative approach for research on quantum software engineering.

## 3 An Iberoamerican Seedbed for Quantum Software Engineering

Quantum computing is still in its early stages, and the development of software that exploits the advantages of quantum computing is still a challenge. Existing software development methodologies and tools are based on the computer model proposed by Von Neumann and Turing's conception of a program, which may not be adequate to the quantum computing paradigm. The Iberoamerican Seedbed for Quantum Software Engineering initiative aims to foster research, innovation, and teaching in the intersection of software engineering and quantum computing and explore alternative methods and tools for quantum software engineering. The primary outcomes of this initiative will be:

- Foster the creation of joint research and innovation projects in quantum software engineering, focusing on developing new methodologies and tools that take into account the specific features of quantum computing.
- Promote staff and student exchange programs between Iberoamerican countries to facilitate the exchange of knowledge and expertise in quantum software engineering and alternative methods and tools.
- Conduct joint seminars and workshops to share knowledge and expertise in quantum software engineering and alternative methods and tools.
- Establish a network of quantum simulators in supercomputing centers or other computing infrastructures able of offering to other researchers and to the industry the possibility of exploring the use of quantum technologies.
- Create a learning program for quantum software engineering that includes practical training using simulators or real quantum hardware.
- Disseminate the results of the research and innovation projects and the exploration of alternative methods and tools to the scientific community and the general public.

With this goals in mind, a three stage plans is proposed: planning, implementation, and consolidation. During the planning stage (first year of project) the following activities will be carried out:

- Create a website and social media accounts for the Iberoamerican Seedbed for Quantum Software Engineering.
- Define the research lines and projects that will be developed in the next two years, focusing on developing new methodologies and tools that take into account the specific features of quantum computing.
- Identify and recruit researchers, professionals, and students with expertise in computer science, physics, and engineering to participate in the initiative.
- Define the guidelines for the exchange programs and joint seminars, focusing on the exchange of knowledge and expertise in quantum software engineering and alternative methods and tools.
- Organize a first joint seminar to share knowledge and expertise in quantum software engineering and alternative methods and tools.
- Secure funding for the implementation of staff and student exchange during the second year of the project, and to implement multi-year collaborative research projects.

## Acknowledgements

This work has been financially supported by the Ministry of Economic Affairs and Digital Transformation of the Spanish Government through the QUANTUM ENIA project call - Quantum Spain project, by the Spanish Ministry of Science and Innovation under project PID2021-124054OB-C31, by the Regional Ministry of Economy, Science and Digital Agenda, and the Department of Economy and Infrastructure of the Government of Extremadura under project GR21133, and by the European Union through the Recovery, Transformation and Resilience Plan - NextGenerationEU within the framework of the Digital Spain 2026 Agenda.

## References

1. Barzen, J., Leymann, F., Falkenthal, M., Vietz, D., Weder, B., Wild, K.: Relevance of near-term quantum computing in the cloud: A humanities perspective. In: Ferguson, D., Pahl, C., Helfert, M. (eds.) Cloud Computing and Services Science. pp. 25–58. Communications in Computer and Information Science, Springer International Publishing (2021). https://doi.org/10.1007/978-3-030-72369-9˙2
2. Enrique Moguel, Javier Berrocal, Jose Garc´ıa-Alonso, Juan Manuel Murillo: A roadmap for quantum software engineering: applying the lessons learned from the classics. In: Short Papers Proceedings of the 1st International Workshop on Soft- ware Engineering & Technology (Q-SET'20) co-located with IEEE International Conference on Quantum Computing and Engineering (IEEE Quantum Week 2020). vol. 2705. CEUR (2020)
3. Garcia-Alonso, J., Rojo, J., Valencia, D., Moguel, E., Berrocal, J., Murillo, J.M.: Quantum software as a service through a quantum api gateway. IEEE Internet Computing **26**(1), 34–41 (2021)
4. Grover, L.K.: A fast quantum mechanical algorithm for database search (1996). https://doi.org/10.48550/ARXIV.QUANT-PH/9605043, https://arxiv.org/abs/quant-ph/9605043
5. Moguel, E., Rojo, J., Valencia, D., Berrocal, J., Garcia-Alonso, J., Murillo, J.M.: Quantum service-oriented computing: current landscape and challenges. Software Quality Journal **30**(4), 983–1002 (2022)
6. Rojo, J., Valencia, D., Berrocal, J., Moguel, E., Garcia-Alonso, J., Rodriguez, J.M.M.: Trials and tribulations of developing hybrid quantum-classical microser- vices systems. arXiv preprint arXiv:2105.04421 (2021)
7. Shor, P.W.: Algorithms for quantum computation: discrete logarithms and factoring. In: Proceedings 35th annual symposium on foundations of computer science. pp. 124–134. Ieee (1994)
8. Valencia, D., Garcia-Alonso, J., Rojo, J., Moguel, E., Berrocal, J., Murillo, J.M.: Hy- brid classical-quantum software services systems: Exploration of the rough edges. In: Quality of Information and Communications Technology: 14th International Con- ference, QUATIC 2021, Algarve, Portugal, September 8–11, 2021, Proceedings 14. pp. 225–238. Springer (2021)

# Biomechanical Analysis of Basketball Free Throw Shooting Technique in Athletes of The Pedagogical andTechnological University of Colombia

**Jorge Armando Abril[1], Jose Charris and Marco Javier Suarez Baron**

[1] Universidad Pedagogica y Tecnologica de Colombia-UPTC, Sogamoso, Colombia[2] Faculty of Systems and Computing marco.suarez@uptc.edu.co

**Abstract.** Biomechanics is one of the most complete fields of the study of the metric composition of the human body and its relationship with physical and me-chanical variables characteristic of motion kinematics analysis, which is a great advantage when conducting studies on the performance of athletes and the anal-ysis of their technical and tactical skills, since it gives us a more concrete per- spective on them. For this research process, it has been decided to carry out a series of tests to launch free basketball shots in two athletes from the Pedagogicaland Technological University of Colombia.

## 1 Introduction

This section will specify everything related to the development of the field work, such as the classification and selection of the biomechanical variables to be analyzed, in addition to this, the population sample that will be studied for the analysis of the freethrow mechanics will be delimited and the techniques and procedures that will be usedfor the development of the biomechanical study will be mentioned.

In order to successfully perform the biomechanical analysis of the technical skill ofbasketball free throw shooting in athletes of the Pedagogical and Technological Uni- versity of Colombia, within the fieldwork, data collection, analysis and procedures, it is proposed:

- To characterize in a specific and adequate way the biomechanical variables thataffect the development of the technical-tactical skill in basketball free-throw shooting.
- Establish through a biomechanical analysis methodology, a classification of rou-tines suitable for improving basketball free-throw shooting technique.
- Design a visualization system for anthropometric data and results using MLframeworks

The analysis of the basketball free throw shooting technique is essential to under- stand the success of some athletes, because the effectiveness of this technique has a high impact on this sport. However, the studies that have been conducted on the anthro-pometric and biomechanical relationship to reach an optimal characterization of the variables involved in the technical ability of an athlete are usually focused on other disciplines or are very ambiguous, so that also the use of technological tools and data analytics in such studies is very small, despite being tools that give us great reliability to make forecasts and analysis accurately and effectively.

Throwing is a movement that determines the action of sending an object to another place, this action has its own characteristics of the intention with which it is performed,from this consideration there have been several studies on the biomechanical factors ofthrowing depending on the specialty to be analyzed, such as the sports and military area, among others. The investigations carried out allow to evidence the incidence of variables such as angles, planes, positions and articulations that adjust to the character-istics of the human body and how they interact with the mechanics required to fulfill the objective of the action [1]

In order to make a biomechanical description of the throwing action, it is necessaryto identify the phases that make up the throwing action, and how the athlete's body partsbehave within each of these phases, that is, to identify the behavior patterns of the upperlimbs and their joints, their measurements, and their variation as the movement is exe-cuted [2]. Is essential to bear in mind that the execution of free throws does not have arigid and predetermined standard, however, there are techniques associated with the effectiveness of the free throw [3]. In the anthropometric study of the upper limbs, thevalues corresponding to the dimensions of the arms, forearms, hands, and their length and extension, among other variables, are treated [5]

The specific study of upper limbs has been widely used for the creation of prosthesesas it allows through the results obtained to make products appropriate for the needs of people who have suffered malformations, injuries, or loss of any of their limbs, as it allows to establish ideal characteristics in terms of optimal materials for use, and dif- ferent components involved in the design and production of prostheses according to ISO 13405-1 standard, such as interface components (interaction with the patient), functional, alignment, structural and cosmetic (appearance as close as possible to a nat-ural human limb) [6]. In this way, the relationship between these dimensions and the execution of movements can be established by means of a dynamic anthropometric analysis of the upper limbs.

## 2 Materials and methods

### 2.1 Method

**Phase 1. Information collection.**
Analyses were carried out based on the study of a population sample composed of play-ers of the men's basketball team of the Pedagogical and Technological University of Colombia Sogamoso Sectional Faculty, whose ages range between 17 and 24 years, who also meet the requirement of being active students of an undergraduate career.
For the purpose of this research, the process was segmented into phases as follows.

**Phase 2. Information processing**

• Capture of participants movements during the execution of the free-throw shooting technique, using high-definition cameras and motive optitrack software.

• Processing of images collected from high-definition cameras using the Python pro- gramming language.

• Storage of data extracted through image processing, this will be done by integratingit with the documentation in flat files obtained in the measurement and anthropometriccharacterization of the participants.

**Phase 3. Learning model**

• Extraction of data obtained in flat files with the motive optitrack software, and sub- sequent integration with the flat file containing the anthropometric characterization anddigital processing data.

• Classification of variables according to successes during motion capture sessions.

• Use of Weka software for data analysis and statistical processing of the data stored in the flat file, according to the classification assigned to the variables. This will allowus to demonstrate by means of correlation matrices the interaction of the variables dur- ing the execution of the movement, and its relation with an optimal technique accordingto the successes or deficiencies in the attainment of scores by the players according to their throwing technique.

• From the above, using the software, a classification tree is generated, which will giveus results of the behavior of the variables to achieve an optimal launching technique.

**Phase 4. Evaluation and training**

• Finally, a classification of the adequate behavior of the variables in the throwing technique will be made, for the construction of an optimal training methodology, by means of routines, based on the results obtained in the classification tree, for the im- provement of the free-throw throwing technique.

• In addition, a teaching/learning process of free throw shooting will be designed, through sessions and routines established according to the optimal characterization made in the study, this will serve as a guide to adequately develop the sports training process, thus making a contribution to basketball instructors and players.

**Phase 5. Design and codification**

• Finally, the methodological design of development, generation of schemes of ac- tions, packages and others corresponding to a software development is planned, wherethe deductive model is already established with the results in accordance with the ob- jectives, it will be embodied in coding to obtain as an expected result a system where the capture of movement and level of assertiveness of the linguistic sign of the exposedmodel can be observed

Detailed description of the method steps can be seen in figure 1.

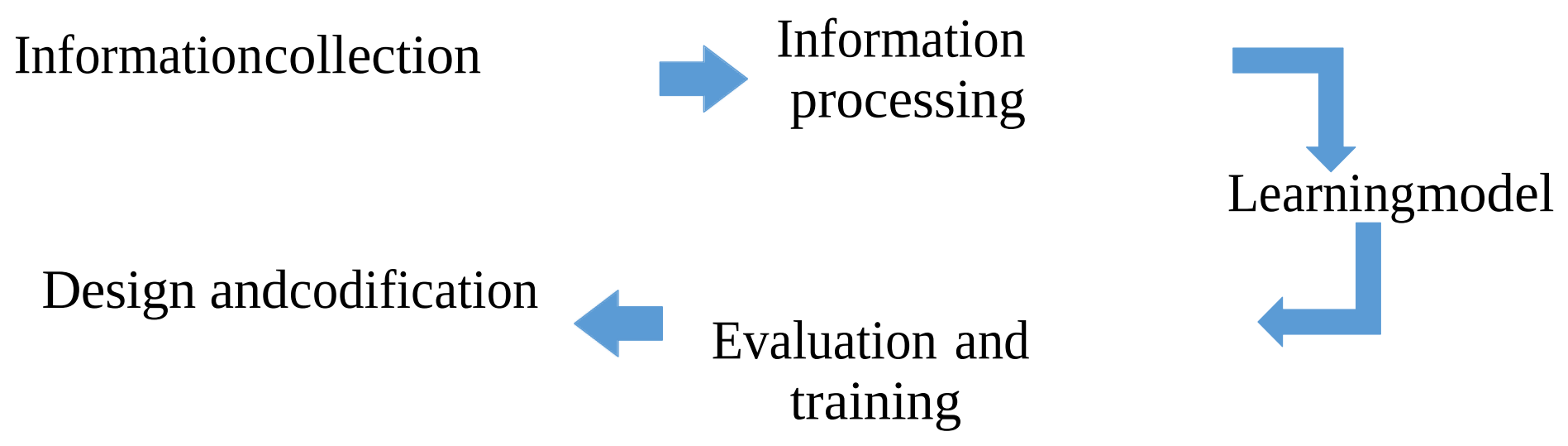

***Fig 1.** Applied Method*

## 2.2 Measuring instruments and techniques

• High-definition camera: Which will allow us to perform motion capture in natural environments (basketball court)

• Software Motive Optitrack: The Motive Optitrack software allows us to capture movements by means of a hardware consisting of 12 cameras, and an integration with a system of capture and digital processing of video material, from which a data extrac-tion is performed through flat files that give us detailed information of the behavior ofthe variables mentioned above.

• Python programming language for digital image processing.

• Anthropometric and goniometric measurement techniques: To obtain the data and characteristics of the upper limbs of each player in the sample under study.

• Weka Software: For analysis of measures of central tendency, correlational analysisof variables (to establish how variables interact and their relationship to an effective technique), and classification trees to analyze results. Motion capture is performed withMotive Optitrack software and hardware or with a high-definition camera as can be seen on figure 2.

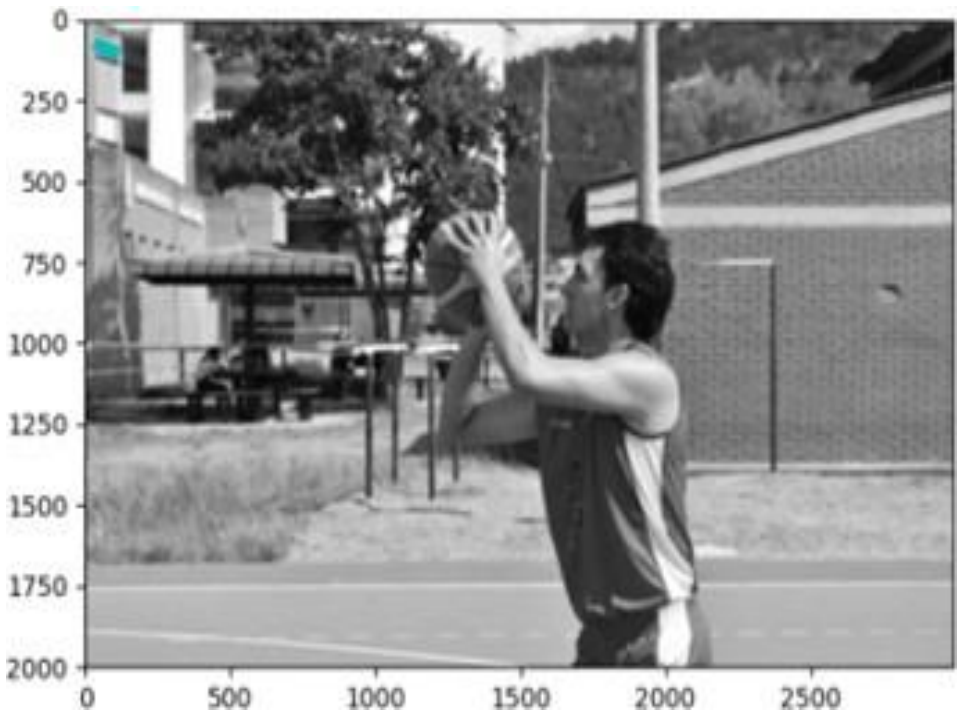

***Fig 2.*** *Motion capture.*

## 2.3 Analysis procedure sequence

• Capture of participants movements during the execution of the free-throw shooting technique, using high-definition cameras and motive optitrack software.

• Processing of images collected from high-definition cameras using the Python pro- gramming language.

• Storage of data extracted through image processing, this will be done by integratingit with the documentation in flat files obtained in the measurement and anthropometriccharacterization of the participants.

• Extraction of data obtained in flat files with the motive optitrack software, and sub- sequent integration with the flat file containing the anthropometric characterization anddigital processing data.

• Classification of variables according to successes during motion capture sessions.

• Use of Weka software for data analysis and statistical processing of the data stored in the flat file, according to the classification assigned to the variables. This will allowus to demonstrate by means of correlation matrices the interaction of the variables dur-ing the execution of the movement, and its relation with an optimal technique accordingto the successes or deficiencies in the attainment of scores by the players according to their throwing technique.

• From the above, using the software, a classification tree is generated, which will giveus results of the behavior of the variables to achieve an optimal launching technique. Based on the data collected from the free throw capture as seen in figure 3.

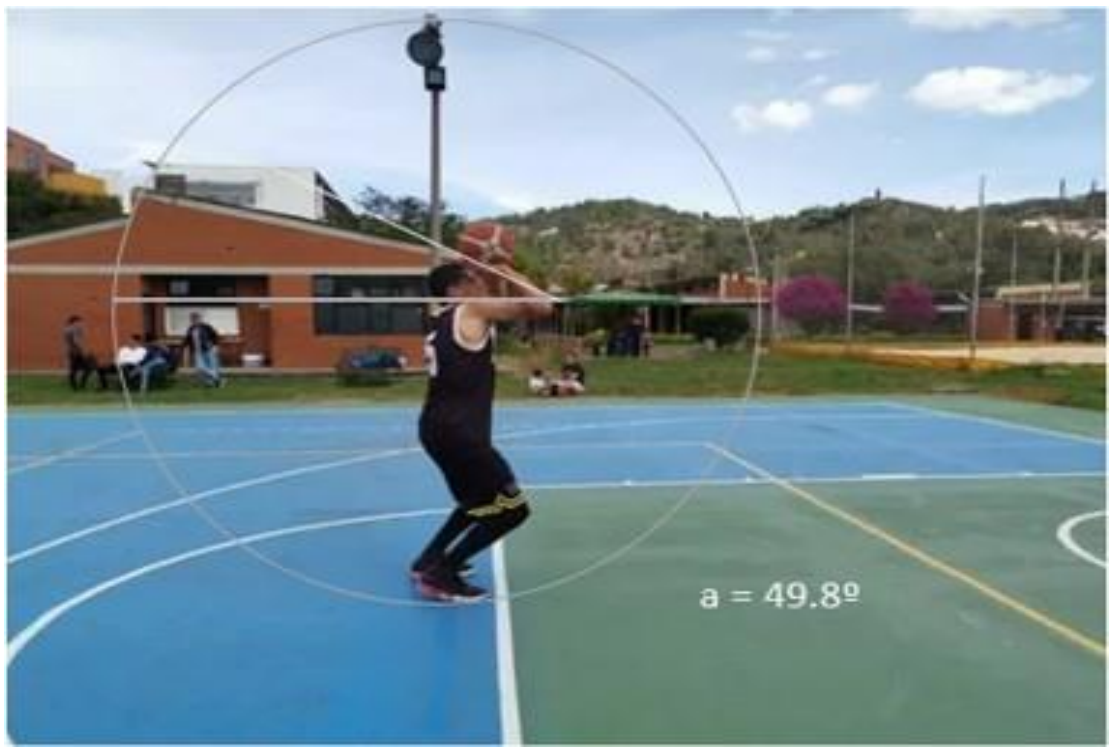

***Fig 3.*** *Free throw capture*

### 2.4 Data structure for analysis

The data structure presented is non-linear, therefore, it behaves as a tree within which different components are correlated and whose trends and projec-tions in terms of behavior are related to external variables [8].

For this particular case, the level structure of the data is reflected in a first level that represents the general data of the athlete, the next level corresponds to the variables to be analyzed, after this analysis a result of data that affect thebehavior of the throwing mechanics is obtained (Incident variables) and those that do not meet this condition are discarded. From this, a weight is assigned to each variable depending on its incidence in the launching technique, from which a probabilistic estimation is made by means of a Poisson distribution, because the result of a free throw launching is subject to several external con- ditions and the probability of occurrence of a high effectiveness must be takenas an event that occurs, but not in a constant way [9]. Finally, we have the finalresult corresponding to the execution of the free throws by the athletes. Data structure can be seen in figure 4.

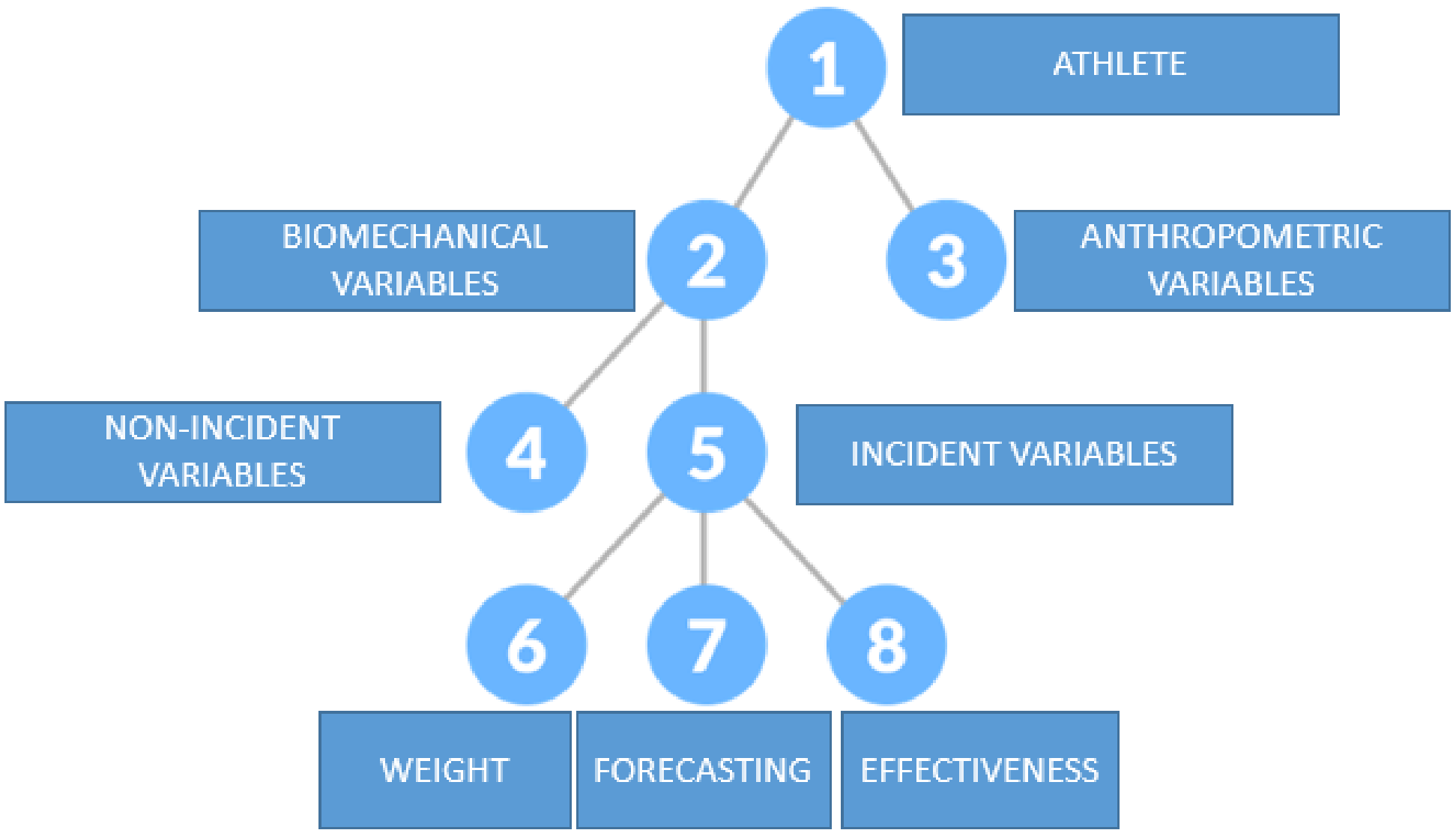

***Fig 4.*** *Data Structure*

## 3 Results and discussion

Anthropometric characterization
An anthropometric characterization of the length of the upper limbs that interact withinthe mechanics of free-throw shooting was carried out, which provides us with an indi-cator of the variables analyzed in order to interpret their correlation with the effective-ness of free-throw shooting. The characterization was performed as shown in table 1.

***Table 1.*** *Characterization*

| Athlete | Height | Arms length | Hands length | Forearm length | Elbow-Shoulder |
|---|---|---|---|---|---|
| **Player 1** | 175 cm | 168 cm | 17 cm | 32.5 cm | 40 cm |
| **Player 2** | 184 cm | 189 cm | 20 cm | 28 cm | 35 cm |
| **Player 3** | 173 cm | 175 cm | 20 cm | 27 cm | 28 cm |
| **Player 4** | 178 cm | 181 cm | 20 cm | 27 cm | 35 cm |
| **Player 5** | 191 cm | 190 cm | 20 cm | 32 cm | 49 cm |
| **Player 6** | 188 cm | 183 cm | 20 cm | 27 cm | 37 cm |

**Arm-forearm angle**

The analysis of the angle formed between arm and forearm gives us information aboutthe correct execution of the throwing technique because the angle called a, can oscillatebetween 45º at the beginning of the movement and 85º at the end of it.

**Wrist flexion**

The flexion of the wrist is analyzed in the final sequence of the free throw, what we intend to measure for the purpose of this study are the initial and final angle correspond-ing to the flexion from its beginning until the moment in which the player gets rid of the ball.

**Effectiveness analysis by classification**

A classification analysis was performed by means of the J48 algorithm, taking as a reference class the effectiveness presented by the players during the free throw sessions,in which the classification gave us results about the behavior of the variables accordingto the value of the class, the class was defined as high, medium and low according to the percentage of successes for every 10 free throws made as can be seen in table 2.

***Table 2***. *Standard definition of classification*

| Launches | Successful | Effectiveness (Class defined) |
|---|---|---|
| Quantity 10 | **0 - 4** | **Low** |
| Quantity 10 | **5 - 6** | **Regular** |
| Quantity 10 | **7 - 10** | **High** |

Based on this definition, a classification is made with respect to each of the variables, which yields the following results from the analysis carried out with respect to the height of the athletes, which shows a higher effectiveness in athletes with greater stat-ure, and establishes a relationship between those who meet the denomination of effec-tiveness (Low/Regular/High) for every 5 athletes as shown in table 3.

***Table 3.*** *Analysis of the height variable by classification*

| Height | Effectiveness | Relation (Compliant/Total) |
|---|---|---|
| <= 175 cm | **Low** | **3/5 Low – 2/5 High** |
| > 175 cm | **High** | **1/5 Low – 4/5 High** |

In the same way, analyses were performed with respect to the other variables corre- sponding to upper limb extension such as elbow-shoulder length, forearm length, and arm length. These variables do not yield results that indicate a direct correlation with respect to the effectiveness of. As described in table 4.

***Table 4.*** *Analysis of the elbow/shoulder length variable by classification*

| Elbow/shoulder length | Effectiveness | Relation (Compliant/Total) |
|---|---|---|
| <= 35 cm | **High** | **2/5 Low – 3/5 High** |
| > 35 cm | **Low** | **3/5 Low – 2/5 High** |

In addition, a classification analysis was performed for the biomechanical variables es-tablished for the study, such as the angle formed between the arm/forearm, and wrist flexion. This can be seen in table 5 and table 6.

***Table 5.*** *Result of analysis by classification of the arm/forearm angle*

| Arm/forearm angle Initial | Arm/forearm angle Final | Effectiveness | Relation (Compliant/Total) |
|---|---|---|---|
| 45° | **85°** | **High** | **1/5 Low – 4/5 High** |
| 55° | **85°** | **Regular** | **2/5 Low – 3/5 High** |
| >55° | **>= 85°** | **Low** | **5/5 Low – 0/5 High** |

***Table 6.*** *Result of analysis by classification of the wrist angle*

| Wrist angle Initial | Wrist angle Final | Effectiveness | Relation (Compliant/Total) |
|---|---|---|---|
| 110° | **95°** | **High** | **1/5 Low – 4/5 High** |
| <110° | **>°95** | **Low** | **3/5 Low – 2/5 High** |
| <100° | **> 90°** | **Low** | **4/5 Low – 1/5 High** |

## 4 Conclusions

The results obtained from the analysis allow us to infer that athletes with a greater height tend to have better effectiveness in terms of free-throw shooting, however, it is perhaps the anthropometric factor that has more impact in relation to the effectivenessand efficiency of the technique, because the anthropometric measurements of upper limbs do not show such a direct relationship and represent only a characterization of the athlete.

The biomechanical variables analyzed, such as the initial and final angles of the wrist,arm and forearm at the moment of executing the technique, have a direct relationship with the effectiveness of the throw, that is, if the technique is effective it is reflected inthe behavior of these variables through the angles, this allows us to demonstrate the displacement of the upper limbs during the execution of the technique and subsequentscores.

## References

1. Arora, C., Singh, P., and Varghese, V., "Biomechanics of core musculature on upper extremity performance in basketball players," Journal of Bodywork and Move- ment Therapies, vol. 27, pp. 127-133, 2021.
2. Lam, W. K., Kan, W. H., Chia, J. S., and Kong, P. W., "Effect of shoe modifica- tions on biomechanical changes in basketball: A systematic review," Sports Biome- chanics, vol. 21, no. 5, pp. 577-603, 2022.
3. Lempke, L. B., Oh, J., Johnson, R. S., Schmidt, J. D., and Lynall, R. C., "Single-versus dual-task functional movement paradigms: a biomechanical analysis," Journal of Sport Rehabilitation, vol. 30, no. 5, pp. 774-785, 2021.
4. Martínez, S. P. A., Quintero, Y. J. V., Huertas, L. M. R., and Salazar, L. G., "Métodos ergonómicos observacionales para la evaluación del riesgo biomecánico asociado a desordenes musculoesqueléticos de miembros superiores en trabajadores 2014-2019," Revista Colombiana de Salud Ocupacional, vol. 10, no. 2, pp. 31-42, 2020.
5. Medina Cabrera, M. L., Toledo Ríos, R., and Sánchez Oms, A. B., "Procedimiento para el análisis biomecánico de la variabilidad del movimiento en el lanzamiento de disco."
6. Pan, H., Li, J., Wang, H., and Zhang, K., "Biomechanical analysis of shooting performance for basketball players based on Computer Vision," in Journal of Physics:Conference Series, vol. 2024, no. 1, p. 012016, Sept. 2021.
7. Pérez Toasa, R. D., "Evaluación biomecánica en el proceso de aprendizaje de la técnica del lanzamiento libre del baloncesto en escolares de la Unidad Educativa TeresaFlor en el periodo abril–agosto 2021," Bachelor's thesis, Universidad Técnica de Ambato-Facultad de Ciencias Humanas y de la Educación-Carrera de Pedagogía de laActividad Física y Deporte, 2021.
8. Srinivas, T. A. S., Somula, R., Aravind, K., and Manivannan, S. S., "Pattern Pre-diction Using Binary Trees," in Innovations in Computer Science and Engineering: Proceedings of 8th ICICSE, vol. 171, pp. 43, 2021.
9. Yang, C., and Jia, M., "Health condition identification for rolling bearing basedon hierarchical multiscale symbolic dynamic entropy and least squares support tensor machine–based binary tree," Structural Health Monitoring, vol. 20, no. 1, pp. 151-172,2021.
10. Zhang, J., Chi, B., Singh, K. M., Zhong, Y., and Ju, C., "A binary-tree elementsubdivision method for evaluation of singular domain integrals with continuous or discontinuous kernel," Engineering Analysis with Boundary Elements, vol. 116, pp. 14-30, 2020.

# A Bayesian Approach to Duplicate Detection forReal Estate Listings

**Felipe Dioguardi[1], Leandro Antonelli[1], Marcos May[3], Juan Pablo del Río[3], and Diego Torres[1,2]**

[1] LIFIA, CICPBA-Facultad de Informática, UNLP, Argentina
{name.surname}@lifia.info.unlp.edu.ar
[2] Departamento de Ciencia y Tecnología, UNQ, Argentina
[3] LINTA-CICPBA, FaHCE, UNLP, Argentina
[4] CONICET, Argentina

**Abstract.** The availability of large amounts of real estate data on the internet presents a great opportunity for analysts and statisticians to derive insights. However, ensuring the quality of the data can be chal- lenging due to the presence of duplicate listings. This study proposes a duplicate detection strategy for a real estate knowledge graph contain- ing listings scraped from different web pages. The presented approach involves using Duke to discover the implicit owl: sameAs links between records, which achieved a precision of 66.8%, a recall of 70.4%, and an F-measure of 68.6%. Although many strategies are based on using a single metric function to compare records, it was found that segmenting this comparison according to the attributes compared and giving each seg- ment a different weight on the matching result is a successful method forsolving this problem. The strategy was evaluated through a ground truthdataset created by domain experts, which consisted of a real estate list-ing knowledge graph with duplicate and unique entities. This approach effectively cleaned the knowledge graph of noisy data and can be use- ful for making accurate statistical analyses in real estate domains. The suggested approach can be used to create a real estate observatory thatanalyses real estate listings knowledge from different sources to generate useful statistics. Analysts can use this tool to extract valuable informa- tion and make data-driven decisions in the real estate market.

## 1 Introduction

The sheer volume of online information poses advantages and disadvantages when extracting insights from it. While more data can provide new ways of measuring and understanding the domain, poor-quality data can lead to inaccu- rate conclusions and bad decision-making [3]. Therefore, prioritizing data qualityover quantity is crucial when managing a large knowledge base. For this reason, duplicate detection is a key task during the preprocessing stage of information.

Over the years, many researchers have proposed strategies to detect duplicaterecords [1,9]. Still, their effectiveness varies depending on factors such as the amount of knowledge, the type of information, and the domain it belongs to. This paper evaluates the performance of a duplicate detection strategy based ona Bayesian Classifier over a real estate listings knowledge base built from recordsextracted from different websites.

Duplicate detection, often called record linkage, entity resolution, and data deduplication, is the process of identifying different records that refer to the same entity [5]. Such a problem is not trivial, as it involves some challenging steps, from the careful selection of record pairs to be compared to the methodof comparison of these records and the final linking process. Removing duplicate records from a knowledge base is critical for ensuring reliable and trustworthy data [10], as their presence can impair statistical analysis and undermine their results.

Numerous tools and strategies have been proposed to address duplicate de- tection. For instance, vectorization techniques [13] and fuzzy matching [7] have been suggested to identify potential matches. Likewise, both machine learning models [2] and rule-based systems [11] have been explored to compare potentially matching records. Finally, comprehensive frameworks that automate all aspects of instance matching have also been developed [15]. However, the success of these methods heavily relies on their adaptability to the specific char- acteristics of each domain, and the selection of an appropriate method can beinfluenced by factors such as the size of the dataset, the data quality, the natureof the matching functions, and the requirements of the application.

Specifically, in real estate listings, the challenges of deduplication intensify further. The real state market involves having the same building appear in mul- tiple listings. This can happen if, for example, more than one real estate agent offers the building, and each of the agents publishes it in separate listings on the same site. The same logic can apply when the agents publish the same build-ing on different sites. Additionally, completely different listings can have similar descriptions due to templates used by real estate agents or websites. The geo- graphic coordinates of the offered properties can also pose challenges, depending on who publishes them and how. On top of that, naming conventions can vary across sources, making it difficult to identify duplicates without a comprehensive approach that considers all relevant attributes of each listing.

This research aims to assess the efficacy of a duplicate detection strategy thatleverages Bayes' Theorem to aggregate probabilities, applied to a knowledge baseof real estate listings. The performance of this approach is evaluated using the precision, recall, and F-measure metrics, and an unbiased test is ensured by usinga knowledge base with a manually curated ground truth dataset.

The remainder of this paper is organized as follows. Section 2 provides an overview of related work in the field of duplicate detection. Section 3 defines theproblem of duplicate detection and outlines the challenges specific to real estatelistings. Section 4 describes the knowledge graph used in the study, including its sources and characteristics. In Section 5, the paper presents the metrics used to evaluate the effectiveness of our duplicate detection strategy. Section 6 describesin detail the duplicate detection tool used in this study. Section 7 presents theresults of the evaluation. Finally, in Section 8 the paper concludes the evaluation and provides insights into the relative strengths and weaknesses of the different approaches while discussing potential areas for future work.

## 2 Related Work

This section provides an overview of the related work in the area of duplicate de-tection. Several works have surveyed the literature on duplicate detection tools, such as Elmangarmid et al. [5] in the context of databases, Assi et al. [1] with their survey on instance matching systems, and Huaman et al. [9] with their review of the state-of-the-art for duplicate detection in knowledge graphs.

Niu et al. [14] conducted a thorough analysis of the literature on duplicate record detection, covering various techniques, algorithms, and tools used to de- tect duplicate records in databases. They also discussed common errors thatcan make duplicate matching difficult and presented methods for improving effi-ciency and scalability in approximate duplicate detection algorithms. Zong et al. [16] proposed an approach for deduplication that involves extracting semantic information from the material record names and representing it in a form that allows for effective comparisons to be made.

Manku et al. [12] developed a near-duplicate detection system for a multi- billion-page repository, demonstrating the applicability of Charikar's fingerprint- ing technique for this goal, and evaluating it over real data to confirm the prac- ticality of their design.

Dong et al. [4] studied how to improve truth discovery by detecting de-pendence between sources and analysing the accuracy of sources. They devel- oped Bayesian models that discover copiers by analysing values shared between sources, and the results of their models can be considered as a probabilistic database, where each object is associated with a probability distribution of var- ious values in the underlying domain. Experimental results showed that their algorithms can significantly improve the accuracy of truth discovery and arescalable when there are numerous data sources.

## 3 Problem Definition

This section will explore the problem of detecting duplicates in knowledge graphs, examining its complexity and the common structure of potential solutions. Firstly,essential concepts related to knowledge representation and the definition of du- plicates will be introduced.

A knowledge base is a collection of concepts related to a specific domain thatcan be stored in various formats, such as XML or RDF. These formats allow for the creation of a knowledge graph, which represents knowledge as a network of entities linked by relations, capturing their semantic relationships and enabling the discovery of new knowledge through implicit connections. Ontologies are a key component of knowledge bases, providing a formal and structured way of representing concepts and their relationships within a domain. They define the vocabulary and rules for describing entities and their relationships, allowing for interoperability between different knowledge bases and the sharing of knowledgeacross domains.

Duplicates in knowledge graphs occur when two nodes representing the sameentity have different identifiers. Nodes that share the same attributes but differ in their identifiers are called exact duplicates. On the other hand, nodes that differ in some or all of their attributes but still refer to the same entity, are callednear-exact or fuzzy duplicates. For example, consider the two nodes shown in Figure 1. Despite having identical attributes (e.g., title, location, price), theyhave different identifiers. Similarly, in Figure 2, the nodes show some attribute differences (e.g., title, price), but still refer to the same real estate. To capture this fact, it is important to establish an owl: sameAs relationship between these nodes as well, indicating they both represent the same entity. All these duplicatesshould be linked through the owl: sameAs relationship to indicate that they describe the same entity, forming the basis for detecting duplicates in knowledgegraphs.

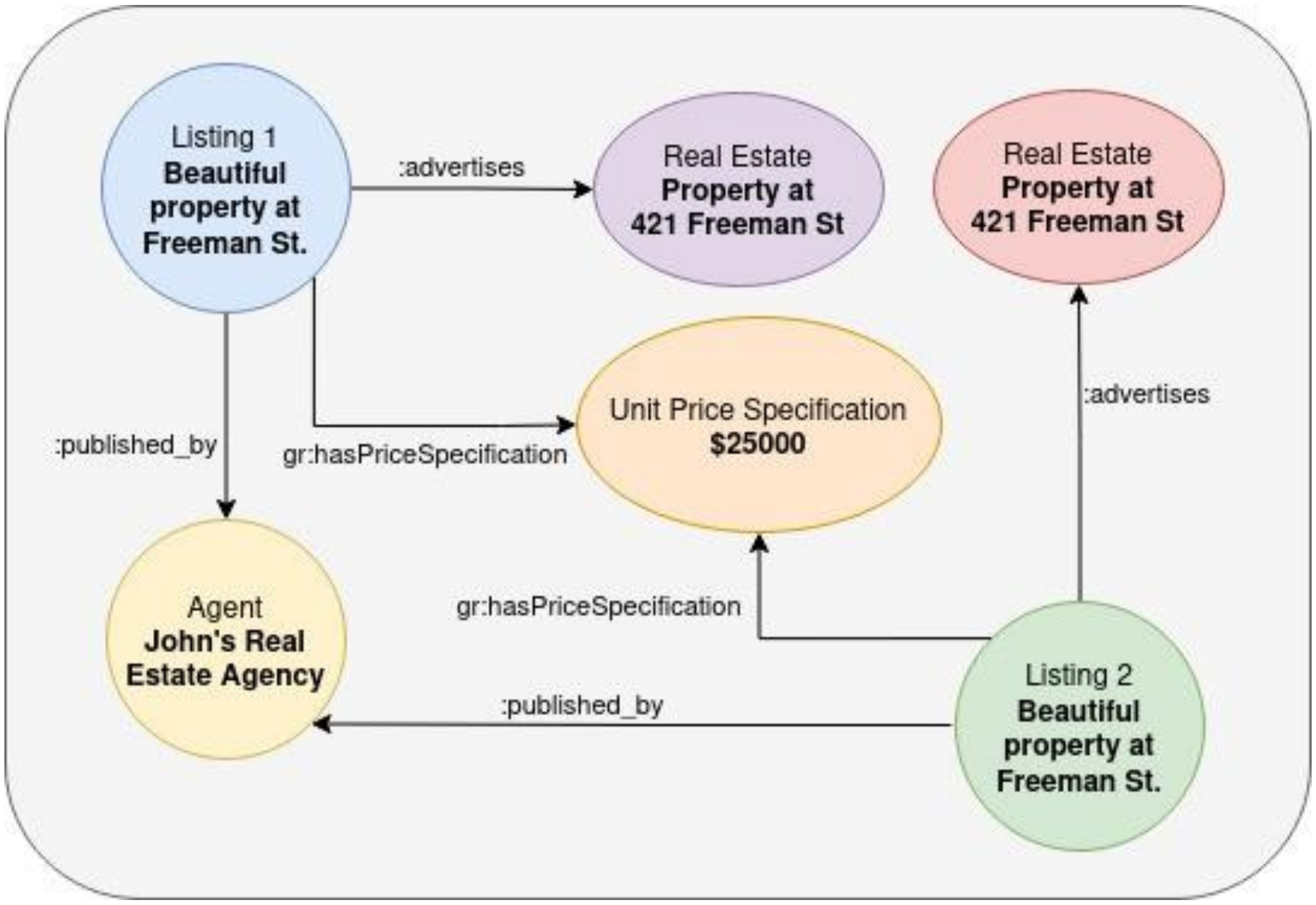

***Fig. 1.*** *Exact duplicates in a knowledge graph.*

The problem of duplicate detection has different names depending on thecontext, such as record matching, record linkage, deduplication, entity resolution, and instance matching. Instance matching (IM) aims to discover duplicated entities between two knowledge graphs. Assi et. al [1] defined it as follows:

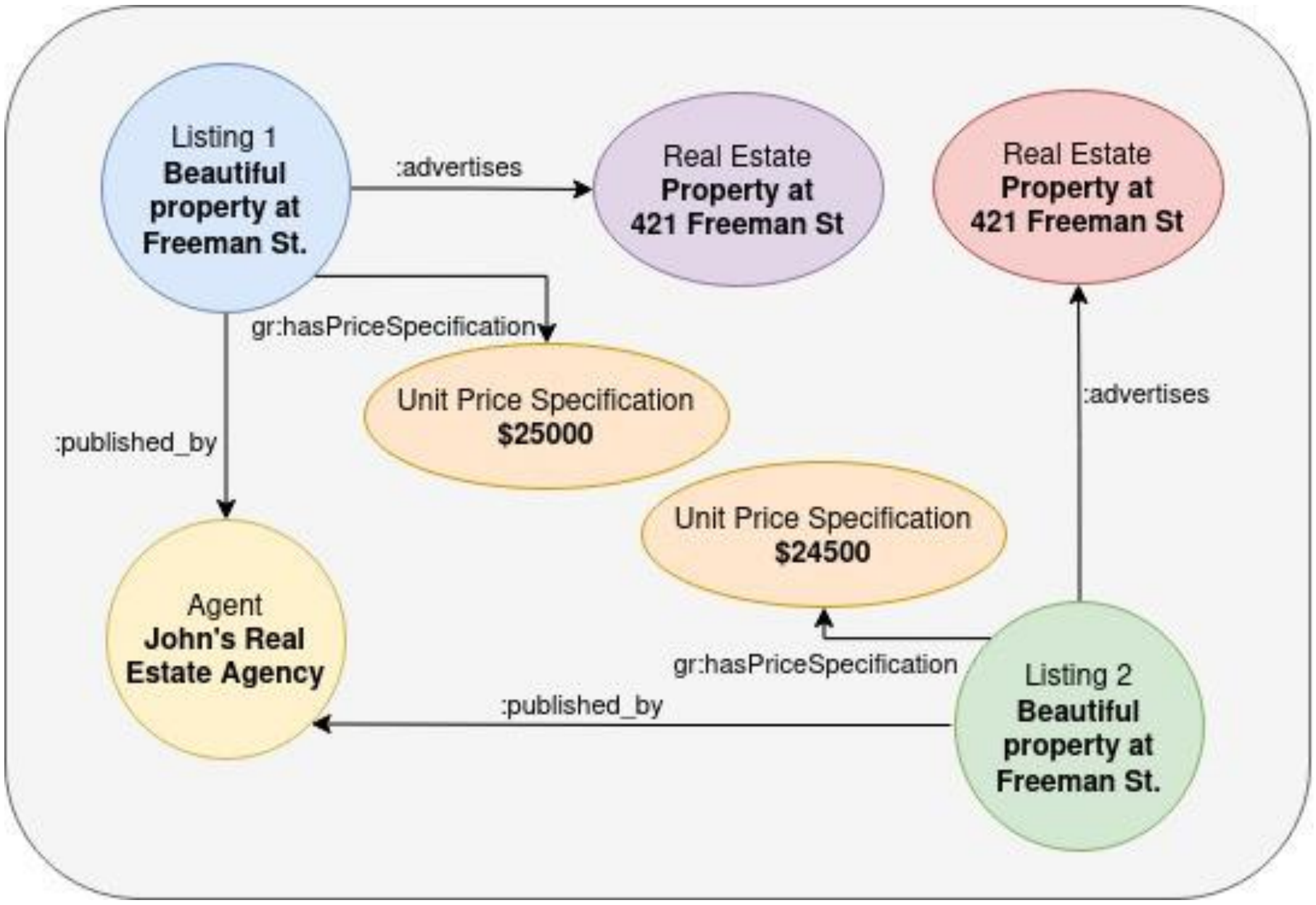

***Fig. 2.*** *Fuzzy duplicates in a knowledge graph.*

**Definition 1 (Instance matching).** *Given two sets of instances* S *and* T *belonging to two KBs* ($KB_1$ and $KB_2$)*, the aim of IM is to discover the set* M *of identity links owl: sameAs, according to a chosen identity criterion, whichare not already defined in the KBs. Formally,* $M = \{(i_1, i_2) \mid (i_1, i_2) \in S \times T, \langle i_1, \textit{owl: sameAs}, i_2\rangle, \langle i_1, \textit{owl: sameAs}, i_2\rangle \notin KB_1 \cup KB_2\}$.

To detect duplicates within a knowledge graph, duplicate detection can be defined as an instance matching problem where the graph is compared to itself, i.e., where $KG_1 = KG_2$.

Detecting duplicates in knowledge graphs is complex and depends on the sizeof the graph and the nature of the duplicates. Detecting duplicates efficiently involves four steps: preprocessing, blocking, matching, and clustering. Prepro- cessing cleans and normalizes the data to remove syntactical and formatting differences. Blocking divides, the data into smaller subsets to reduce the search space. Matching compares the attributes of candidate duplicates and determineswhether they refer to the same entity. Clustering groups together duplicates thatrefer to the same entity.

Detecting and resolving duplicates is crucial to ensure trustworthy and reli- able data in knowledge graphs. By improving the accuracy and effectiveness of duplicate detection in knowledge graphs, more informed decisions can be madeand better outcomes can be achieved.

## 4 Knowledge Graph

This section will discuss the knowledge graph used in the context of the duplicatedetection system for real estate listings.

The knowledge graph is a critical component of the duplicate detection sys-tem, as it stores and organizes real estate listings from various sources andcaptures the semantic relationships between them. The real estate listings in theknowledge graph come from different websites and agents, resulting in a highlevel of heterogeneity in the information. This heterogeneity can make it chal-lenging to detect duplicates due to the differences in data format and structure.

The knowledge base is stored in two different formats, CSV and RDF. CSV isa simple tabular format that is easy to work with, while RDF is a more complexgraph-based format that allows for semantic relationships to be captured. The choice of format depends on the specific requirements of the duplicate detectionstrategy being evaluated.

The schema defines the variables used in the CSV and ensures that all listings have a consistent set of attributes. Additionally, an ontology is used to createthe RDF graph and define relationships between the entities in the knowledge graph. The ontology is derived from established vocabularies and ontologies, such as GoodRelations[1], Schema.org[2], FOAF[3], and RealEstateCore[4] [8]. These resources provide a rich set of definitions and relationships leveraged to de- velop the knowledge graph. Specifically, entities such as RealEstateListing, RealEstate, and Agent are defined within the ontology, along with their corre-sponding relationships.

The attributes for the real estate listings include price, location, numberof bedrooms, and square meters. These variables are standardized across all listings to facilitate comparison and detection of duplicates. The ontology, onthe other hand, defines the relationships between entities in the graph, such as the relationship between a property and its location or between a property and its agent.

The knowledge graph example depicted in Figure 3 provides an illustrationof how the ontology can be effectively used to capture the semantic relationshipsbetween different entities in real estate listings. For example, the graph demon- strates how a particular real estate listing can be associated with a specific real estate agency, such as John's Real Estate Agency. Additionally, the listing can have a relationship with its address, which is represented by the Feature entity. If two listings are advertising real estate located at the same address, theywould both share a relationship with that instance. By effectively capturing and representing these semantic relationships, the knowledge graph plays a critical role in identifying potential duplicates by comparing not only the attributes of the listings themselves but also the context in which they exist.

---

[1] GoodRelations: http://purl.org/goodrelations/
[2] Schema.org: https://schema.org/
[3] FOAF: http://xmlns.com/foaf/0.1/
[4] RealEstateCore: https://realestatecore.io/

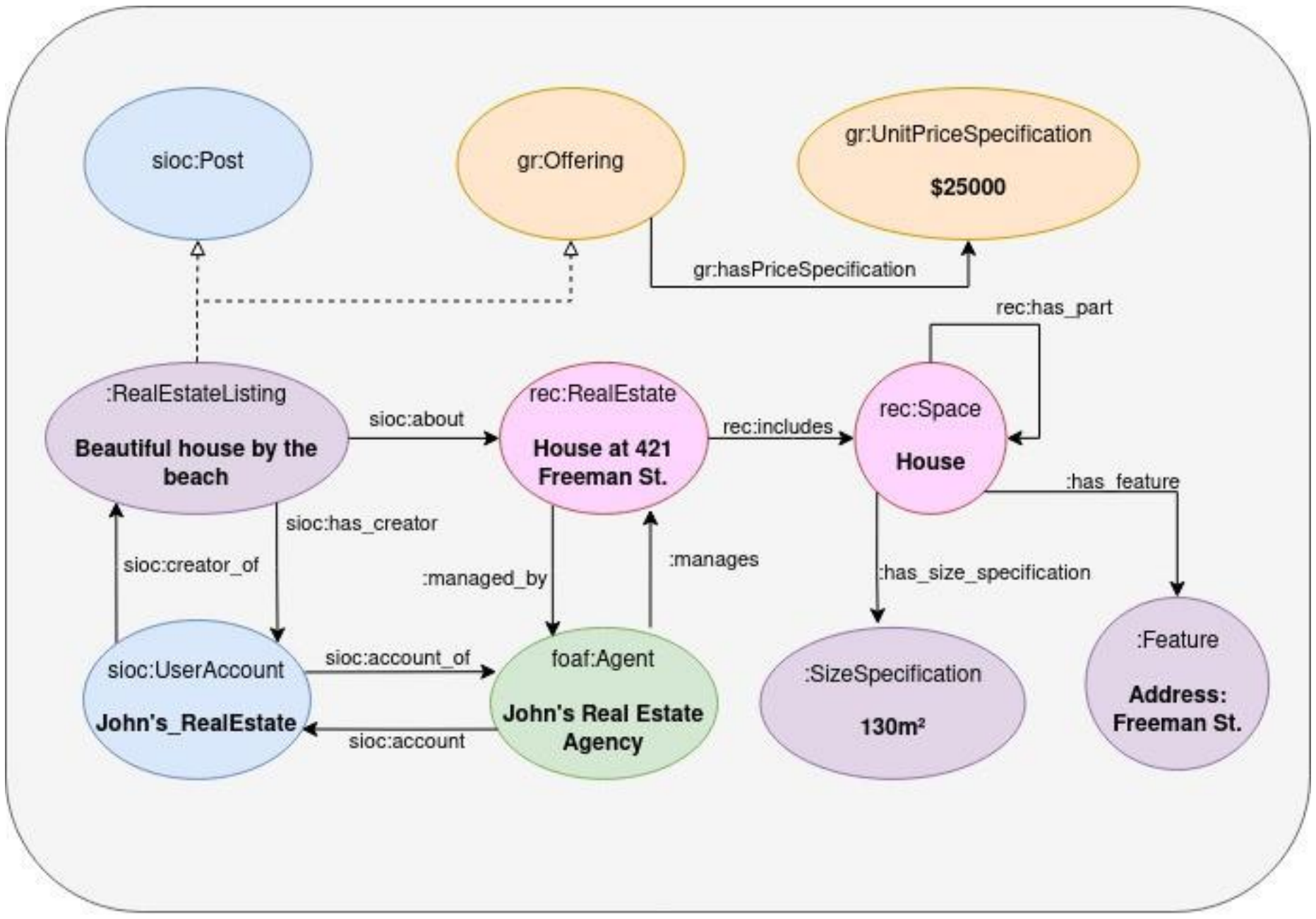

***Fig. 3.*** *Example knowledge graph featuring a real estate listing.*

## 5 Metrics

This section defines the metrics used to evaluate the performance of duplicate detection algorithms. It will explain the insights provided by each metric andhow they will be calculated.

Duplicate detection is a binary classification problem by definition, wherethe algorithm is provided with a pair of records and must classify them as eithera match or a non-match. Each of the classifications done by the algorithm falls into one of the following categories:

- True Positive (TP): the algorithm correctly identifies a match between tworecords.
- False Positive (FP): the algorithm incorrectly identifies a match between tworecords.
- True Negative (TN): the algorithm correctly identifies that two records arenot a match.
- False Negative (FN): the algorithm incorrectly identifies that two recordsare not a match.

The number of matches in each category will determine how well the algo- rithm performs overall.

Three metrics will be used for evaluation: precision, recall, and F-measure[6]. Precision is the fraction of true positive results among all positive results, thus measuring how accurate the algorithm is at classifying a pair of records asa match. It is defined as:

$$\text{Precision} = \frac{TP}{TP + FP} \quad (1)$$

Recall, on the other hand, is the proportion of true positives among all true positives. It is useful to describe how comprehensive the algorithm is in detecting duplicate pairs.

$$\text{Recall} = \frac{TP}{TP + FN} \quad (2)$$

For example, if a deduplication tool measures *Recall* = 0.3 and *Precision* =0.9, then it means it is 90% accurate in its detection of duplicate records, but only identifies 30% of all duplicate records in the database.

The F-measure is a way to combine precision and recall into a single metric, defined as:

$$\text{F-measure} = 2 * \frac{Precision * Recall}{Precision + Recall} \quad (3)$$

## 6 Bayesian Approach

The following section presents the Bayesian strategy employed for duplicate detection, along with the tool chosen to implement the pipeline. Each componentof the classification pipeline is described in detail, specifying how they contributeto solving the problem.

Many existing tools for record comparison in duplicate detection often relyon a single similarity function and treat all attributes uniformly as strings or character vectors. However, the evaluated approach suggests a departure from this conventional method by using a segmented matching process. This strategy advocates for segmenting the comparison based on the attributes of the listings being evaluated, assigning distinct weights to each segment to obtain an overall matching result.

To determine whether two listings are duplicates, the approach matches theirrespective attributes. Then, the two values of each attribute get compared to calculate a similarity score, which is assigned to that attribute. Finally, these scores are combined using Bayes' Theorem to generate an overall similarity scorethat represents the similarity between the two records. This approach allows for the computation of a probability, based on the available evidence, indicatingthe likelihood that the records are duplicates. By comparing this probability toa similarity threshold specified by the user, the determination of a match or non-match between the records can be made.

By adopting this approach, a more nuanced and comprehensive evaluation of record similarity is achieved, enabling the generation of probabilistic evidence to support the duplicate detection process. This methodology empowers users with greater control over the threshold for considering records as duplicates.

## 6.1 Duke

Duke[5] is a powerful and versatile entity resolution engine written in Java. Al- though it was developed as part of the document archive system Sesam [7],Duke is domain-agnostic, meaning it can be used with data from any domainand structure.

While Duke operates on records as the data entities for deduplication, its out-put consists of links that represent owl: sameAs connections between duplicated records. Duke's duplicate detection process involves three steps, as described below.

First, Duke loads each record in the knowledge base and preprocesses its attributes using Cleaners. These Java objects implement the Cleaner inter- face and standardize the data by performing operations such as removing extra spaces, converting cases, and removing stop words.

Next, Duke employs a Database implementation to index each record. This allows it to call the function findCandidateMatches(record) for each one of the records, and match them with their most likely pair according to the index criterion.

After the blocking step, Duke matches each record pair by comparing their attributes using a Comparator. Duke's Comparator interface allows for the def- inition of custom comparison functions that assign similarity scores to record attributes, which are then aggregated into an overall score that represents the similarity between two records. Duke comes with several built-in comparators, including string distance measures like Jaro-Winkler and Levenshtein, and token-based measures like the Dice coefficient and Jackard index. As with the other interfaces, Duke's Comparator interface can be extended to include custom com-parison functions that consider domain-specific knowledge and characteristics.

To determine whether two records are duplicates, Duke computes a similarityscore between them using the approach described for matching. This similarity score represents the likelihood of the records being duplicates. Subsequently, Duke compares the resulting probability with a user-defined threshold. If the probability exceeds the threshold, the records are classified as duplicates.

## 6.2 Configuration and Similarity Strategies

Detecting duplicates with this Bayesian approach involves employing a combi- nation of similarity functions that consider the nature of the attributes being compared. Different similarity functions are used depending on the attribute type to ensure accurate matching. For instance, when comparing long stringslike descriptions, the Levenshtein distance metric proves to be effective. This metric measures the number of character replacements required to transform one text into another. In the context of the real estate knowledge graph, eachlisting comprises several semantic properties associated with some real estate, such as the description, number of bedrooms, and location. The Bayesian ap- proach leverages the concept of semantic similarity by composing the syntacticsimilarities among each specific property. This enables a comprehensive evalu- ation of the complex listings by integrating the results of individual attribute comparisons.

[5] Duke: https://github.com/larsga/Duke/

In configuring the real estate deduplication strategy, an essential component is the utilization of a Lucene[6] database during the blocking step. This step involves executing highly efficient full-text queries to identify potential duplicate candidates. Lucene's indexing and searching capabilities enable the strategy to effectively partition and search through the data, aiding in the identification of records with similar textual information.

In the preprocessing stage, the default cleaners were applied, primarily fo- cusing on tasks such as removing excess spaces and converting text to lowercase.This preparation step ensures consistent and standardized data for further analysis.

During the matching process, a combination of specialized comparators was employed to handle different types of data. For long strings of text, such as descriptions or addresses, the Levenshtein comparator was utilized to measure the similarity between records. Shorter strings like district names were evalu- ated using the Jaro-Winkler comparator. To handle numerical attributes suchas surface area and the number of rooms, dedicated numerical comparators wereemployed. These comparators use the ratio between the largest and smallest numbers, thus taking into account the numerical significance and proportional- ity of the attributes. Furthermore, a geolocation comparator was employed to assess the distance between two coordinates. If the distance exceeded 100 meters, the records were deemed completely apart, indicating a significant spatial sepa- ration. The specific values for the configuration are described in Table 1, which also shows the range of the similarity threshold selected for each variable. Theserepresent the probability of a pair of records being duplicates, given that the values for that attribute are exactly the same or completely different, according to the comparator used. For instance, given that the titles of two real estate listings are exactly the same, the probability that the listings refer to the same entity is 70%. Similarly, if the titles are completely different, that probability reduces to 19%. The probabilities calculated for each attribute are the ones that get reduced to a single number using Bayes' Theorem.

The ranges for each attribute were selected based on how helpful they arein distinguishing between listings. For instance, if an attribute such as the titleor price is very informative, it was given a wider range. This means that if two listings have the same value for that attribute, they are likely to be the same,and if they have very different values, they are likely to be different listings.

In the Bayesian approach used, probabilities closer to 0 or 1 are considered more important than those closer to 0.5. For example, the fact that two listings have the same covered surface value does not provide much information to deter-mine if they are referring to the same or a different entity. However, if they havecompletely different covered surface values, they are more likely to be different listings. For that reason, the similarity threshold for the covered surface ranges from 0.09, which means that the probability of two listings being the same is 9%,to 0.55, which means the algorithm should not give it much importance. This range is set based on the degree of impact each attribute has on the similarity calculation.

[6] Lucene: https://lucene.apache.org/

***Table 1.*** *Configuration used to detect duplicates in the knowledge graph.*

| Attributes | Comparator | Min. similarity threshold | Max. similarity threshold |
|---|---|---|---|
| Title | Levenshtein | 0.19 | 0.90 |
| Description | Levenshtein | 0.39 | 0.70 |
| Price | Numeric | 0.05 | 0.80 |
| Maintenance fee | Numeric | 0.05 | 0.60 |
| Property type | Exact | 0.10 | 0.56 |
| Age | Exact | 0.07 | 0.50 |
| Coordinates | Geoposition | 0.20 | 0.80 |
| Address | Levenshtein | 0.09 | 0.90 |
| District | Jaro-Winkler | 0.39 | 0.51 |
| Total surface | Numeric | 0.10 | 0.67 |
| Covered surface | Numeric | 0.09 | 0.55 |
| Land surface | Numeric | 0.07 | 0.60 |
| Amount of rooms | Exact | 0.10 | 0.61 |
| Amount of bathrooms | Exact | 0.17 | 0.51 |
| Amount of garages | Exact | 0.10 | 0.60 |
| Amount of bedrooms | Exact | 0.10 | 0.60 |

# 7 Evaluation

This section will evaluate the performance of the duplicate detection strategy presented previously over a real estate listing knowledge base. The creation ofa ground truth dataset using expert labelling will be explained first. Then, the results of the evaluation will be presented, followed by a discussion of the performances achieved by the Bayesian strategy.

## 7.1 Ground Truth Dataset Generation

The term ground truth refers to a dataset that contains reliable information used as a standard for evaluating different algorithms. The ground truth dataset used in this study was generated to measure the performance of the duplicate detection strategy presented earlier. It comprises two separate files: a knowledge base that contains all the information about each real estate listing, and a linkfile that specifies which listings are duplicates of one another.

To create this ground truth dataset, a team of experts began by using a web scraping tool to generate a knowledge base of real estate listings. To minimize the risk of missing any duplicates, the dataset was divided into smaller subparts based on the listings' proximity, as determined by their latitude and longitude. The experts manually reviewed each subpart to identify duplicates, comparing various attributes such as title, description, price, number of bedrooms, age, typeof property, and address. They recorded the duplicates found in a spreadsheet, resulting in 3688 rows, each representing a set of listings that were duplicates of each other. Each set may contain two or more listings since multiple listings can represent the same property. The experts then removed any listings that were not labelled on the spreadsheet, resulting in a smaller knowledge base of 9333 listings.

To create the ground truth dataset, a script was developed to randomize the spreadsheet and split it in half. One half represented duplicates, while the other half represented non-duplicates. The first half was used to create a new knowledge base containing listings known to be duplicated, and the link file recorded which listings were duplicates of each other. For the second half, one random listing was selected from each set to be part of the new knowledge base,while the other listings were deleted. This ensured that the selected listings had no known duplicates in the new database.

### 7.2 Results

The performance of the duplicate detection strategy was evaluated using the ground truth dataset described above. Duke's output was compared with the links file in the ground truth dataset, and precision, recall, and F-measure wereused as performance metrics.

The ground truth file contained 6543 records, and our strategy was able to find 3139 out of 4455 correct links, resulting in a recall of approximately 70.4%. However, it's important to note that the strategy also found 1554 links thatwere not in the ground truth, resulting in a precision of only 66.8%. Overall,that leads to an F-measure of 68.6%.

This finding is interesting and could be attributed to various factors. Forinstance, if the real estate listings were published by the same agency, it's pos- sible that they used text templates for the description and title of the listing, leading to a higher number of false positives. In this case, it may be worth evaluating a different configuration with lower weights on these variables. An- other approach could be to use a deduplication strategy that's more flexible and doesn't rely solely on Bayesian classification, which assumes that the probabili- ties given by each variable are independent. For example, such a strategy could assign a smaller weight to the description variable, since listings published bythe same agent are likely to have similar descriptions.

Regarding the coordinates of the listings, it was found that real estate agents may not use the actual latitude and longitude of the property, either marking a point nearby or using the location of the agency itself. This can lead to many records having similar coordinate values, although they refer to different prop- erties. Therefore, the fact that two coordinates are different or the same doesn't necessarily confirm or discard a match, respectively.

However, it cannot be confirmed that missed duplicates are actually wrong, as it is also possible that the strategy was able to find duplicate listings thatthe experts who created the ground truth dataset overlooked. Although this file can serve as a representative sample of the larger real estate listings knowledge base, identifying this many duplicates can be challenging, which may have led to errors in the ground truth dataset. This underscores the importance of refining the duplicate detection strategy for this domain to improve the accuracy of the results.

## 8 Conclusions and Future Work

This study assessed a duplicate detection strategy applied to a real estate knowl-edge graph compiled from various sources. The approach involved computing a probability score indicating the likelihood of two records being duplicates, al- lowing for the identification of implicit owl: sameAs links between them. The evaluation of this approach resulted in a precision of 66.8%, recall of 70.4%, and F-measure of 68.6%. By segmenting the record comparison process based on the considered attributes and assigning weights to each segment, the strategy effec- tively addressed the challenge of duplicate detection. Furthermore, the approach demonstrated its ability to cleanse the knowledge graph by eliminating noisy data, thereby enhancing the quality and reliability of the real estate information available for analysis within the domain.

The strategy was evaluated by employing a ground truth dataset created bya domain expert, which consisted of a real estate listing knowledge graph with duplicate and unique entities, as well as a links file that identified the missing owl: sameAs links in the dataset.

In the future, alternative configurations will be explored to enhance the over-all accuracy of the presented approach and to identify more duplicates. Addition-ally, other techniques for detecting duplicate real estate listings, such as graph structural analysis, window sliding, graph embedding, and machine learning-driven solutions, will be experimented with.

## References

1. Assi, A., Mcheick, H., Dhifli, W.: Data linking over RDF knowledge graphs: A survey. Concurrency and Computation: Practice and Experience **32**(19), e5746 (2020).
2. Barlaug, N., Gulla, J.A.: Neural Networks for Entity Matching: A Survey. ACM Transactions on Knowledge Discovery from Data **15**(3), 1–37 (Jun 2021). https://doi.org/10.1145/3442200
3. Batini, C., Scannapieco, M.: Data Quality: Concepts, Methodologies and Tech-niques. Springer Science & Business Media (Jan 2006).
4. Dong, X.L., Berti-Equille, L., Srivastava, D.: Integrating conflicting data: the role of source dependence. Proceedings of the VLDB Endowment **2**(1), 550–561 (Aug 2009). https://doi.org/10.14778/1687627.1687690, https://dl.acm.org/doi/10.14778/1687627.1687690
5. Elmagarmid, A., Ipeirotis, P., Verykios, V.: Duplicate Record Detection: A Survey. Knowledge and Data Engineering, IEEE Transactions on **19**, 1–16 (Feb 2007). https://doi.org/10.1109/TKDE.2007.250581
6. Ferrara, A., Lorusso, D., Montanelli, S., Varese, G.: Towards a Benchmark for Instance Matching. The 7th International Semantic Web Conference p. 13 (Jan2008)
7. Garshol, L.M., Borge, A.: Hafslund Sesam – An Archive on Semantics. In: The Semantic Web: Semantics and Big Data, vol. 7882, pp. 578–592. Springer, Berlin,Heidelberg (2013).
8. Hammar, K., Wallin, E.O., Karlberg, P., Hälleberg, D.: The RealEstateCore Ontology. In: Ghidini, C., Hartig, O., Maleshkova, M., Svátek, V., Cruz, I., Hogan, A., Song, J., Lefrançois, M., Gandon, F. (eds.) The Semantic Web – ISWC 2019, vol. 11779, pp. 130–145. Springer International Publishing, Cham (2019). https://doi.org/10.1007/978-3-030-30796-7 9, https://link.springer.com/10.1007/978-3-030-30796-7 9, series Title: Lecture Notes in Computer Science
9. Huaman, E., Kärle, E., Fensel, D.: Duplication Detection in Knowledge Graphs: Literature and Tools. arXiv:2004.08257 [cs] (Apr 2020), arXiv: 2004.08257
10. Huang, Y., Chiang, F.: Refining Duplicate Detection for Improved Data Quality. TDDL/MDQual/Futurity@ TPDL (2017)
11. Koumarelas, l., Papenbrock, T., Naumann, F.: MDedup: duplicate detection with matching dependencies. Proceedings of the VLDB Endowment **13**(5), 712–725 (Jan2020).
12. Manku, G.S., Jain, A., Das Sarma, A.: Detecting near-duplicates for web crawling. In: Proceedings of the 16th international conference on World Wide Web. pp. 141–150. ACM, Banff Alberta Canada (May 2007). https://doi.org/10.1145/1242572.1242592, https://dl.acm.org/doi/10.1145/1242572.1242592
13. Mel, A., Kang, B., Lijffijt, J., De Bie, T.: FONDUE: A Framework for Node Disambiguation and Deduplication Using Network Embeddings. Applied Sciences **11**(21), 9884 (Jan 2021). https://doi.org/10.3390/app11219884, number: 21 Pub- lisher: Multidisciplinary Digital Publishing Institute
14. Niu, J., Niu, P.: An intelligent automatic valuation system for real estate based on machine learning. In: Proceedings of the International Conference on Artificial Intelligence, Information Processing and Cloud Computing. pp. 1–6. ACM, Sanya China (Dec 2019). https://doi.org/10.1145/3371425.3371454, https://dl.acm.org/doi/10.1145/3371425.3371454
15. Zhu, H., Wang, X., Jiang, Y., Fan, H., Du, B., Liu, Q.: FTRLIM: Distributed Instance Matching Framework for Large-Scale Knowledge Graph Fusion. Entropy **23**(5), 602 (May 2021). https://doi.org/10.3390/e23050602
16. Zong, W., Wu, F., Chu, L.K., Sculli, D.: Identification of approximately duplicatematerial records in ERP systems. Enterprise Information Systems **11**(3), 434–451 (Mar 2017). https://doi.org/10.1080/17517575.2015.1065513, https://www.tandfonline.com/doi/full/10.1080/17517575.2015.1065513

# Identification of Collaboration issues during the adoptionof Scrum in Software development teams

**Ingrith C. Muñoz[1], César A. Collazos[1] and Julio A. Hurtado[1]**

[1] IDIS Research Group, University of Cauca, 190001, Popayán, Colombia.
{ingrithc, ccollazo, ahurtado}@unicauca.edu.co

**Abstract.** Scrum is one of the most popular agile methodologies for software development, and its use in software development teams has been increasing in recent years. This increased use of Scrum is due in part to the benefits it offers, such as greater flexibility, greater collaboration among team members, and greater ability to adapt to changing project requirements.

Scrum is based on collaborative work among team members to achieve project goals, since team members are expected to work together interactively and col- laboratively to achieve these goals. Due to the above, collaboration issues ac- quire a relevant role, since they can have a significant impact on the Scrum adoption process. This article presents the collaboration problems encounteredin software development teams during the Scrum adoption process identified through semi-structured interviews conducted with leaders of software devel- opment teams adopting Scrum.

## 1 Introduction

Agile methodologies are a software product development approach characterizedby its flexibility [1] [2]. One of the most used agile methodologies is Scrum [3], which is a framework for the agile management of software projects; the use of agile methodologies and specifically Scrum has become more and more frequent in small software development organizations [4] [5]. However, the lack of understanding of Scrum during its adoption has led to multiple drawbacks or obstacles [4], even when the team has received training in this framework [6]. Due to the above, the scientific community has tried to establish the determining factors in the adoption of Scrum [7],in order to serve as support in an adoption initiative.

Among the critical factors associated with Scrum adoption difficulties are those re-lated to human behavior (personality, beliefs, culture, etc.), hereinafter the person factor [8] [7]; which is critical in the agile context considering that one of its princi- ples promulgates: "Individuals and their interactions, about processes and tools. This implies that great importance should be given to factors such as communication, in-teraction, integration, organizational culture, companionship and collaboration [9].

Particularly, collaboration is an important factor in software development, since itis an effective means to achieve business objectives and increase performance, for thisreason, collaborative practices acquire great value today for the software industry. Collaboration issues in software development teams can have a significant impact on the Scrum adoption process. In Scrum, collaboration is critical to team success, as team members are expected to work together interactively and collaboratively to achieve project goals [7]. If there are team collaboration issues, this can affect the effectiveness of Scrum, as team members may find it difficult to work together and achieve project goals. For example, if there is a lack of communication in the team, there can be coordination and cooperation problems, which can delay the develop- ment process. Similarly, if team members are not committed to the Scrum process, there may be a lack of motivation and enthusiasm, which can affect the quality of work and customer satisfaction. Therefore, it is important to address team collabora- tion issues to ensure a successful Scrum adoption.

Due to the above, there is a need to understand and identify the collaboration issuesfaced by software development teams during the adoption of Scrum.

This article presents in section 2 a state of the art and related works, in section 3 presents a study for the identification of collaboration factors that affect the adoption of Scrum and in section 4 the conclusions and future work.

## 2 Background

### 2.1 Agile methods

Agile methods are a set of techniques for project management and software devel- opment. They emerged as an effort to improve perceived and real weaknesses of con- ventional software engineering [8], but have also been extended to other types of projects. Agile methods comply with a number of principles and values stated in the agile manifesto [8]. In addition, agile methods are attentive to the project environmentand stakeholders, offer a more open working environment and a much more flexible frame-work [1] [2]

**Scrum**

Scrum is an agile project management methodology specifically referred to soft- ware engineering [3]. Since it has a flexible adaptation to change, it is recommended for use in complex projects with changing or poorly defined requirements, where early and highly complex deliverables are needed. Scrum focuses on how the team members should work so that the system is flexible and adapts to constantly changing conditions, therefore, its operation is based on their roles, events, artifacts and associ- ated rules.

**Scrum adoption**

Scrum adoption refers to the process of implementing and using the Scrum agile methodology in a software development team [10]. Scrum adoption may involve ma- king changes to the organizational culture, training team members, implementing newtools and processes, and managing change at the organizational level [10].

The adoption of Scrum gives greater importance to people and their relationships, i.e., adoption must be a voluntary and conscious decision to internalize not only the Scrum process and methodology, but also the agile values and principles, so that the team's thinking changes, and to influence change in the organizational culture. Conse-quently, adoption generates a shared vision of the process, the agile values and princi-ples, and the organizational culture.

### 2.2 Collaboration

Collaboration is working together to achieve a task or objective that would not be achieved individually under the same time, effort and cost constraints [11]. A collabo-rative team is a group of people with complementary skills who share tasks, re- sources, responsibilities and leadership, to achieve a common understanding that al- lows them to achieve a common goal [12], which is achieved if and only if all mem-bers achieve their individual goal [13].

Achieving collaboration in software engineering teams includes (1) adequate communication of requirements, design decisions, tasks, changes, among others; (2) group awareness of what each participant is doing and the current status of the pro- ject, their availability and knowledge; and (3) appropriate interactions to coordinate and perform tasks [14].

### 2.3 Existing work

L. V. Araujo and A. N. Castrillón [15] conducted a study of the main factors, pro- blems, challenges, risks and situations that affect the adoption of Scrum through inter-views with groups of experts in order to make a roadmap for the adoption of Scrum.

N. Khan, N. Ikram, and S. Imtiaz [7] focuses on the common problem of software development teams of having a large number of items in their backlog, also known as backlog. The article presents Scrum as a methodology that can help teams solve this problem and improve software development efficiency. The authors describe how Scrum helps teams prioritize and organize their backlog, and how teams can work together more effectively to achieve project goals. In addition, the author highlights the importance of communication and collaboration in the Scrum implementation process. The article concludes that adopting Scrum can be an effective solution to backlog problems and can improve the productivity and quality of the developed software. Overall, they propose a solution to backlog problems during Scrum adop- tion. The study focuses on a development team, and applies interviews for infor- mation gathering, in addition, it uses notes and project management tools for backlog monitoring.

Irum Inayat and Siti Salwah [16], propose a framework to study requirements- driven collaboration in distributed agile teams and determine the impact of their col- laboration patterns during iteration, defining collaboration in terms of communication as information sharing among team members and awareness of each other's knowledge. The authors conducted two case studies on agile teams, whose infor- mation was collected through questionnaires, semi-structured interviews and observa- tion. Furthermore, the authors conclude that the proposed framework may be useful for researching and im- proving collaboration in the future. Overall, the article pro- vides valuable insight into collaboration in agile teams and how to improve it to achieve success in agile projects.

Batra, Weidong and Mingyu [17] explore the concept of collaboration in the con- text of agile software development. The authors define collaboration as an interactive process in which team members work together to achieve common goals and identify four key dimensions of collaboration in agile software development: communication, coordination, cooperation, and commitment. Each dimension is examined in detail, and factors that can affect collaboration in each area are highlighted. The authors also describe the benefits of collaboration in agile software development, including in- creased efficiency, improved product quality, and increased customer satisfaction. The authors note that there is little research examining the nature and dimensions of collaboration in the context of agile software development. Interview data collected from five software development outsourcing providers in China were used to develop this paper.

## 3 Qualitative Study on the collaborative aspects taken intoaccount by development teams during Scrum adoption

In many organizations, drawbacks have been found when adopting Scrum as a framework in their projects [4], these are related to factors associated with human behavior such as culture, organizational, interaction, companionship, communication. etc [10]. Next, the main problems and situations that arise during the adoption of Scrum are consolidated according to the empirical perspective of work team leaders. These problems will make it possible to identify the collaborative aspects that the team recognizes and which of them represent weaknesses in both the collaborative factors and the methodology.

### 3.1 Design

**Research question:** What are the collaborative aspects that software companies iden-tify during Scrum adoption and how well do they cover collaboration-related fea- tures?

**Persons**

The interviews are focused on systems engineers or similar, development team leaders in the context of the agile Scrum methodology in its adoption phase corre- sponding to small software development companies in the department of Cauca, Co- lombia.

At the time of carrying out this study, it was found that in the department of Cauca there are 95 software development companies, 36 of them are small software devel- opment companies, when contacting these companies, it was possible to determine that only 7 have equipment that is in process. of Scrum adoption, of which 2 of them are in training and stated that they have no experience to report, for which only 5 companies can be studied. However, only 4 companies were interested in participat- ing in this study and sharing their experience in adopting Scrum [18].

***Table 1:*** *Software development companies in the department of Cauca*

| | |
|---|---|
| Software development companies | 95 |
| Small software development companies | 36 |
| Small software development companies with Scrum adopter teams | 7 |
| Small software development companies with Scrum adopter teams with experience to report | 5 |

### 3.2 Data collection method

The method of data collection in this study was semi-structured interview. Semi- structured interviews are an intermediate approach between standardized, mostly closed- ended surveys of individuals and free-participation sessions, with groups, are the semi- structured interviews [19].

The semi-structured interview is conducted conversationally with one respondent at a time, employing a combination of closed-ended and open-ended questions, often ac- companied by backward follow-up questions (why and how). The dialogue can zigzagaround agenda items, rather than rigidly adhering to literal questions as in a standardized survey, and can delve into entirely unanticipated issues. Semi-structured inter- views are relaxed, engaging and, being face-to-face, can be longer than telephone surveys, although they last as long as focus groups. One hour is considered a reasona- ble maximum length for a structured interview to minimize fatigue for both inter- viewer and interviewee.

The practical steps for designing and conducting semi-structured interviews are: se- lecting and recruiting respondents, drafting the questions and the interview guide, applying the recommended techniques for this type of interview, and analyzing the information collected.

**Questions**

The questionnaire of this semi-structured interview was applied to two experts whoparticipated as judges in the content validation and presented their opinion, which allowed refining the questions in order to obtain the most relevant information for thisstudy. The interview consisted of 22 questions, divided into four categories: Scrum (3 questions), interaction (6 questions), communication (5 questions) and awareness (8 questions).

| | |
|---|---|
| S | 1. How long have you been using the Scrum agile methodology?<br>2. What were the main problem(s) you encountered during the adoption ofScrum?<br>3. What problems do you see persisting from adoption to the present? |
| Interaction | 1. Could you describe how the team carries out its daily activities?<br>2. Could you describe how you conduct the meetings?<br>3. What benefits do you find in the way the activities are being developed?<br>4. Could you tell us about the problems related to the interaction among the participants during the development of the activities? What do you think are the causes?<br>5. Could you please detail the procedure to follow when encountering and reporting a problem?<br>6. On a scale of 1-5 with 1 being very little and 5 being very much, how do you consider the interaction between the team participants? |
| Communicatio | 1. Which digital channels and media do you use for communication? Whichare the most indispensable?<br>2. What benefits do you find in the way your team's communications aregoing? What are its strengths?<br>3. What do you consider to be the most frequent problems during commu-nication between participants?<br>4. What are the most common communication problems during meetings?What do you think are the causes?<br>5. On a scale of 1-5 with 1 being very little and 5 being very much. How doyou consider the team's communication to be? Why? |
| Awareness | 1. How does the team achieve a shared and ongoing understanding of the problem, the product and progress of the project for the entire team?<br>2. How do you know at this point what the status of the project is and what issue each team member is currently resolving?<br>3. What are your responsibilities and those of your colleagues in software projects?<br>4. To what extent do you feel responsible for the success or failure of the product? Why?<br>5. Do you consider that each team member is responsible for the success or failure of the product? Why?<br>6. On a scale of 1-5 with 1 being very little and 5 being very much. How do you consider your ongoing knowledge of the product's state of progress and the current tasks that team members are addressing?<br>7. Could you describe problems related to unfamiliarity with aspects relatedto the product, the project or the work of the team members?<br>8. Could you describe some strategies for solving the problems identified inyour work team during this interview? |

**Participants**

The participants were four leaders of development teams belonging to small software development companies in the department of Cauca (Colombia), with ages between36 - 51 years old, who voluntarily agreed to participate in the interview to deepen the collaborative aspects of Scrum during its adoption phase.

Regarding the gender of the participants, it can be highlighted that despite the fact thatthe study included a female participant, said participant is the leader of a larger team (15 people) and tends to behave as two teams (8 and 7 people).

***Table 1:*** *Participant information*

| | | |
|---|---|---|
| Leader 1 | Age | 38 |
| | Genre | Male |
| | Years of experience | 10 |
| | Equipment size | 9 people |
| Leader 2 | Age | 51 |
| | Genre | Male |
| | Years of experience | More than 20 |
| | Equipment size | 4 persons |
| Leader 3 | Age | 45 |
| | Genre | Male |
| | Years of experience | 13 |
| | Equipment size | 7 people |
| Leader 4 | Age | 36 |
| | Genre | Female |
| | Years of experience | 7 |
| | Equipment size | 15 people: Team divided into two,8 -7 people |

## 3.3 Results and data analysis

The WEFT QDA software tool [20], a qualitative research tool used for the analysis of textual data such as interview transcripts, documents and field notes, was used for the analysis of the interviews. This tool is free to use and has a public domain license and is available for Windows and Linux.

This tool allows: 1) to save the data in an organized way based on categories defined by the researcher; 2) to search and classify the data (interview transcription) in analyt-ical categories established by the researcher; 3) to establish relationships from the data through searches, 4) to visualize the searches in the form of texts or double-entry tables [19]. Once the information from the interviews was integrated into the tool, 4 categories of analysis were generated (Interaction, Communication, Awareness and Scrum) that generate a linearity between the objectives of this article and the analysis of results.

The WEFT QDA tool classified text fragments into 4 categories:

- Interaction: Aspects related to interaction in the work team are addressed and analyzed, either with the customer or among members of the same team from the Scrum perspective.
- Communication: The tools, techniques and/or communication channels usedby the work team to have a constant communication among them from the Scrum perspective are analyzed.
- Awareness: Aspects related to project awareness, roles and/or responsibilities within the project are addressed and analyzed from a Scrum perspective.
- Scrum: It deals with the Scrum methodology that could not be included in any of the previous categories.

The first three categories represent Collaboration, since, as mentioned above, they are necessary for an activity to be collaborative. This classification presented the fol- lowing results: ***Table 2:*** *Text fragments by category*

| Category | Number of text fragments |
|---|---|
| Interaction | 19 |
| Communication | 40 |
| Awareness | 66 |
| Scrum | 13 |
| **Total** | **138** |

As a result of the interview, a list of activities was identified and grouped by fre- quency, i.e., how many teams coincided in mentioning the activity (1, 2, 3 or 4), and the activities developed by the teams were associated to collaborative activities and toevents or Scrum elements in order to identify activities within the methodology wherethe problems specified by the participants may occur.

The number of activities obtained in each category and the number of activities by frequency are presented below.

***Table 3:*** *Activities vs. Frequency Collaboration*

| Category | Number of activities | Frequency by activity |
|---|---|---|
| Interaction | 3 | 3 Activities with frequency 4 |
| Communication | 5 | 4 Activities with frequency 3<br>1 Activity with frequency 2 |
| Awareness | 6 | 3 Activities with frequency 4<br>2 Activities with frequency 3<br>1 Activity with frequency 1 |

***Table 4:*** *Activities vs. Scrum Frequency*

| Category | Number of activities | Frequency by activity |
|---|---|---|
| Interaction | 1 | 1 Frequent activity 4 |
| Communication | 2 | 2 Activities with frequency 4 |
| Awareness | 5 | 2 Activities with frequency 4<br>3 Activities with frequency 2 |

The activities with the highest incidence (frequency 4) reveal potential problems or risk factors present during adoption that may occur frequently in Scrum adopting teams. On the other hand, activities with lower incidence (frequency 1) reveal less common problems or risk factors.

Participants explicitly identified some of the potential problems they encountered during Scrum adoption based on their experience. Most of these potential problems are associated with the person factor evidenced by culture, beliefs, language, person- ality, and attitude.

For this work, risk is defined as the different factors that affect the correct adoption ofScrum. The main risks identified in this study were found from the text fragments (explicit statements of the participants) and in the associated activities. Below is the list of risks:

- Individual work thinking: some individuals prefer to work alone or develop their activities without relating to the work team.
- Downplaying the purpose of each Scrum event (lack of awareness): They do not know the purpose of the events and how they impact the process.
- Lack of participation or absence during sessions: Some team members do not wish to participate during sessions so they avoid being present, do not pay at- tention and/or present little or no participation.
- The person factor can affect the adoption process: the process can be affected by a person's personality, character, customs, beliefs, etc.
- Absence of collective ownership of information: team members are unawareof the progress of the process and the activities carried out by other team members.
- Lack of effective communication: Communication through multiple messag- ing tools or channels can lead to loss of information or confusion.
- Lack of cooperation and constant communication with the client: The custom- er is taken into account only at the end of the process.
- Lack of discipline in daily meetings: Meetings are extended, the objective is not met, the progress of the process is not clear.
- Lack of constant customer participation: The customer loses or has no interest so does not get involved in the process.
- Lack of understanding of agile values: team members have different under- standing of agile values.
- Lack of understanding of meeting objectives: team members do not have the same perception of the purpose of Scrum meetings.

## 4 Conclusions and future work

Scrum is a popular agile project management methodology used in software devel- opment and other complex projects. Although Scrum focuses on collaboration among team members, there are several collaboration issues that can arise. The semi- structured interviews conducted provide insight into the potential issues that Scrum adopter teams from small software development organizations face during their adop- tion process regarding collaboration.

- A person who wants to work individually, with little or no interaction with other team members, and who is reluctant to change this behavior can threaten the adoption of Scrum, since individual attitudes are detrimental to teamwork and project development.
- The use of several communication channels does not guarantee good commu- nication and can lead to confusion or information not being taken into account in a timely manner. Tools such as Skype, google, even WhatsApp, allow con- stant communication between team members. However, these tools do not of- fer a guide to manage communications.
- Team members have different understandings of agile events and values, which hinders the participation of some team members and in turn creates an absence of a shared understanding of the problem, the product, and the pro- gress of the project.
- Some teams incorporate artifacts from traditional methodologies that they con-sider indispensable to the project, leading to an adaptation of these artifacts to the agile methodology. Teams are applying a modified adoption.
- The person factor, i.e., culture, beliefs, ethnicity, language, personality, atti- tude, etc. can affect the adoption process.
- Teams see the need for Scrum training of its members for a better adoption of the methodology. However, the adoption must be a voluntary and conscious decision to internalize not only the Scrum process and methodology, but also the agile values and principles, so that the team's thinking changes, and influ- ence the change in the organizational culture.

In conclusion, collaboration issues in software development teams directly impact the Scrum adoption process, as they can cause team members to lose confidence in the Scrum process, which can make them reluctant to adopt the methodology in the fu- ture. In addition, they weaken one of the pillars of the methodology and hinder the adoption process, the identification of potential problems during the adoption process can be considered an alert that shallow adoption may result, although Scrum is adapt- able, it requires commitment from all team members and a shared understanding ofthe process, so it is important to proactively address potential collaboration issues to ensure the successful adoption of Scrum.

As future work, it is planned to interview more software development companies adopting Scrum in order to identify more potential problems in a national and interna-tional context. Based on the previously identified problems, a series of recommenda- tions will be made to strengthen collaboration during the adoption of Scrum and a case study to validate these recommendations. This in order to provide tools to Scrum adopter teams to mitigate the potential problems reported in this work.

## References

1. Papadopoulos, G. Z.: Moving from Traditional to Agile Software Development Methodol- ogies Also on Large, Distributed Projects. Procedia - Social and Behavioral Sciences, 175, 455-463. https://doi.org/10.1016/j.sbspro.2015.01.1223. (2015).
2. Rodríguez, C., & Dorado, R.: ¿Por qué implementar Scrum? Ontare, 3(1), 125-144. https://doi.org/10.21158/23823399.v3.n1.2015.1253. (2015).
3. Schwaber, K. SCRUM Development Process. En Springer eBooks (pp. 117-134). https://doi.org/10.1007/978-1-4471-0947-1_11. (1997).
4. González, J. J. M., Pardo, C., & Gómez, O. S.: Revisión sistemática acerca de la imple- mentación de metodologías ágiles y otros modelos en micro, pequeñas y medianas empre- sas de software. Revista Tecnológica - ESPOL, 28(5). (2015).
5. VersionOne. VersionOne 10th Annual State of Agile Report, VersionOne, vol. 6, nº 2, pp. 2-8. (2006)
6. Ali, A., Rehman, M., & Anjum, M.: Framework for Applicability of Agile Scrum Method- ology: A Perspective of Software Industry. International Journal of Advanced Computer Science and Applications, 8(9). https://doi.org/10.14569/ijacsa.2017.080932. (2017).
7. N. Khan, N. Ikram y S. Imtiaz.: SCRUM Adoption: A Solution to Backlog Problems. Fifth Asian Conference on Information Systems, At Krabi, Thailand. (2016)
8. J. López, R. Juárez, C. Huertas, S. Jiménez y C. Guerra.: Problems in the Adoption of Ag- ile-Scrum Methodologies: A Systematic Literature Review. 4th International Conference in Software Engineering Research and Innovation (CONISOFT), Puebla, México. (2016).
9. C. Collazos, S. Ochoa y J. Mendoza.: Collaborative evaluation as a mechanism for im- proving evaluation of classroom learning. Ingeniería e Investigación, vol. 27, nº 2, pp. 72- 76. (2007).
10. Kanane, A. Challenges related to the adoption of Scrum. Case study of a financial IT com- pany. UMEA University, Department of informatics. IT management master program, Umea, Suecia. (2014).
11. R. Owen Briggs, G. Kolfschoten, G.-J. De Vreede y D. L. Dean.: Defining Key Concepts for Collaboration Engineering. Americas Conference on Information Systems Proceedings,Acapulco, Mexico. (2006).
12. D. Mishra, A. Mishra y S. Ostrovska. Impact of physical ambiance on communication, col-laboration and coordination in agile software development: An empirical evaluation. In- formation and Software Technology, vol. 54, nº 10, p. 1067–1078. (2012).
13. S. Werner Knoll, M. Hörning y G. Horton.: Applying a ThinkLet and ThinXel based Group Process Modeling Language: A Prototype of a Universal Group Support System. deProceedings of the 42nd Hawaii International Conference on System Sciences, Hawai. (2009).
14. C. Restrepo, L. Jiménez y J. A. Hurtado.: Integrating collaboration engineering with soft- ware process modeling: a visual approach. In Proceedings of the XVIII International Con- ference on Human Computer Interaction (pp. 1-2), Cancun, Mexico. (2017).
15. L. V. Araujo y A. N. Castrillón.: Una ruta de trabajo para la adopción de Scrum en peque- ñas organizaciones en la industria del software. Universidad del Cauca, Popayán, Cauca. (2017).
16. I. Inayat y S. S. Salim.: A framework to study requirements-driven collaboration among agile teams: Findings from two case studies. Computers in Human Behavior, vol. 51, pp. 1367-1379. (2015).
17. Batra, Dinesh; XIA, Weidong; ZHANG, Mingyu. Collaboration in agile software devel- opment: Concept and dimensions. Communications of the Association for Information Systems, 2017, vol. 41, no 1, p. 20. (2017).
18. DANE, Geovisor Directorio de empresas 2023. http://www.pressure.to/qda/, last accessed 2023/05/27.
19. Laura Díaz-Bravo, Uri Torruco-García, Mildred Martínez-Hernández, Margarita Varela- Ruiz. La entrevista, recurso flexible y dinámico. Universidad Nacional Autónoma de Mé- xico, México D.F., México. (2013).
20. University of Surrey, WEFT QDA. http://www.pressure.to/qda/, last accessed 2023/04/21.

# A literature review to enabling strategic decision support systems for crisismanagement

**Elandaloussi Sidahmed[1] and Zaraté Pascale [1]**

[1] IRIT / Toulouse University
2 Rue du Doyen Gabriel Marty, 31042 Toulouse Cedex 9, Francesid.elandaloussi@irit.fr
Pascale.Zarate@irit.fr, web-page: https://www.irit.fr/~Pascale.Zarate

**Abstract.** This study aims to draw the main concepts in the literature concerning strategic decision support systems (SDSS) which have become widely incorporated in many specialized areas like logistics, medical, transport, trade, education...etc. However, this paper reports a literature review on decision support systems (DSS) for strategic crisis management. The paper deals with SDSS technologies to benefit from their advantages and integrates them into crisis management tasks. Moreover, managing a crisis effectively is an important component of deciding the level of crisis responsibility. In this paper we definea methodology to analyse a literature review. This analysis allows us to define four criteria to distinguish which SDSS is better.

## 1 Introduction

Crisis management is described as the process by which different organizations and firms deal with a sudden emergency situation. The covid-19 crisis that began in early 2020 is a better example for now of crisis management. It has caused the most sudden and deepest global recession since the Second World War. Since thisperiod, there exists a growing body of literature on crisis management. For instance, **[**1**]** presents a report that defines a thorough comparison and in-depth analysis of the emergency and recovery plans announced by European countries.

Our methodology is composed of two steps: first defining the research question consist of analysing the literature review for strategic decision support systems before classifying them and determining which SDSS will be better in similar situations. Although, strategic DSS is based to many DSS features. Some researchers have adopted a decision support system in their own activities in order to improve some technical aspects. However, other surveys have developed Strategical decision Support Systems. Despite the SDSS deployment complexity, previous research in several fields highlight the benefits of using SDSS such as industrial **[**2**],** supplychain management **[**3**],** medical domain [4**],** logistics management [5,6] especially in transport [7**],** crisis management **[**8**]**. The survey **[**9**]** provides some preliminary empirical evidence on what the most promising crisis strategies are to manage the COVID-19 crisis.

This article is divided into four main parts. The first part is devoted to discussing a comprehensive methodology for literature review. The second part is dedicated to outline a brief description of the most important studies in SDSS technologies. Third part is dedicated to analyze and classify the selected literature. The last part summarizes the paper and offers further insights for future studies.

## 2 Research Methodology

In this section we describe the research methodology adopted to collect conferences and journals articles that includes Strategic Decision support systems as keywords. However, we have used an advanced search for articles related to SDSS that appeared only in these 4 data sources: Web of knowledge, Science direct, Scopus and Google scholar by using the following keywords as important search criteria : "SDSS", "Strategic Decision Support Systems", "Crisis management", "SDSS architecture", "Strategic Management models", "Strategic planning models".

After that, we have filtered the papers to retain only those in English related to the SDSSs fields. Then, we scanned titles and abstract for each selected paper. Finally, we reviewed their citations and references.

Figure 1 shows an overview of the used searching and selecting papers methodology to demonstrate the various search efforts that are carried out. Furthermore, some findings papers are presented in this present study and the rest in an extension paper.

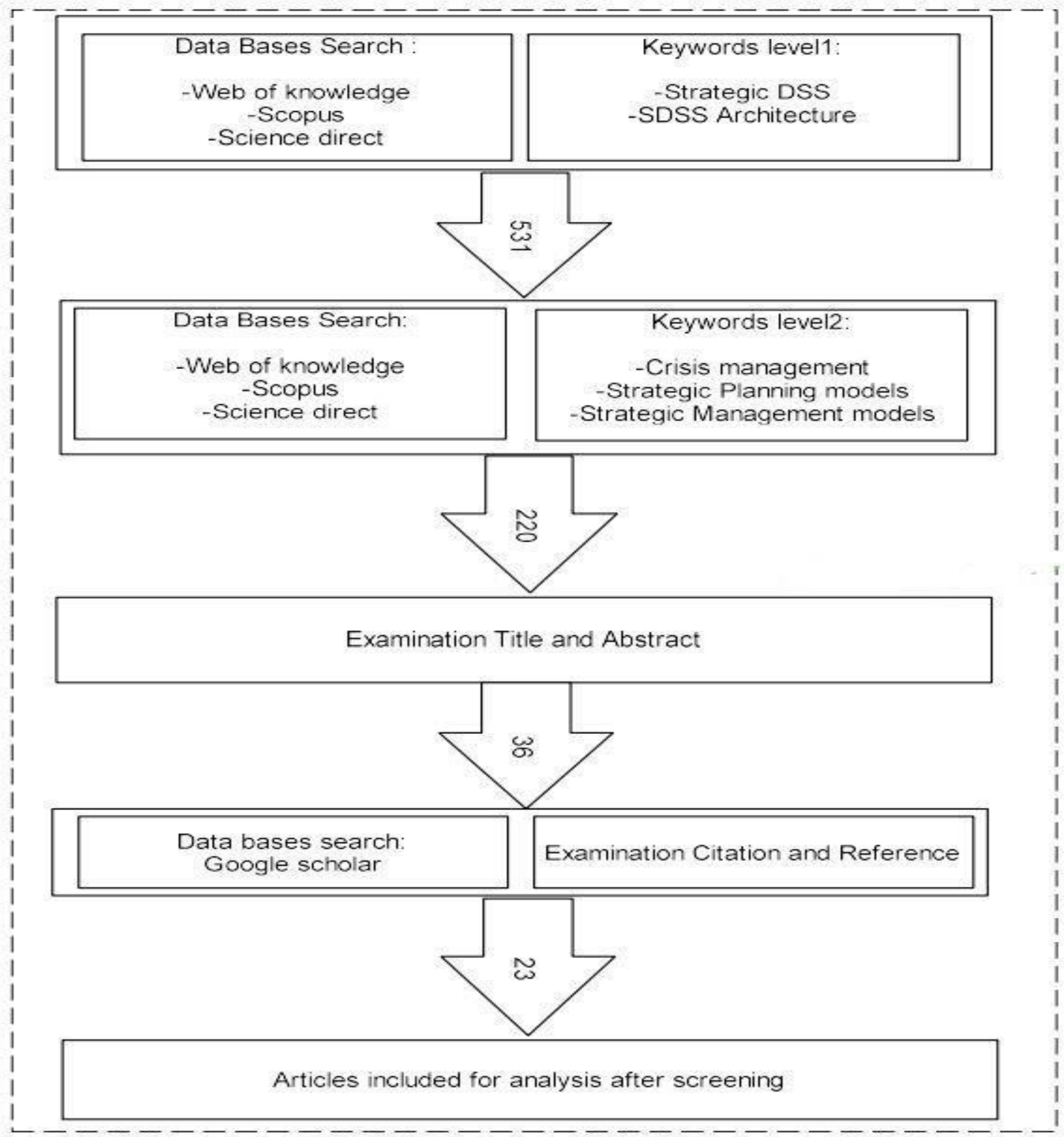

**Fig. 1.** Searching and selecting papers methodology.

## 3 Strategic Decision Support Systems

Decision support systems (DSS) are an important application of management information systems which apply different decision rules and models. [10] Consider a DSS as "an interactive and adaptable Computer Based Information System which helps non-organized management problems".

Strategic DSS provide more explanatory information than DSS by explaining concepts and simulating thinking. According to [11] strategy has also been viewed as consisting of three processes: strategic planning [12], strategic thinking [13], and opportunistic decision making [14].

On the one hand, the technical aspects of DSS are widely considered by several surveys but strategic ones are very neglected. In fact, the combination of strategic actions with DSS is very promising and their literature is large and varies according to the SDSSs functionality. Although, without strategic methods the decision falters toachieve organizational goals.

On the other hand, previous research deals with the benefits of strategic DSS in many specialized areas (supply chain and logistics [15], the agrifood industry [16] ....). For instance, in order to achieve effective decision making in Logistics [15] proposes a strategic DSS framework which combines both the strategic management process and the Strategic Information Systems Planning (SISP) process. Another area of logistics where strategic DSS are used to handle storage systems based on the estimation of parameters is costs for racks and storage equipment [17]. Regarding [15] the existing strategic actions included in DSS are summarized in **Table1**.

***Table. 1.*** *The existing strategic actions including in DSS [15].*

| Strategic Actions including in DSS | References |
|---|---|
| Identification of environmental trends, internal trends and performance trends | [18] |
| Planning responses to the issues | |
| Assessing the impact and urgency of the issues, and prioritizing the issues | |
| Evaluation | [19] |
| Culture Analysis | |
| EnvironmentAnalysis | |
| SWOT Analysis | |
| Strategy Operating Subsystem | [20] |
| Goal-setting Subsystem | |
| EnvironmentalAnalysisSubsystem | |

## 4 Discussion

After examination of the different selected SDSS that have been developed we could use a set of criteria to classify these systems: Single user based SDSS, Group user based SDSS, Methodology based SDSS and Software based

SDSS. Among the selected paper in SDSS some of them are dealt in this survey and classified according to these criteria as below:

***Table. 2.*** *Surveys classification according to the generated criteria.*

| Simple user based SDSS | Group user based SDSS | Methodology based SDSS | Software based SDSS |
|---|---|---|---|
| [21] | [19] | [2] | [22] |
| [23] | [15] | [24] | [11] |
| [26] | [20] | [28] | [29] |
| [30] | [31] | [15] | [19] |
| [32] | [33] | [34] | [20] |
| - | [35] | [17] | [31] |
| - | [36] | [36] | [33] |
| - | - | [37] | [26] |
| - | - | [10] | [38] |
| - | - | [39] | - |
| - | - | [15] | - |
| - | - | [20] | - |

Based on this classification **(See Table2)**, we can assume that SDSS methodology oriented are very largely covered. Then several SDSS software are designed and finally considering the low number of studies for Single decision makers, we can assume that SDSS is more a matter of Group of decision Makers rather than single Decision Maker. Furthermore, to make it very simple the application of all works indicated in this table will be described in detail in an extended study.

## 5 Conclusion

In this study, the existing literature on SDSS is reviewed and their comprehensive assessments are provided. Thus, managing a crisis efficiently is an important component for controlling and promptly preventing itsnegative consequences. In fact, to classify the selected literature four criteria have been introduced: Single user based SDSS, Group user based SDSS, Methodology based SDSS, Software based SDSS.

Based on this analysis, we can see that Strategic DSS is more a matter of group of decision Makers and that they are well supported more by methodologies rather than techniques. Otherwise, as future work, an extended paper of this work will include a detailed analysis of covid-19 pandemic management, will explain the different architecture of defined strategic DSS and we will demonstrate which SDSS is better to deal with each crisis.

## References

1. Aussilloux, V., Mavridis, D., Baïz, A., Garrigue, M.: The effects of the covid-19 crisis on productivity and competitiveness. (2021).
2. Agostino, B., Massei, M., Sinelshnkov, K.: Enabling Strategic Decisions for the Industry of Tomorrow. Volume 42, Pages 548-553,doi.org/10.1016/j.promfg (2020).
3. Ye, F., Liu, K., Li, K., Lai, K., Zhan, Y., Kumar, A.: Digital supply chain management in the COVID-19 crisis: An asset orchestration perspective. Int. J. Production Economics 245 108396, (2022).
4. Shahparvari, S., Hassanizadeh, B., Mohammadi, A., Kiani, B., Lau, K., Chhetri, P., Abbasi, B.: A decision support system for prioritised COVID-19 two-dosage vaccination allocation and distribution. Volume 159,102598, doi.org/10.1016/j.tre.2021.102598, (2022).
5. Kamariotou, M., Kitsios, F., Michael, A., Manthou, V.: Strategic Decision Support Systems for Logistics in the Agrifood Industry, Proceedings of 8th International Conference on Information and Communication Technologies in Agriculture, Chania, Greece (2017).
6. Henrik N., Thompson, A., Lindahl, P., Broman, G.: Introducing strategic decision support systems for sustainable product-service innovation across value chains, Proceedings of Sustainable Innovation, Malmö, Sweden, id: diva2:889466(2008)
7. Barfod, M., Salling, K.: A new composite decision support framework for strategic and sustainable

transport appraisals,Volume 72, Pages 1-15, (2015)
8. Jaziri, R., Miralama, S.: The impact of crisis and disasters risk management in COVID-19 times: Insights and lessons learned from Saudi Arabia, Medicine and Public Health 18 ,100705, (2021).
9. Klyver, K., Nielsen, S.: Which crisis strategies are (expectedly) effective among SMEs during COVID-19, doi.org/10.1016/j.jbvi.2021.e00273, (2021).
10. Alyoubi, B.: Decision support system and knowledge-based strategic management, Procedia Computer Science, 65, 278-284, (2015)
11. Belardo, S., Duchessi P., Coleman, J.: A strategic decision support system at OrellFussli, Journal of Management Information Systems, 10(4), 135-157, (1994).
12. Weihrich, H.: The TOWS Matrix-A Tool for Situational Analysis. In Long Range Planning- Vol. 15, N. 2, (1982).
13. Earl, J.J.: Information systems strategy formulation In R J. Boland and R A. Hirschheim (eds.), Critical Issues in Information Systems Research. New York: John Wiley, pp. 157, (1987).
14. Ward, J., Griffiths, P.; Whitmore, P.: Strategic Planning For Information Systems. New York: John Wilev. (1990).
15. Kitsios, F., Kamariotou, M.: Decision support systems and strategic planning: information technology and SMEs' performance. International Journal of Decision Support Systems, 3(1-2), 53-70, (2018).
16. Hajimirzajana, A., Vahdat, M., Sadegheih A., Shadkam, E., El Bilali, H.: An integrated strategic framework for large- scale crop planning: sustainable climate-smart crop planning and agri-food supply chain management, volume 26, Pages 709-732, (2021)
17. Accorsi, R., Manzini R., Maranesi, F.: A decision-support system for the design and management of warehousingsystems, Computers in Industry, Vol. 65 No 1, pp. 175-186, (2014).
18. Korpela, J., Tuominen, M.: A decision aid in warehouse site selection," International Journal of ProductionEconomics, Elsevier, vol. 45(1-3), pages 169-180, August, (1996).
19. Moormann, J., Lochte-Holtgreven, M.: An approach for an integrated DSS for strategic planning. Decision SupportSystems, 10(4), 401-411, (1993).
20. Yoo, S., Digman, L.A.: Decision support system: a new tool for strategic management', Long Range Planning, Vol.20, No. 2, pp.114–124.
21. Dung, L., Giang, P.: "Strategic responses of the hotel sector to COVID-19: Toward a refined pandemic crisismanagement framework", Volume 94, 102808, (2021)
22. Jiirgen, M,, Lochte-Holtgreven, M., An approach for an integrated DS ~k.J for strategic planning , (1993).
23. Saaty, T. L.: The analytic hierarchy process in conflict management". International Journal of Conflict Management,ISSN: 1044-4068, (1990)
24. KyuLee, J., Lee H., ,Interaction of Strategic Planning and Short-term Planning: An Intelligent DSS by the Post-ModelAnalysis Approach , (1987)
25. Belardo, S., Duchessi, P., Coleman, R. : A strategic decision support system at Orell Fussli. Journal of ManagementInformation Systems, 10(4), 135-157, (1994).
26. Hornby, R., Golder, P., Williams .A: "SDP: a strategic DSS. Decision Support Systems, 11(1), 45-51, (1994).
27. Yoo, S., Digman, L., Decision support system: A new tool for strategic management". Long Range Planning, 20(2),114-124, (1987).
28. Shahrooz, S., Hassanizadeh, B., Mohammadi, A., Kiani, B., Lau, K., Abbasi, B.: A decision support system forprioritised COVID-19 two-dosage vaccination allocation and distribution , (2022)
29. P. Beraldi , A. Violi , F. De Simone A decision support system for strategic asset allocation, (2011)
30. Cooper, R., Kaplan, R.: The Design of Cost Management Systems: Text and Cases.
31. Korpela, J., Tuominen, M.: A decision support system for strategic issues management of logistics. Internationaljournal of production economics, 46, 605-620.
32. Sprague, R., Carlson, E.: Building effective decision support systems. Prentice Hall Professional Technical Reference.(1982).
33. Zviran, M.: ISSPSS: a decision support system for information systems strategic planning. Information & management,19(5), 345-359, (1990)
34. Fanti, M., Iacobellis, G., Ukovich, W., Boschian, V., Georgoulas, G., Stylios, C.: A simulation based DecisionSupport System for logistics management. Journal of Computational Science, Vol. 10, pp. 86-96, (2015)
35. Finlay, P., Marples, C.: Strategic group decision support systems a guide for the unwary. Long Range Planning,25(3), 98-107, (1992).
36. Eden, C.: Perish the thought!. Journal of the Operational Research Society, 36(9), 809- 819. (1985)

37. Lee, M., Chen, I., Chen, R., Chung, C.: A target-costing based strategic decision support system. Journal of Computer Information Systems, 43(1), 110-116, (2002).
38. Thierauf, R.: User-oriented decision support systems: Accent on problem finding, International Journal of Business Intelligence Research. **9**:1. (38-52), (1988).
39. Pietrzak, M., Jałosiński, K., Paliszkiewicz, J., Brzozowski, A.: A case study of strategic group map application used as a tool for knowledge management, Journal of Computer Information Systems, 55(2), 68-77, (2015).

# *Rayuela*: a pluggable adaptive gamificationapproach for CLCS

**Dalponte Ayastuy María[1,2], Alejandro Fernández[1], and Diego Torres[1,2]**

[1] Depto CyT, Universidad Nacional de Quilmes
Roque Saenz Peña 352, Bernal, Buenos Aires, Argentina. {mdalponte}@unq.edu.ar
LIFIA,[2] CICPBA-Facultad de Informatica, Universidad Nacional de La Plata50 y 120, La Plata, Buenos Aires, Argentina.
alejandro.fernandez,diego.torres@lifia.info.unlp.edu.ar

**Abstract.** Collaborative location-based collecting systems (CLCS) is a particular case of collaborative systems where a community of users col- laboratively collects data associated with a geo-referenced location and timestamp. This paper proposes *Rayuela* as a framework to lever existing CLCS with adaptive game challenges through a cross-project approach.This tool allows incorporating a gamification engine in a transparent wayto a set of CLCS-supported citizen science projects. The game challengesare generated and adapted considering the user's profile, the project's coverage goals, area priority configuration, and global system status. Additionally, the article describes a set of extension points to combine different challenge generation and challenge recommendation strategies.

## 1 Introduction

Collaborative location-based collecting systems (CLCS) is a particular case of collaborative systems where a community of users collaboratively col-lects data associated with a geo-referenced location and timestamp. Partic- ularly, the citizen science collecting systems that convene users to collect location-based data with a scientific goal, can be seen as CLCS. The con- tributions of the volunteers are known as *check-in tuples* with the structure $< position, timestamp, sample_data >$[5].

Examples of CLCS-supported citizen science projects are AppEar project [3], GeoVin [4], or iNaturalist [12]. The collection task could require a mobile de-vice to fulfill a survey form or capture multimedia (images, video, sound) being located in a specific area or point. In addition, this may need to be done at a particular time or time interval.

Citizen science projects depend on the sustained participation of many peo- ple, and gamification can facilitate the engagement and retention of users [10].

Gamification is the application of game elements and mechanics (like points, badges, leaderboards, storytelling, etc) in contexts where people's task is notsimply to play [6]. Nevertheless, this paper will focus on game challenges, whichare tasks or problems whose difficulty depends on the user's skills, abilities, mo-tivation, and knowledge [11]; considering each collection task (or a set of them) asa game challenge and the areas of CLCS's territory as a game board. In addition,the combination of the spatial and temporal constraints of the collection tasks builds up an idea of difficulty and reward that are considered game mechanics. Despite the advances in gamification research and its demonstrated benefits related to user engagement and activity, these benefits are considered contextspecific and not generalizable to all individuals [8]. The *one-size-fits-all* approach presents several limitations because of the users' different motivations, personali-ties, needs, and playing styles [2, 9, 1]. Currently, the research stream on adaptivegamification is considering how to dynamically adapt the game elements andmechanics each user needs in each context. The gamification approach could beadapted to the community members and the project's goals for better partici-
pation and sustained user engagement.

Assessing the impact of a given adaptive gamification strategy empirically presents several challenges. On the one hand, it is necessary to demonstratethat a scenario with adaptive gamification is better than one where gamification is not adaptive (and it is possibly also better than one without gamification). Therefore, the evaluation should develop two parallel case studies on a real-worldCLCS. On the other hand, considering that CLCS can have support tools (digital or not), it is necessary to propose a mechanism that allows the applicationof an adaptive gamification approach transparently. Also, researchers need to agilely and with low effort change the key aspects of adaptive gamification to generate the various case studies (by tunning the challenge generation and the recommendation strategies).

With this objective in mind, this article proposes *Rayuela*, an approach toadaptively gamify existing CLCS citizen science projects as an external appli-cation. Moreover, it can be set up with more than one existing CLCS project,thus consolidating a platform of gamified projects. In addition, it offers extensionpoints so researchers can plug in new game challenges generation and recommen-dation strategies. It allows managers of citizen science projects to monitor theprogress of game players and assess the impact of gamification in their projects.The following sections are organized as follows. In Section 2 the approach is presented and the uses cases are explained. Finally, Section 3 shares conclusions
and future work.

## 2 Approach

An overview of the challenge-based adaptive gamification approach is provided by Figure 1. An existing citizen science project (gray box in the Figure) defines tasks for volunteers to complete, and is not yet gamified. The tasks are collec- tion tasks characterized by time and space restrictions. Volunteers complete the tasks using the tools provided by the citizen science project managers (gray ar- row), which can be a mobile application, a web application, or some non-digital technology.

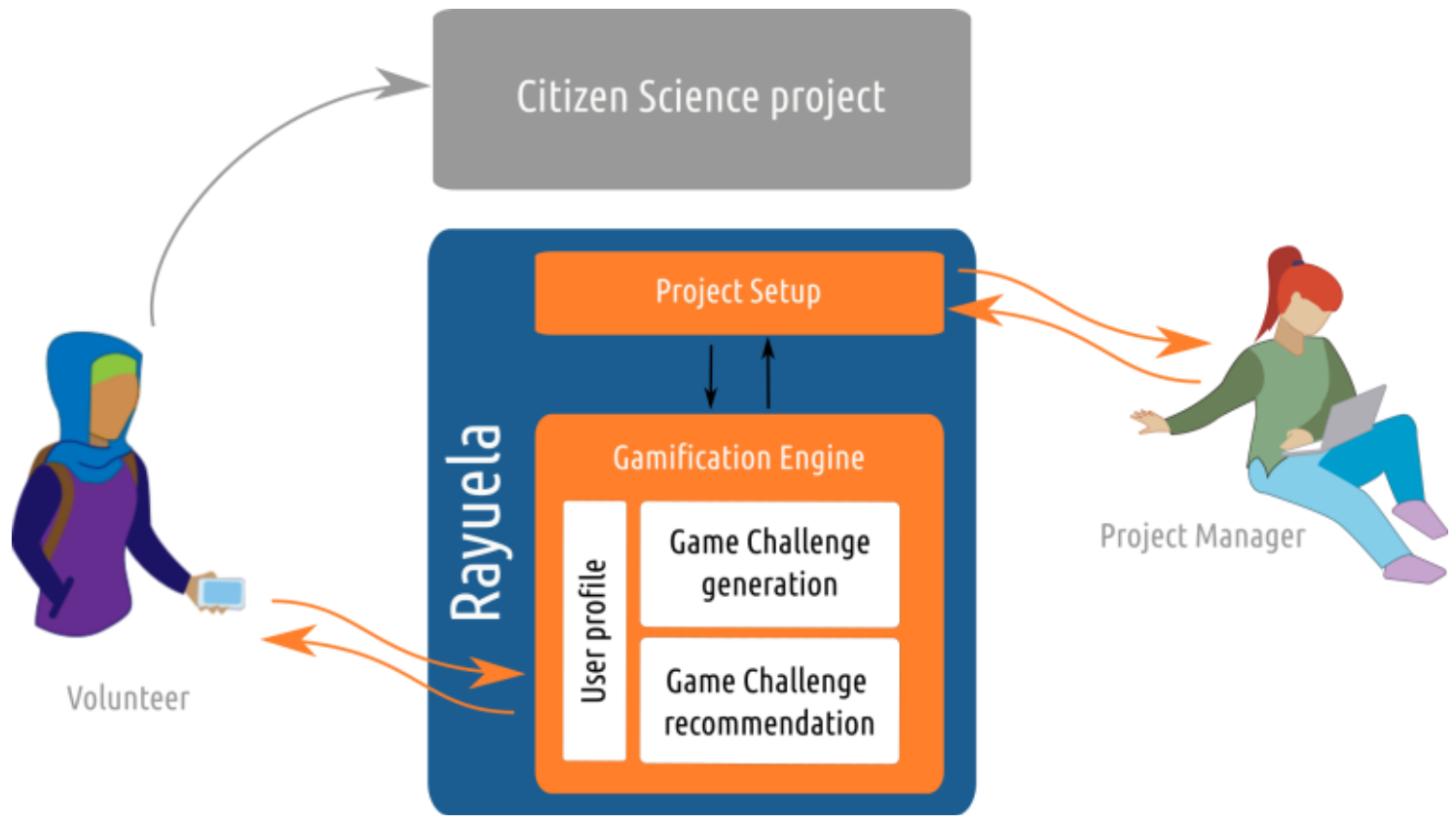

**Fig. 1.** Approach architecture

*Rayuela* platform adds an adaptive gamification layer to the citizen science project. Using it, the volunteer enriches her participation in the project, partic- ularly by selecting in *Rayuela* the tasks to be completed and login them when finished. Playful activities are proposed to accomplish the project tasks basedon criteria established by the project manager -who configures the spatial and temporal restrictions and task types-. In proposing activities, *Rayuela* also con- siders the profile of the volunteer (the history of solved and unsolved tasks and traveling behavior).

*Rayuela* architecture has two main modules, the project setup, and the gameengine. The first one stores the configuration provided by the project manager. The second one has the function of providing game challenges to the volunteer. Likewise, the game engine has 3 extensible components. Firstly, the user profile gathers all the explicit and implicit information about the user's preferences, characteristics, and behavior. Secondly, the game challenge generator builds a set of game challenges by applying strategies based on the project manager configurations. Lastly, the game challenge recommender builds an ordered listof game challenges from the user's profile and the project's goals.

These components can be seen as framework extensions. For instance, the generation of the challenges can be done manually by the manager of the gam- ified project or automatically, applying an exhaustive strategy or taking intoaccount both the user's historical behavior and the status of coverage of the project's goals. Also, the recommendation of challenges can be approached withdifferent strategies, more or less computationally demanding to consider more or fewer aspects of the information retrieved about the user. For example, a rec-ommendation system can consider multiple criteria, prioritizing tasks according to project goals, or also the spatiotemporal behavior of the user.

The framework use cases are presented through the interaction among the users and the framework, considering two user roles: on the one hand, the project manager role and, on the other hand, the volunteer role.

## 2.1 Project manager role interactions

As was previously mentioned in Section 2, this role is in charge of configur-ing the project setup that relates to the needs and objectives of the project.Each gamified project's manager must be able to describe the project, definetask types (e.g. survey, photo or video capture, etc.), define work areas, define time constraints for tasks, and configure game challenges. Moreover, the defini- tion of these aspects is a precondition for the generation/configuration of game challenges (as seen in the bottom timeline of Figure 2).

**Project setup** The setup of the adaptive gamified project needs the definitionof the work areas, the time restriction and the task types.

**Game Challenges Generation** The generation of game challenges can be ap- proached with different levels of suitability. The most straightforward approach is manual loading, with a choice of area, time restriction, type and the numberof tasks. Another option is the exhaustive generation of game challenges result- ing from combining all areas, time restrictions, and task types. Moreover, other approaches related to the project's coverage goals can be considered. A more complex approach can use as input, the coverage setting in terms of sample amount -by task type-, and the area priority.

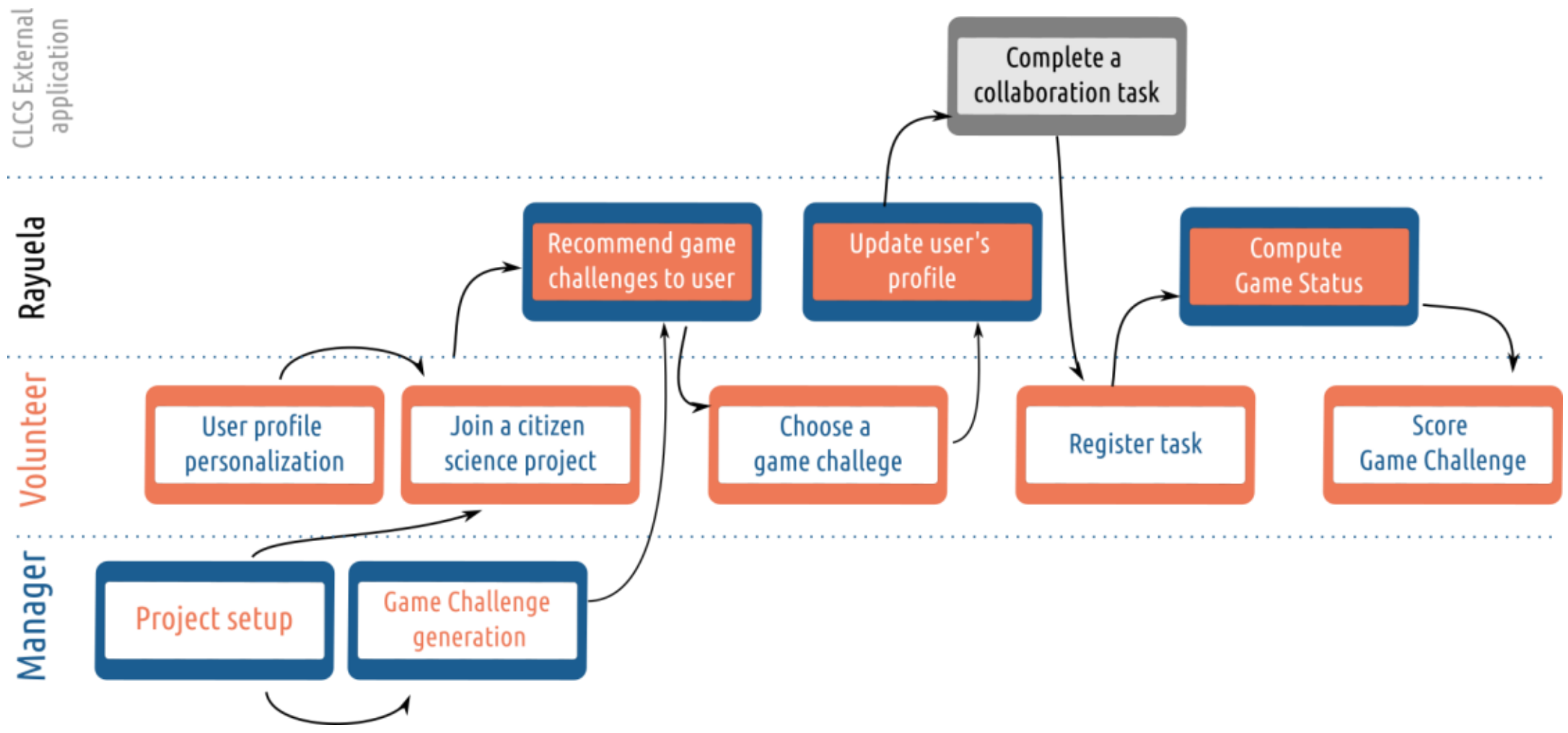

**Fig. 2.** *Rayuela* swimlanes

## 2.2 Volunteer role interactions

By interacting with *Rayuela* the volunteer can browse a set of game challengesthat were generated from the project setup and recommended based on her profile. By choosing one game challenge from the list, she commits to performingthe tasks requested by the challenge, which are carried out externally using the citizen science project. Each completed task must be logged in *Rayuela* to show progress in the game. In addition, the volunteer can express her opinion about the challenge she completed by providing a score.

The volunteer's gamified experience is composed of the interactions describedin volunteer's swimlane of Figure 2.

**User's registration and personalization The** user registers to the applica- tion *Rayuela* and personalizes her profile. The profile personalization -the changeof information, adjustments, or other parameters to a player's personal prefer- ences, abilities, needs, or requirements [7]- is an important stage as it allows the user to take ownership of the platform. Her profile can include data about whereshe lives if she has a car, or how old she is. Personal information can be an input for the recommendation process.

**Subscription to a project** The user can browse the active citizen science projects in the platform and join one of them to enable the game challenges assignment (see Screen (A) in Figure 3). This can be seen as she wants to playthat project's game.

**Game challenge recommendation** The user is offered a set of game elements based on her game profile and spatiotemporal behavior in relation to the CLCS. CLCS needs to optimize volunteers' work by combining the defined objectivesfor the CLCS with the completed tasks in the territory and the implicitly ex- pressed volunteers' preferences. This recommendation stage can be seen as a multi-criteria recommendation since the game challenges list is built consider- ing, on the one hand, what is needed to achieve the project objectives and, on the other hand, the user's behavior profile, preferences, and characteristics. See an example game challenge recommendation in Figure 3 B. Each game challengesets a goal for the user, expressed by an area, a time restriction, and a typeand number of tasks. Also, the game challenge presents a difficulty level and a reward to the user.

## Game challenge assignment

The user chooses a game challenge from the recommended list, that representsa commitment to perform a certain task in a specific area and a specific time interval. By choosing an element from the ordered list, the user expresses her preferences. The underlying idea is to derive the user's preferences by having herchoose a game element from a list sorted by the estimated user score. Asuming the example in Figure 3 (B), the user may select the third game challenge (with Area: *Reserva El destino*, time restriction: week days, task type: survey and 1 sample), which has the lowest score in the recommended list. This will trigger a profile update that will improve the recommendations.

**Check-in registration** The user moves to a sampling area and completes the sampling task, generating a check-in tuple through the **citizen science ap- plication**. The user does a collection task (that can be of different types -e.g. completing a scientific survey, capturing photos or videos, among others- using an external application (see the upper swimlane in Figure 2) for a specific cit-izen science project. The user registers the newly generated check-in throughthe *Rayuela* application. This action must take the data from the mobile de-vice through the GPS and clock to ensure the user is located in the informed geographic coordinate and timestamp (as seen in Figure 3-C).

**Game status feedback** The user receives feedback from the gamification layer regarding the state of the game. The recently completed check-in can contributeto the progress in more than one game challenge within the project, and therefore the registration of the check-in should compute the progress in all possible activegame challenges. Users must be aware of their game progress by means of somekind of gamification resource, e.g. a progress bar (see Figure 3-D).

**Game challenge scoring** The user can **give explicit feedback** about a com- pleted game element through a multi-criteria scoring device. When the user completes a challenge (see bottom challenge in Figure 3-D), the possibility of rating the game challenge is enabled in three ways: through a choice between like and dislike, through an overall score for the game challenge (single criteria rating) or through a detailed score in several criteria: the area, the time restric- tion, the number of samples, the difficulty or the obtained reward (multi-criteria rating). These expressed preferences must be used to update the user profile andcontribute to better recommendations.

## 3 Conclusion and Future work

*Rayuela* is a gamification platform that can be set up with several citizen sci- ence projects and is transparent to the collaboration tools they already have. It allows participating in more than one citizen science project, choosing among a project-specific list of generated and recommended game elements for the player,recording the completion of sampling tasks, and being aware of the progress in the game.

Citizen science projects may apply adaptive gamification to have greaterparticipation from the general public and reach a higher project efficiency.

This tool allows incorporating a gamification engine in a transparent wayto a set of CLCS-supported citizen science projects. It also allows materializinga platform of gamified projects that has several advantages. On the one hand, people's linkage for a project can be transferred to other projects, since people discover new projects through the platform. On the other hand, by joining anew project, the user does not start from scratch. This project capitalizes on the experience and knowledge the person acquired by participating in other projects.

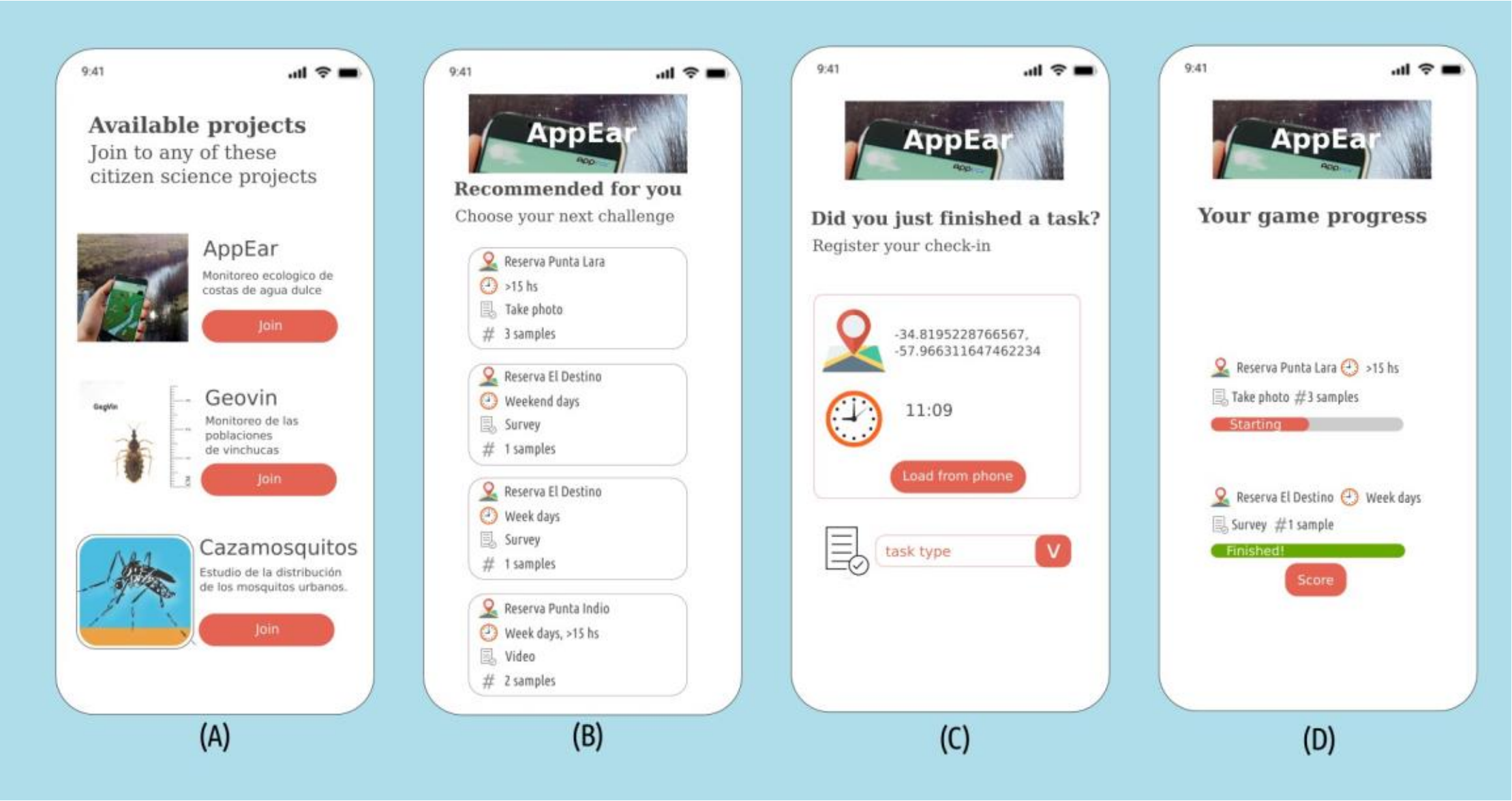

**Fig. 3.** *Rayuela* screens: gamified projects list (A), recommended challenges for user (B), check-in registration (C), and game layer feedback (D)

The challenge-based gamification approach considered the user's profile, theproject's coverage goals, area priority configuration, and global system status to adapt the game challenges.

As further work, the inclusion of other game elements and mechanics will be analyzed.

## Acknowledgements

This article publication is based upon the work of the Project *RUN Citi-zen science and co-creation for regeneration of urban rivers and mitigation of risks*, which has received funding from the CYTED *Programa Iberoamericano deCiencia y Tecnologıa para el Desarrollo*, under grant agreement No 420RT0008, http://cyted.org/es/run.

## References

1. Busch, M., Mattheiss, E., Hochleitner, W., Hochleitner, C., Lankes, M., Fröhlich, P., Orji, R., Tscheligi, M.: Using player type models for personalized game design - an empirical investigation. Interaction Design and Architecture(s) Journal **28**,145–163 (01 2016)
2. Bockle, M., Novak, J., Bick, M.: TOWARDS ADAPTIVE GAMIFICATION: A SYNTHESIS OF CURRENT DEVELOPMENTS. Research Papers (Jun 2017), https://aisel.aisnet.org/ecis2017_rp/11
3. Cochero, J.: Appear: A citizen science mobile app to map the habitat quality of continental waterbodies. Ecologia Austral **28**, 467–479 (08 2018)
4. Cochero, J., Pattori, L., Balsalobre, A., Ceccarelli, S., Marti, G.: A convolutional neural network to recognize chagas disease vectors using mobile phone images. Ecological Informatics **68**, 101587 (2022)
5. Dalponte Ayastuy, M., Torres, D.: Relevance of non-activity representa- tion in traveling user behavior profiling for adaptive gamification. In: Pro- ceedings of the XXI International Conference on Human Computer Inter- action. Interaccion ’21, Association for Computing Machinery, New York, NY, USA (2021). https://doi.org/10.1145/3471391.3471431, https://doi.org/ 10.1145/3471391.3471431
6. Deterding, S., Dixon, D., Khaled, R., Nacke, L.: From Game Design El- ements to Gamefulness: Defining ”Gamification”. In: Proceedings of the 15th International Academic MindTrek Conference: Envisioning Future Me- dia Environments. pp. 9–15. MindTrek ’11, ACM, New York, NY, USA (2011). https://doi.org/10.1145/2181037.2181040, http://doi.acm.org/10.1145/ 2181037.2181040, event-place: Tampere, Finland
7. Gobel, S., Wendel, V.: Personalization and Adaptation, pp. 161–210. SpringerInternational Publishing, Cham (2016). https://doi.org/10.1007/978-3-319-40612- 1 7, https://doi.org/10.1007/978-3-319-40612-1_7
8. Hamari, J., Koivisto, J., Sarsa, H.: Does gamification work? a lit- erature review of empirical studies on gamification (01 2014). https://doi.org/10.1109/HICSS.2014.377
9. Heeter, C., Magerko, B., Medler, B., Lee, Y.H.: Impacts of forced serious game play on vulnerable subgroups. Int. J. Gaming Comput. Mediat. Simul. **3**(3), 34–53 (Jul 2011). https://doi.org/10.4018/jgcms.2011070103, https://doi.org/ 10.4018/jgcms.2011070103
10. Huotari, K., Hamari, J.: Defining gamification: a service marketing perspective. In: Proceeding of the 16th international academic MindTrek conference. pp. 17–22(2012)
11. Iversen, S.: In the double grip of the game: Challenge and Fallout 3. Game Stud- ies **12** (2012), http://www.gamestudies.org/1202/articles/in_the_double_grip_of_the_game
12. Nugent, J.: inaturalist, citizen science for 21st-century naturalists. Science Scope **41**(7), 12–13 (2018)

# Named Entity Extraction in Requirement Specification: A Comparison

**Luciana Tanevitch[1], Leandro Antonelli[1] and Diego Torres[1,2]**

[1] LIFIA, CICPBA-Facultad de Informática, UNLP, Argentina
{name.surname}@lifia.info.unlp.edu.ar
[2] Departamento de Ciencia y Tecnología, UNQ, Argentina

**Abstract** Software requirements specifications are often written in nat- ural language, but extracting the main concepts from these documents can be challenging for computer agents. Named entity extraction is a task that involves recognizing entities in a text and linking them to a knowledge graph for disambiguation. In the domain of requirements engineering, applying this task can facilitate the representation and effi- cient management of complex information. This work compares different named entity extraction tools in the task of extracting entities from a requirements specification, considering technical aspects but also per- formance in terms of precision, recall, f1-score and accuracy.

## 1 Introduction

Software Requirements Specifications (SRS) define in an unstructured way the requirements that a system should satisfy. Identifying and extracting the main concepts involved in a requirements specification could be useful for several auto-matic tasks, such as classification and categorization tasks [15]. However, com- puter agents are not able to process and immediately understand the content and information included in the natural language documents such as SRSs. Nev- ertheless, research lines introduced techniques to enable machines to convert text into information that can be processed automatically and to deal with the ambiguity of natural language [4,7].

Natural Language Processing (NLP) is a branch of Artificial Intelligence that enables computers to understand texts written in natural language[13,16]. NLP can be used to extract entities from a text using a specific technique called Named Entity Recognition (NER), which allows recognition and classification of named entities in a text into predefined classes [13]. As natural language allows for multiple meanings of the same concept, once entities are detected, it is necessary to disambiguate them to determine their true meaning according to the context in which they occur.

Knowledge Graphs (KG) allow structuring complex information in a proper format for computers [11]. Moreover, knowledge graphs are suitable for auto- matic data processing. Each node in the graph represent a unique concept, so linking a named entity in a text to its corresponding node enables meaning dis- ambiguation. Al-Moslmi et. al propose a pipeline for transforming texts into KGs [3]. This process is known as Named Entity Extraction (NEE) and involves three main tasks: Named Entity Recognition (NER), Named Entity Disambiguation (NED) and Named Entity Linking (NEL).

The aim of this paper is to compare various NEE tools applied to require-ments domain, considering core features (supported languages, KG used for link- ing and disambiguation technique), and their performance in requirements spe-cification analysis (measured in terms of precision, recall, f1-score and accuracy). This paper is organized as follows. In Section 2, previous work in the liter- ature in this area is introduced. Then, Section 3 defines the evaluation methodincluding the metrics, the tools to be compared, and the data used for the eval-uation. The results of the evaluation are described in Section 4. Finally, Section 5 summarizes the paper's conclusions and suggests some possible future work.

## 2 Related Work

Tedeschi et al. [21] evaluate NER-based strategies that help narrow down the performance gap between systems trained on limited data and those trained on massive corpora. This work is relevant to our study as it explores approaches for improving NER system performance. Checco et al. [6] develop a tool that de- tects named entities in fashion blogs and links them to a fashion ontology. While their focus is on a different domain, their work demonstrates the potential of en- tity extraction and linking techniques to discover domain-specific information. Hosseni and Bagheri [12] consider different entity linking techniques for detectingimplicit entities, particularly in tweets. Given the potential ambiguity in require- ment specifications, their approach is valuable for detecting and disambiguating implicit references. Vychegzhanin and Kotelnikov[23] conduct a comparison of domain-independent NER tools, focusing on their characteristics and perform- ance in recognizing specific entity types on pre-trained news datasets. Rizzo and Troncy[19] propose a framework for evaluating NEE tools based on named entitydetection, entity type detection, and entity disambiguation criteria. Abdallaha et al. [2] build a framework for evaluating entity extraction tools applied to texts from various domains. Their criteria consider generic, technical, and advanced features, as well as efficiency measures.

## 3 Evaluation Method

### 3.1 Metrics

**Core features** Following Abdallaha et al. [2], we considered the following core features to evaluate the tools: supported language, disambiguation technique, and knowledge graph used.

The "language" feature specifies supported languages, crucial for multicul- tural requirements, as precise terminology may not translate well into inter- mediate languages. The "disambiguation technique" determines entity disam- biguation based on context, impacting accuracy and performance. The tool's "knowledge graph", such as DBpedia and Wikidata, is essential for entity link- ing.

**Performance** According to 4 parameters are used for evaluating the performance: precision, recall, f1-score and accuracy.

Precision assesses the model's ability to accurately identify entities by con- sidering both true positives and false positives. Recall measures the model's cap-ability to identify all correct entities, with false negatives representing missed entities. The F1-score balances precision and recall. Accuracy verifies if the linkedresource is contextually correct for the named entity.

### 3.2 Data

To choose the tools, a systematic review was conducted to gather various NamedEntity Extraction (NEE) tools from studies in a similar domain as this work. Ad- ditionally, we performed searches in public code repositories using the keywords *named entity extraction*, *named entity linking*, and *named entity disambiguation* to identify additional tools. Once collected, we began narrowing down the num- ber of tools due to various reasons, such as tools no longer being hosted, unavail- able source code, or limited entity recognition in the specific domain of applica- tion. Consequently, the tools we will analyze are those that are freely available, support the English language, and have shown promising results in preliminary testing using different small-scale requirement datasets (i.e., high entity detec- tion rates). The selected tools for this work are: Wikifier [5], DBpedia Spotlight [17], Babelfy[18], TagMe [10] and Spacy.

The following text was used as input for the tools, which it was writtenby a functional analyst, and describes a functional requirement that involvesvarious technical concepts related to the named entity extraction pipeline: *Asa news agency, we want to automatically extract and disambiguate mentionsof places, people, and organizations from newspapers. When a new article issubmitted, the system should identify all named entities in the text and link them to their corresponding entities in a knowledge graph such as DBpedia orWikidata. The system should use contextual information such as the surrounding words and the sentence structure to disambiguate entities that have multiplepossible meanings. The system should also be able to handle entity coreference,where different mentions in the text refer to the same entity. The disambiguatedentities should be stored in a structured format, such as RDF or JSON, for further processing and analysis. The system should be scalable and able to handlea large volume of articles in real-time. The accuracy of the system should beevaluated against a manually labeled dataset to ensure high precision and recall.* An expert analysis can be performed on this data to determine which entities should be automatically detected. It is called *ground truth* dataset because it contains the entities expected to be found by the evaluated tools, and is defined as follows: *news agency, places, persons, organization, article, system, named en-tity, newspapers, knowledge graph, DBpedia, Wikidata, contextual information, sentence structure, meanings, coreference, RDF, JSON, dataset, accuracy, ana-lysis, precision, recall* A table with expected entities for each named entity is shown in this document.

## 4 Results

Table 2 compares the core features of the tools mentioned in Section 3. Sinceall the tools support several languages, Babelfy is the one with the most cov- erage. NER performance may differ in each language, as each uses a different model. They use machine learning approaches for disambiguation phase, each approach has its owns advantages and disadvantages. The evaluated tools can link mentions to Wikidata, DBpedia or BabelNet. As these graphs do not de- scribe specific concepts from the domain of software requirements, some namedentities are likely to remain unlink. The importance of BabelNet is that it in- cludes lexical resources, which provide a foundation of structured knowledge, sousing lexical and semantic knowledge could improve disambiguation tasks.

Table 3 compares the performance of the tools. DBpedia Spotlight achieves perfect precision and accuracy scores, but it has a low recall, indicating that it recognized fewer entities compared to the ground truth. On the other hand, Ba- belfy recognized a high number of entities but suffered from low precision due tooverfitting, detecting irrelevant concepts unrelated to the requirements domain.TagMe and Spacy kept a good balance in all their measures: they recognized al- most all the expected words with a precision higher than 65%. Wikifier performs well in entity detection but has a lower recall compared to the aforementioned tools. TagMe obtained the higher results for the used data, but Spacy was so close: it has similar values for accuracy and precision, keeping the same recall. Sowe can say that even if TagMe had a good performance in requirement domain, better results could be obtained using an specific-domain knowledge graph for training Spacy.

## 5 Conclusions

The domain of requirements is highly ambiguous and complex. In order to enable automatic processing of requirement specifications, techniques such as knowledge graphs can be employed. Linking a concept to a knowledge graph allows for dis- ambiguation of its meaning, and entity extraction tools can be used for this pur- pose. Evaluation results exhibit considerable variability in accuracy, precision, and recall. While precision and recall are useful metrics, accuracy is critical in determining whether the tool is performing correctly.

It is possible that the combination of different tools can improve results and provide greater accuracy in identifying requirement entities. Moreover, using a KG specific for requirements domain could help to solve the problem about thoseunrecognized entities due there is not an entry in a general-purpose KG.

This evaluation may be applied to different domains to determine whether thetools have the same ability to recognize entities as they do in the requirements domain. The next step could be to detect relationships between the identified entities in order to build a knowledge graph. A more specific study could be performed using massive datasets of the requirement domain, what it would re- quire techniques for annotating massive data. In addition, automatic techniques should be considered to properly choose the expected entities for each of the knowledge graphs to be evaluated.

**Table 2.** Core features comparison results

| **Tool** | **Language** | **Disambiguation technique** | **Knowledge graph** |
|---|---|---|---|
| DBpedia Spot- light | Among the supported languages are German, English, Spanish, French, Italian, and Portuguese. The full list can be found at [9] | They modeled DBpedia resource occurrences in a Vector Space Model (VSM) using a variant of the TF-IDF algorithm to weigh words based on their ability to distinguish between candidates for a given surface form. Therefore, they use cosine similarity to compare the similarity between context vectors and the context surrounding the surface form.In an improved version [8], they use a generative model to calculate the probability that a candidate is correct for a mention, which improves disambiguation accuracy, as well as time performance and required space | DBpedia |
| Wikifier | It supports around 100 languages among which are German, English, Spanish, French, Italian, Portuguese, and Chinese. | They build a bipartite graph where each node on the left represents mentions and each node on the right represents Wikipedia entries for those mentions. The graph is augmented by edges between concepts based on their semantic relatedness. They use the graph to calculate the PageRank score for each vertex, and after some iterations, they obtain the relevant concepts for each mention. A threshold value can be specified by the user to discard all candidates that were scored below that value. | The entities are linked to wikipedia pages, although the service includes information from the Wikidata ID and DBpedia URI in the response. |
| Babelfy | Provides support for 271 languages [1], including English | They build a directed graph which relates mentions in the text with their Babel candidate They construct a graph that relates mentionsin the text to their possible candidates in Babel, also linking those candidates that are semantically similar. In this way, all possible interpretations are obtained for each mention, and the heuristic of the densest subgraph is applied to obtain the best candidate for a men- tion (based on lexical and semantic coherence), and thus arrive at the most coherent semantic interpretation. | BabelNet |
| TagMe | English, Italian and German | They apply a voting scheme where the goal is to reach a “collective agreement” on the real meaning of some mentions. To select the best candidate, they tested two algorithms: Disambiguation by Classifier (DC), which computes the probability of correct disambiguation for every candidate of a mention and then selects the best one; and Disambiguation by Threshold (DT), which selects the top n-best candidates. Then other works improved that initial version by topical classification and clustering [22] [20]. | Entities are linked to a Wiki- pedia page, but Wikidata URIs could be inferred from them |
| Spacy | English, although a model can be trained to recognize other languages | Starting from a defined knowledge graph, candidates for entities can be generated. Then, a machine learning model selects the most suit- able candidate. The power of Spacy lies in the fact that each compon- ent is customizable, allowing the user to improve accuracy by com- bining techniques for candidate generation and scoring. | It allows configuring the desired knowledge graph, although in this work, an implementation with Wikidata was used. |

**Table 3.** Performance evaluation results

| Tool | Accuracy | Precision | Recall | F1-score |
|---|---|---|---|---|
| DBpedia Spotlight | 1 | 1 | 0.27 | 0.42 |
| Wikifier | 0.85 | 0.66 | 0.63 | 0.65 |
| Babelfy | 0.9 | 0.39 | 0.95 | 0.55 |
| **TagMe** | **0.75** | **0.71** | **0.9** | **0.80** |
| Spacy | 0.68 | 0.66 | 0.9 | 0.76 |

## References

1. (Apr 2015), http://babelfy.org/javadoc/it/uniroma1/lcl/jlt/util/Language.html
2. Abdallah, Z.S., Carman, M., Haffari, G.: Multi-domain evaluation framework for named entity recognition tools. Computer Speech & Language **43**, 34–55 (May 2017)
3. Al-Moslmi, T., Gallofré Ocaña, M., Opdahl, A., Veres, C.: Named Entity Extraction for Knowledge Graphs: A Literature Overview. IEEE Access **8**, 32862–32881 (Feb 2020)
4. Antonelli, L., Delle Ville, J., Dioguardi, F., Fernández, A., Tanevitch, L., Torres, D.: An iterative and collaborative approach to specify scenarios using natural language. In: Workshop on Requirements Engineering (WER22),(Modalidad virtual, 23 al 26 de agosto de 2022) (2022)
5. Brank, J., Leban, G., Grobelnik, M.: Annotating documents with relevant wikipedia concepts. Proceedings of SiKDD **472** (2017)
6. Checco, A., Demartini, G., Löser, A., Arous, I., Khayati, M., Dantone, M., Koop- manschap, R., Stalinov, S., Kersten, M., Zhang, Y.: Fashionbrain project: A vision for understanding europe's fashion data universe. arXiv preprint arXiv:1710.09788(2017)
7. Corral, A., Sánchez Crespo, L.E., Antonelli, L.: Building an integrated requirements engineering process based on intelligent systems and semantic reasoning on the basis of a systematic analysis of existing proposals. JUCS - Journal of Universal Computer Science **28**, 1136–1168 (11 2022)
8. Daiber, J., Jakob, M., Hokamp, C., Mendes, P.N.: Improving efficiency and accuracy in multilingual entity extraction. In: Proceedings of the 9th International Conference on Semantic Systems. pp. 121–124. ACM, Graz Austria (Sep 2013) Named Entity Extraction in Requirement Specification: A Comparison 7
9. DBpedia Spotlight - Shedding light on the web of documents: FAQ, https://www.dbpedia-spotlight.org/faq (accessed March 8, 2023)
10. Ferragina, P., Scaiella, U.: TAGME: on-the-fly annotation of short text fragments (by wikipedia entities). In: Proceedings of the 19th ACM international conference on Information and knowledge management. pp. 1625–1628. ACM, Toronto ON Canada (Oct 2010)
11. Hogan, A., Blomqvist, E., Cochez, M., d'Amato, C., de Melo, G., Gutierrez,C., Gayo, J.E.L., Kirrane, S., Neumaier, S., Polleres, A., Navigli, R., Ngomo,A.C.N., Rashid, S.M., Rula, A., Schmelzeisen, L., Sequeda, J., Staab, S., Zim- mermann, A.: Knowledge Graphs. ACM Computing Surveys **54**(4), 1–37 (May 2022), arXiv:2003.02320 [cs]
12. Hosseini, H., Bagheri, E.: From Explicit to Implicit Entity Linking: A Learn to Rank Framework. In: Goutte, C., Zhu, X. (eds.) Advances in Artificial Intelligence,vol. 12109, pp. 283–289. Springer International Publishing, Cham (2020), series Title: Lecture Notes in Computer Science
13. Khurana, D., Koli, A., Khatter, K., Singh, S.: Natural language processing: state of the art, current trends and challenges. Multimedia Tools and Applications **82**(3), 3713–3744 (Jan 2023)
14. Malik, G., Cevik, M., Bera, S., Yildirim, S., Parikh, D., Basar, A.: Software re- quirement specific entity extraction using transformer models. Proceedings of theCanadian Conference on Artificial Intelligence (May 2022)
15. Malik, G., Cevik, M., Khedr, Y., Parikh, D., Başar, A.: Named Entity Recognition on Software Requirements Specification Documents. Proceedings of the Canadian Conference on Artificial Intelligence (Jun 2021)
16. Martinez, A.R.: Natural language processing. Wiley Interdisciplinary Reviews: Computational Statistics **2**(3), 352–357 (2010)

17. Mendes, P.N., Jakob, M., Garc´ıa-Silva, A., Bizer, C.: DBpedia spotlight: shedding light on the web of documents. In: Proceedings of the 7th International Conference on Semantic Systems. pp. 1–8. ACM, Graz Austria (Sep 2011)
18. Navigli, R., Ponzetto, S.P.: Babelnet: Building a very large multilingual semantic network. In: Proceedings of the 48th annual meeting of the association for compu-tational linguistics. pp. 216–225 (2010)
19. Rizzo, G., Troncy, R.: NERD: Evaluating Named Entity Recognition Tools in the Web of Data (2011)
20. Scaiella, U., Ferragina, P., Marino, A., Ciaramita, M.: Topical clustering of search results. In: Proceedings of the fifth ACM international conference on Web search and data mining. pp. 223–232. ACM, Seattle Washington USA (Feb 2012)
21. Tedeschi, S., Conia, S., Cecconi, F., Navigli, R.: Named Entity Recognition for Entity Linking: What Works and What's Next. In: Findings of the Association for Computational Linguistics: EMNLP 2021. pp. 2584–2596. Association for Compu- tational Linguistics, Punta Cana, Dominican Republic (Nov 2021)
22. Vitale, D., Ferragina, P., Scaiella, U.: Classification of Short Texts by Deploy- ing Topical Annotations. In: Baeza-Yates, R., de Vries, A.P., Zaragoza, H., Cam- bazoglu, B.B., Murdock, V., Lempel, R., Silvestri, F. (eds.) Advances in Informa-tion Retrieval. pp. 376–387. Lecture Notes in Computer Science, Springer, Berlin, Heidelberg (2012)
23. Vychegzhanin, S., Kotelnikov, E.: Comparison of Named Entity Recognition Tools Applied to News Articles. In: 2019 Ivannikov Ispras Open Conference (ISPRAS). pp. 72–77 (Dec 2019)

# Formulation of a model to determine current andpotential areas (*Persea americana* Mill) of Hassvariety avocado crops, in the department of Risaralda, based on edaphoclimatic variables and fruit quality.

**Gloria Edith Guerrero, Julio César Chavarro, Cesar Manuel Castillo, César Augusto Jaramillo, Juan Pablo Arrubla and Andrés Alfonso Patiño**

[1]School of Chemistry, Universidad Tecnológica de Pereira, Colombia.
[2]Faculty of Engineering, Universidad Tecnológica de Pereira, Colombia.

*E-mail(s) Authors: gguerrero@utp.edu.co; jchavar@utp.edu.co;cesar.castillo@utp.edu.co; swokosky@utp.edu.co; juanpablo77@utp.edu.co; andrespatinomartinez@gmail.com;

**Abstract**

Agriculture is one of the fundamental pillars of any worldwide population, and the proper management of information allows timely decisions to move any company forward. National and departmental government agencies support agricultural sectors that emerge as an excellent opportunity to increase production levels andproduct commercialization [1], such as the Hass avocado (*Persea americana* Mill)crop. One of the challenges facing this type of crop is to find potential planting and productivity zones to contribute to technological developments in the agricul- tural sector, benefiting the Hass avocado growers of the department of Risaralda. Therefore, this study proposes the formulation of a model to determine current and potential cultivation areas of Hass avocado (*Persea americana* Mill) in the department, based on edaphoclimatic variables and fruit quality, taking advan- tage of current trends in precision agriculture, including techniques derived from Machine Learning, such as the use of Supervised Learning algorithms, among which is Random Forest.

## 1 Introduction

Food security has become a pressing challenge due to rapid population growth, climatechange and water scarcity, especially in developing countries [2]. Nowadays, modifica- tions and adaptations are made in agricultural practices to improve soil fertility andthe different climatic changes that are currently occurring. Therefore, it is necessaryto evaluate the environmental impacts of agriculture in terms of water, nutrients (soil) and atmospheric components [3].

Avocado (*Persea americana* Mill) Hass variety is the most common commercial avo- cado crop in the world due to the number of essential nutrients and important phytochemicals [4]. This fruit is grown in Colombia, where the Hass avocado pro- duction system has recently increased due to the excellent economic opportunitiesand the high unsatisfied domestic demand [5]. The Hass avocado-producing depart- ments are Tolima, Antioquia, Caldas, Santander, Bolivar, Quind´ıo, Cesar, Valle del Cauca, Risaralda, and Cundinamarca; in Risaralda, avocado is grown in 13 of the 14 municipalities [6]. To address the problems of avocado cultivation, information and communication technology (ICT) approaches are adopted, such as precision agricul- ture or real-time crop data collection [7]. In this sense, ICT topics for the agricultural sector include procedures for the digital management of geographic information on crops, decision-making systems for mechanizing processes based on georeferencing, and epidemiological early warning information systems, among others [7] [8].

However, difficulties are observed with crop management given the variability in soil and edaphoclimatic conditions of the department. This situation causes effects suchas heterogeneity of fruit quality, besides the lack of scientific work to understand the behavior of the crop in the region. This study proposes to formulate a model to deter- mine current and potential zones of Hass avocado (*Persea americana* Mill) cultivation in the department of Risaralda, based on edaphoclimatic variables and fruit quality, using the current trends of precision agriculture and machine learning, to contributeto technological developments in the agricultural sector, benefiting the Hass avocado growers of the department.

## 2 Bibliographic Review

Machine learning (ML) is an important decision support tool in fields such as crop yield prediction, including supporting decisions on what agricultural products to grow and what to do during the crop growing season; therefore, several machine learning algorithms have been applied to support research aimed at crop yield prediction or suggestion, where the most commonly used characteristics are temperature, precipi- tation and soil type [9]; In that sense, the different uses of the data obtained from the soil, such as the selection of the dataset for training, as well as the choice of soil envi- ronmental covariates, could boost the accuracy of automatic learning techniques [10]. One of the tools used within automatic learning are the crop-oriented recommendation systems where, based on provided variables, a model is created to predict or suggest which crop can be cultivated; to establish this, the models use historical data, such as climatic data (temperature, humidity, pH, rainfall) and fertilizer values (nitrogen, potassium, phosphorus) [11]. Machine learning employs methods such as supporting processes in charge of data analysis [12]. In [13] it is mentioned that different Artifi- cial Intelligence techniques have been proposed and among these techniques that are part of Precision Agriculture, more specifically in the fields of crop recommendation systems, it is observed that algorithms such as k nearest neighbors (KNN), similarity-based models, set-based models, neural networks, among others, are included. These algorithms consider several external characteristics, such as meteorological data, and others, such as the soil profile, to provide the best recommendations, by examining the data, the most important attributes are obtained using techniques such as principal component analysis (PCA) and linear discrimination analysis (LDA). These extracted features are used to train models such as Na¨ıve-Bayes (NBC), Random Forest, KNN; using training data and the performance is evaluated on test data using techniques such as cross-validation, RMSE or accuracy. Reliable predictions of crop yields are dif- ficult for the development of agriculture; crop production varies according to different climatic conditions, having conditions such as dry periods and increasing tempera- tures; these forces strengthen the need for analysis of crop production in different climatic conditions, in this sense, [14] analyzes the automatic learning method, the Random Forest supervised algorithm can analyze the growth of crops concerning the current climatic conditions and biophysical change. Similarly, in [10], the ability of the Random Forest algorithm to predict soil classes from different training data sets and extrapolate such information to a similar area was evaluated.

Another aspect to consider is the usage of automatic learning models/algorithms and their possible applications to geospatial data, where special attention is given to the models used that are based on artificial neural networks (multilayer perceptron, general regression neural networks, self-organizing maps), statistical learning theory (support vector machines) [15], geo-statistical echniques such as Ordinary Kriging [16], or algo- rithms such as Random Forest Spatial Interpolation (RFSI) [17].

Obtaining the visualization of different crops and their associated characteristics, high resolution yield maps are used, which are an essential tool in modern agriculture, and these are obtained by spatial interpolation, however, spatial interpolation is gener- ally performed using methods that can be computationally demanding [18]; therefore, some works have been carried out to explicitly take into account the spatial compo- nent in machine learning, where observations on the prediction location are included, performed using Random Forest Spatial Interpolation, comparing it with determinis- tic interpolation methods, such as ordinary kriging, regression kriging, Random Forest and Random Forest for spatial prediction (RFsp) where it is observed that in the case studies of precipitation and temperature, RFSI mostly outperformed regression kriging, inverse distance weighting, Random Forest and RFsp, moreover, RFSI was substantially faster than RFsp, mainly when the training data set was large and high resolution prediction maps were made [17].

## 3 Methodology

This study aims to formulate a recommendation model based on the Random Forest algorithm complemented with Random Forest Spatial Interpolation (a methodology for spatial interpolation used in Machine Learning) involving three data sources: cli- matic and edaphic variables and fruit quality. Considering the formulated model, developing an information system that allows establishing current and potential fruit production zones is proposed in the Department of Risaralda, based on edaphoclimaticand fruit quality variables, thanks to the ongoing execution of the project "Develop- ment of an information system to determine current and potential zones for avocado (Persea americana Mill) Hass variety crops in the Department of Risaralda, based on edaphoclimatic and fruit quality variables (contract 424-201 MinCiencias)"[1].

### 3.1 Description of the Data Set

Sampling will be conducted for a study of soils, fruit quality and climatic data collec-tion in Hass avocado-producing farms in the municipalities of Pereira, Dosquebradas, Santa Rosa de Cabal, Marsella, Apía, Belén de Umbría, Guática and Quinchía. For the data collection of the first data set (dataset), seven (7) LynkBOX CLIMA PLUS climate stations have been installed in the different farms located in the municipalitiesselected for the study, allowing the recording of data on climate variables Tempera- ture (°C), Relative humidity (%), Precipitation (mm), Solar radiation (W/m2), Wind direction and Wind speed, and the building variables humidity, soil temperature and electrical conductivity, which will be recorded for one year long. The second set of datawill be based on a soil fertility analysis, where the parameters, pH, organic matter, elemental analysis (K, Ca, Mg and Na), and phosphorus (P) will be evaluated accord-ing to Colombian technical standards NTC 5264, 5403, 5349 and 5350, respectively. Additionally, in the third set of data, variables such as nitrogen content will be calcu- lated based on organic matter content, aluminum using the KCI IM-EAA volumetrictechnique and texture to the touch Bouyoucos with sodium pyrophosphate - USDA triangular diagram classification - Cl (clayey), L (loamy), S (sandy).

### 3.2 Data Preparation

It will be considered the generated datasets presented in a different format, as observedin the datasets described above; for this reason, it is imperative to clean and normalizethe data for subsequent use with the algorithms. Techniques from the Random Forest algorithm [19] will be used for missing data. A data scaling or normalization processis also conducted to convert the dates and times of the different data sources.

### 3.3 Random Forest Spatial Interpolation

When the data sets are configured, the prediction process will be carried out, which is framed in the use of the Random Forest Spatial Interpolation algorithm, using the localspatial information, it is to say, spatial variables to select the spatial dependencies and the complex spatial patterns that are presented [20]; a first training of the data will be done using the algorithm Random Forest, since being an algorithm based on decision trees, a prediction is made through a series of partitioning rules; the spatial correlationbetween the data obtained is not included in Random Forest standard, it will be taken into account that the nearby data contain information about a prediction location, for this purpose, additional spatial variables will be incorporated in the Random Forest model.

[1]Project supported by MinCiencias, Gobernación de Risaralda, and Alcaldía de Pereira, work executed by the research groups GIA and Oleoquímica of Universidad Tecnológica de Pereira

### 3.4 Accuracy Evaluation

The following accuracy metrics will be used to verify the predictions: coefficient of determination, Accuracy, mean absolute error (MAE) and root mean square error(RMSE) [17].

## 4 Final Considerations

The information system proposed in the execution of the project (contract 424-201 MinCiencias) will allow establishing by means of a suggestion if the current production areas of avocado cultivation are the most appropriate and to determine the potential areas of cultivation.

## References

1. Perfetti, J.J., Bravo-Ureta, B.E., García, A., Pantoja, J., Delgado, M., Blanco,J., Jara, R., Moraga, C., Paredes, G., Naranjo, J., *et al.*: Adecuación de tierras y el desarrollo de la agricultura colombiana: políticas e instituciones. Fedesarrollo **447**, 456 (2019)
2. El-Bendary, N., Elhariri, E., Hazman, M., Saleh, S.M., Hassanien, A.E.: Cultivation-time recommender system based on climatic conditions for newlyreclaimed lands in egypt. Procedia Computer Science **96**, 110–119 (2016)
3. Singh, R., Kumari, T., Verma, P., Singh, B.P., Raghubanshi, A.S.: Compatible package-based agriculture systems: an urgent need for agro-ecological balance andclimate change adaptation. Soil Ecology Letters **4**(3), 187–212 (2022)
4. Dreher, M.L., Davenport, A.J.: Hass avocado composition and potential health effects. Critical reviews in food science and nutrition **53**(7), 738–750 (2013)
5. Ramírez-Gil, J.G., Ramelli, E.G., Osorio, J.G.M.: Economic impact of the avo- cado (cv. hass) wilt disease complex in antioquia, colombia, crops under different technological management levels. Crop protection **101**, 103–115 (2017)
6. MADR, M.d.A.y.D.R.: CADENA DE AGUACATE, Indicadores e instrumentos. Lect Econ. 2019;52(52):165–94 (2019)
7. Martínez, D.H.F., Galvis, C.P.U.: Tic para la investigación, desarrollo e inno- vación del sector agropecuario (2018)
8. Vite Cevallos, H., Carvajal Romero, H., Barrezueta Unda, S.: Aplicación de algoritmos de aprendizaje automático para clasificar la fertilidad de un suelo bananero. Conrado **16**(72), 15–19 (2020)
9. Katarya, R., Raturi, A., Mehndiratta, A., Thapper, A.: Impact of machine learn- ing techniques in precision agriculture. In: 2020 3rd International Conference on Emerging Technologies in Computer Engineering: Machine Learning and Internet of Things (ICETCE), pp. 1–6 (2020). IEEE
10. Geetha, V., Punitha, A., Abarna, M., Akshaya, M., Illakiya, S., Janani, A.: An effective crop prediction using random forest algorithm. In: 2020 International Conference on System, Computation, Automation and Networking (ICSCAN), pp. 1–5 (2020). IEEE
11. Pozdnoukhov, A., Kanevski, M.: Machine learning algorithms for analysis and modeling of geospatial data. In: Annual Conference of International Accociation for Mathematical Geology (IAMG 07), Beijing, China, 25-31 August (2007)
12. Carranza, J.P., Salomón, M.J., Piumetto, M.A., Monzani, F., MONTENE- GRO CALVIMONTE, M., Córdoba, M.A.: Random forest como técnica de valuación masiva del valor del suelo urbano: una aplicación para la ciudad de río cuarto, córdoba, argentina. In: Congresso Brasileiro de Cadastro Técnico Multifinalitário-COBRAC (2018)
13. Sekulić, A., Kilibarda, M., Heuvelink, G.B., Nikolić, M., Bajat, B.: Random forest spatial interpolation. Remote Sensing **12**(10), 1687 (2020
14. Mariano, C., Monica, B.: A random forest-based algorithm for data-intensive spatial interpolation in crop yield mapping. Computers and Electronics in Agriculture **184**, 106094 (2021)
15. Kuhn, M., Johnson, K., Kuhn, M., Johnson, K.: Classification trees and rule-based models. Applied predictive modeling, 369–413 (2013)
16. Talebi, H., Peeters, L.J., Otto, A., Tolosana-Delgado, R.: A truly spatial ran-dom forests algorithm for geoscience data analysis and modelling. Mathematical Geosciences **54**, 1–22 (2022)

# Multidimensional Energy Analysis in Agricultural CoffeeProduction Systems to Assess Sustainability.

**Cristian Méndez-Rodríguez[1,2], Cristian Camilo Ordoñez Quintero[2], Claudia SofiaIdrobo[3], Luis Freddy Muñoz Sanabria[3].**

[1]Environmental Studies Group (GEA), Environmental Sciences, Department of Biology, Univer-sity of Cauca, Popayán 190002, Colombia. cristianmendez@unicauca.edu.co.
[2]Intelligent Management System (IMS), Faculty of Engineering, University Foundation ofPopayán, Popayán 190002, Colombia.
[3]LOGICIEL, Faculty of Engineering, University Foundation of Popayán, Popayán 190002, Co-lombia.

**Abstract.**

A coffee farm is sustainable when it uses the system inputs (energy/materials) and transforms them into high quality beans, avoiding the loss of energy in wasteand pollutants. The aim of the research is to evaluate sustainability in coffee ag-ricultural production systems, through multidimensional energy analysis (with the emergy tool). This research was conducted in the farm "La Angostura" in Popayán, Colombia, during the years 2018-2020. The results show that the farm achieved its best energy performance and was most sustainable in 2020, produc-ing 8750 kg/ha of cherry coffee, with the lowest Transformity (1.35E+06 seJ/J) and the highest Emergy Sustainability Index (0.97), for the three years analyzed. In addition, natural inputs contribute approximately 57% of total emergy, and those coming from the economy 43%. This study provides a precise analysis of the energy flows that interact in the system, the significant uses of energy and thesources of energy at each stage of the production process. This research providesa basis for the management and planning of the territory in relation to its agricul-tural strengths.

## 1 Introduction.

In 2016, Colombia was the third largest coffee producer in the world with approxi- mately 14 million bags (equivalent to 9.23% of world production), followed by Brazil and Vietnam [1]. These data consolidate coffee as the main agricultural product of thecountry and demonstrate its great contribution to the Colombian rural economy. The Cauca department is a region that stands out for its high-quality coffee and denomina-tion of origin; where the aim is to generate a product recognized as excellent at an international level [2].

Coffee is a common crop in the traditional farms of the department of Cauca. According Comité de Cafeteros del Cauca, the department occupies the fourth place nationally in coffee production, with 93.000 families, 31 municipalities of the department of Cauca have 64% of the land for agricultural production to coffee cultivation [1]. A significantpart of the department's coffee production is carried out on the "Popayán plateau", which includes the municipalities of Popayán, Piendamó, Morales, Cajibío, El Tambo and Timbío, with traditional and technified production systems. This research will be carried out at the "La Angostura" farm in the village Clarete Bajo of the Popayán city, where there is a traditional coffee plantation (small producers), with approximately 800coffee trees of the Castillo variety.

In recent years, the agricultural sector of Cauca has focused on the development of strategies for the generation of value chain, with coffee being a key link, focusing on the production of specialty coffee [3], where the improvement of efficiency and sus- tainability constitutes an innovative and necessary bet for the growth of the productivechain. In addition to the importance of coffee in the region, and the scientific advancesin this area, it is necessary to study the use of conceptual and methodological tools thatallow the modeling of a productive chain in relation to its multidimensional energy flows (coming from the natural, social and economic systems) in order to have an inte- gral evaluation of the sustainability of coffee production systems [4]–[6].

A coffee farm is efficient when it uses the energy/materials input to the system and transforms them into high quality beans [7], [8], avoiding the loss of energy in waste and pollutants. The objective of this article is to model the stages of the production process and identify the multidimensional energy flows involved, such as: natural re- sources (sun, wind, rain, soil, etc.) and resources from the socioeconomic component (fertilizers, equipment and machinery, infrastructure, labor, technical services, etc.), inorder to evaluate the sustainability of the farm. This paper uses the methodological tool proposed in [9], where the processes of the production chain are modeled and an Emergy evaluation is carried out to estimate the sustainability of the farm. The study'smain contribution is to provide baseline information on the sustainability of the coffeefarm and the most significant uses of resources (in energy terms).

The article is organized as follows: section 2 presents the methodology (the general context of the research, the first modeling phase, and the Emergy tool to evaluate sus- tainability in agricultural systems); section 3 shows the main results of the research; and finally, the conclusions are presented.

## 2 Methodology.

### 2.1 Agricultural coffee production system description – "La Angostura" farm.

The research was carried out taking as a reference the data provided by the TraditionalFarm "La Angostura", located in the village Clarete Bajo, north of the city of Popayán,Cauca, Colombia, and has an area of 5.8 Hectares. This farm has a traditional polycul-ture of Castillo variety coffee, which is characterized for being one of the Colombian coffee varieties resistant to rust, and stands out for its aroma and citric acidity. The cropis planted in semi-shade, with bushes such as orange, mandarin, lemon, and "guamos"trees. Currently there are 800 coffee plants, it is a young coffee plantation (6 years old), and has a planting density equivalent to 5000 trees/ha. The geolocation of the farm (average elevation 1800 m.a.s.l), the tropical climate of the region and the biodiversityfavor coffee production. Three stages were identified on the farm: planting, harvesting,and processing, where the final product obtained is dry parchment coffee. The analysisof energy flows for the "La Angostura" farm was carried out for the years 2018, 2019 and 2020, due to the unavailability of information from the owners.

### 2.2 Phase 1: Modeling – ANSI/ISA 88 Standard.

This tool allows modeling industrial processes, by describing equipment and proce- dures [9]. The ISA-88 committee has published 5 parts of the standard. The Part 1: Models and Terminology, which seeks to model the production process in relation to the chronological procedures and associated resources, is very useful (Figure 1) [10]. Phase 1 is applied at each stage of an agricultural production system. This process mod-eling using ISA 88 helps to characterize reliably the energy flows (considering conven-tion proposed by Emergy Analysis) present in the agricultural production system.

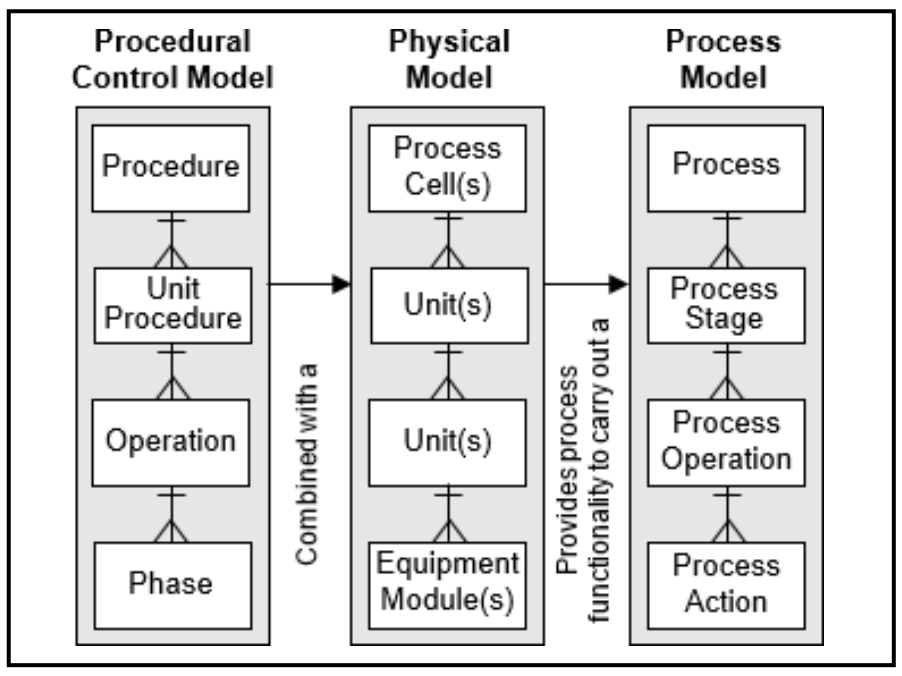

**Fig. 1.** Process modeling - ANSI/ISA 88 Standard.

### 2.3 Phase 2: Emergy

The emergy is defined as the available energy previously required, which has been used directly or indirectly in the transformations necessary to manufacture a product or ser-vice (Y = R + N + F, Figure 2). The unit of measurement of emergy is solar emjoules (seJ). Energy flows are represented in three categories of resources: (i) R as renewable,

(ii) N as non-renewable (R and N flows are provided by the environment) and (iii) F the inputs of the economy (provided by the market and flows coming from the econ- omy) [11]. Emergy is a quantitative tool, based on the law of conservation of energy, thermodynamics, systems theory and systems ecology [12], achieving an integral anal-ysis of the production system [13].

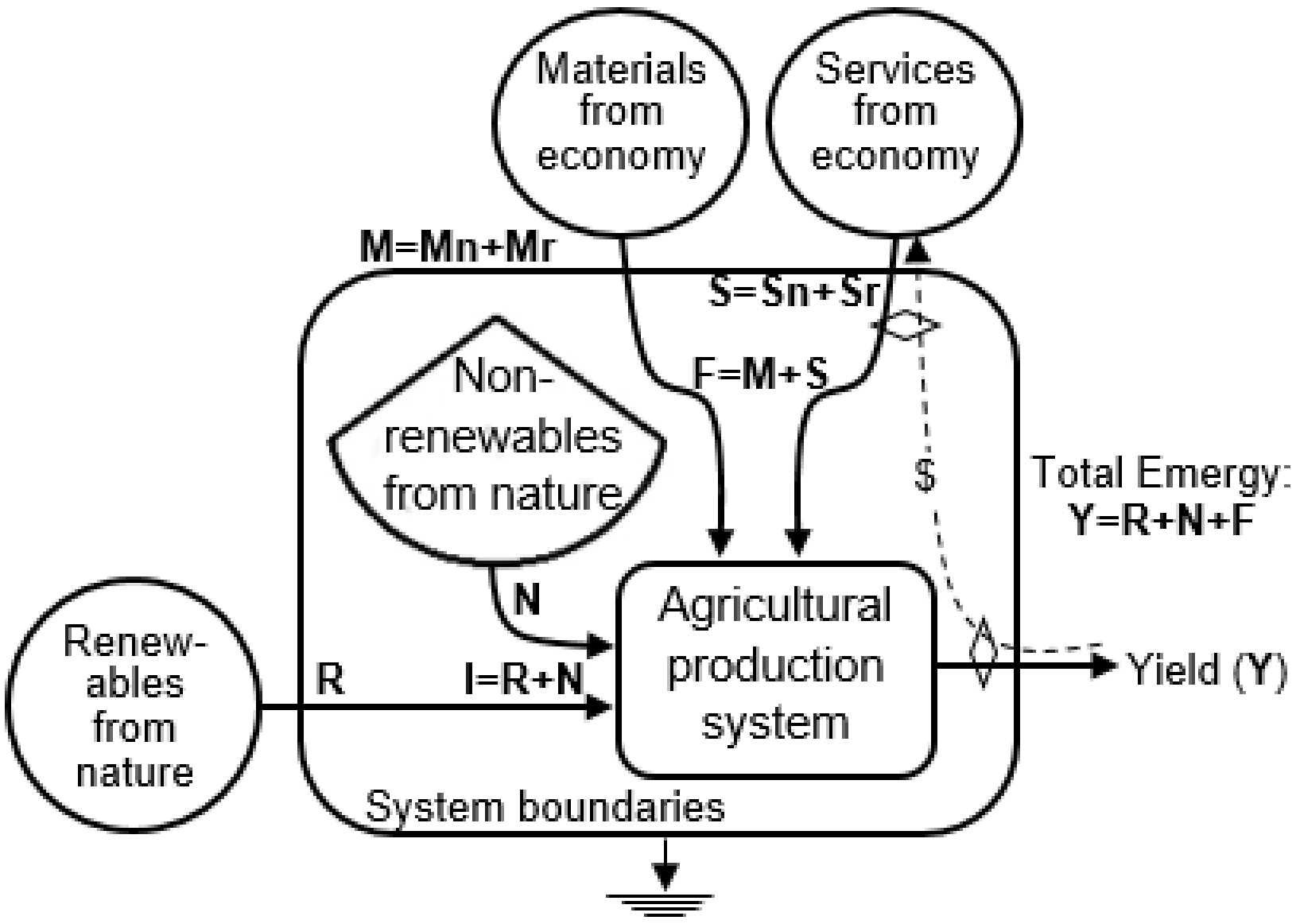

**Fig. 2.** Agricultural production system model, based on the Emergy analysis [14].

## 3 Results and Discussion.

### 3.1. Stages modeling - production process.

The unit procedure, operations, by-products flows, and phases of these stage are pre-sented in Figure 3, and Table 1.

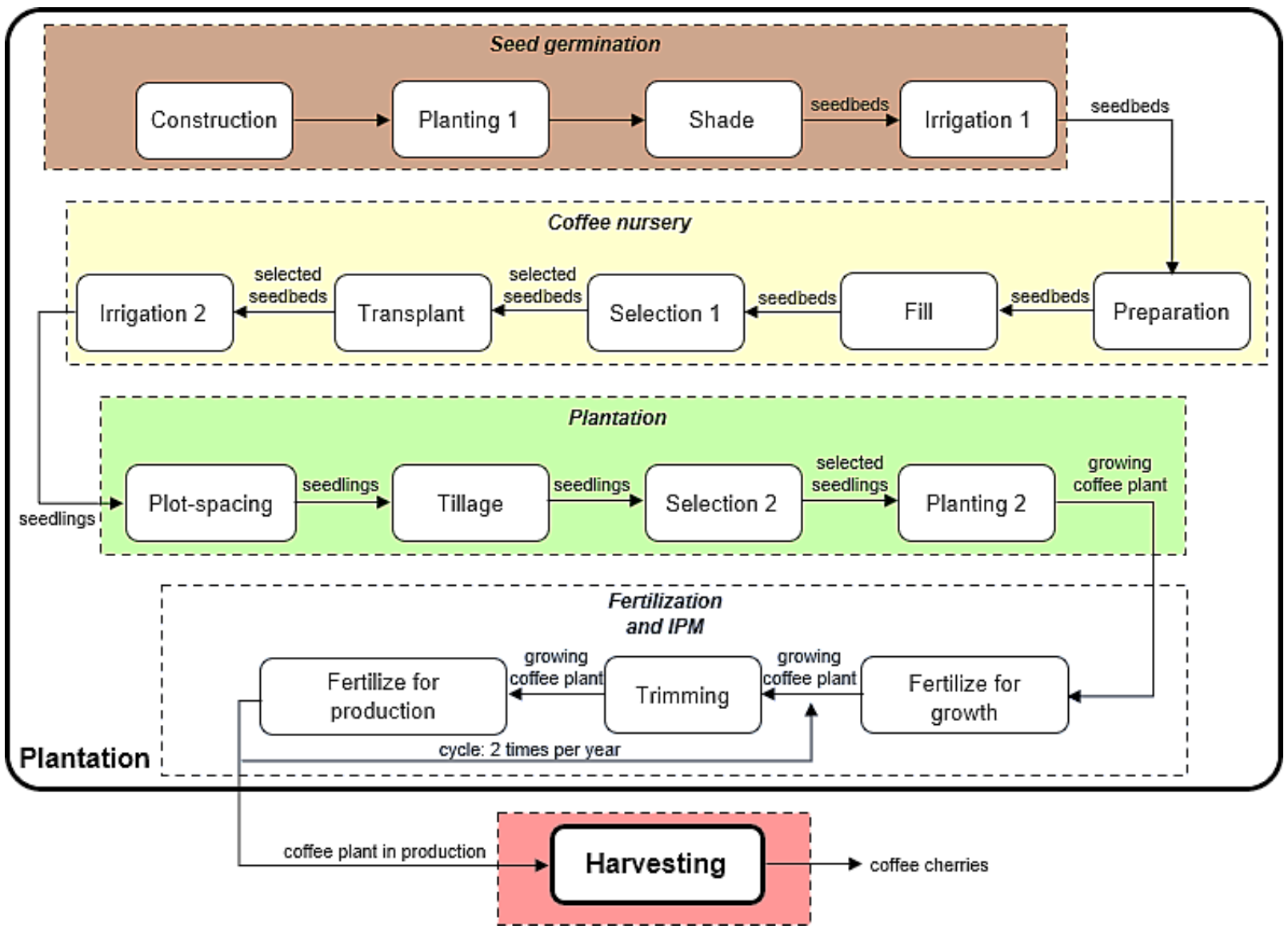

**Fig. 3.** Operations and flows of plantation and harvesting stages.

**Table 1.** Description of plantation and harvesting stages.

| Jnit | Pro-cedure Procedure | Operation | Phase |
|---|---|---|---|
| | Seed germination | Construction | Building the germinator |
| | | Planting 1 | Sowing seeds in germinator |
| | | Shade | Protecting seeds from sunlight |
| | | Irrigation 1 | Keeping the germinator moist |
| | Coffee nursery | Preparation | Preparing substrate organic fertilize |
| | | Fill | Filling coffee bags |
| | | Selection 1 | Selecting seedbeds |
| **Plantation and harvesting cof-fee** | | Transplant | Transplanting seedbeds to bags |
| | | Irrigation 2 | Keeping bag soil moist |
| | Plantation | Plot-spacing | Establish planting density |
| | | Tillage | Digging holes for each seedling |
| | | Selection 2 | Select seedlings |
| | | Planting 2 | Sowing the seedlings in the pits |
| | Fertilization and Pest/weeds Man-agement | Fertilize growth | Fertilizing coffee in growth |
| | | Trimming | Pruning coffee plants |
| | | Fertilize production | Fertilizing coffee in production |
| | Harvesting | Harvesting | Harvesting coffee beans |

## 3.2. Emergy Analysis.

In order to identify the energy flows in a coffee agricultural system, it is necessary to characterize the stages that are carried out in this productive process. In this way, it is possible to detail the energy flows that are contributing to the system in each of these stages. This detailed inventory is shown in Appendix A, where all the energy flows involved in each stage of the production process are presented (for the year 2018). Thistable is designed according to Emergy's methodology.

Figure 4 shows the energy contributions of nature to the farm (green bars) and the con-tribution of the economy (orange bars) for the year 2018. In general, it can be seen thatmost of the energy comes from non-renewable resources, mainly from energy providedby the soil (nature) and energy from labor (economy). The data in Table 2 show that the stage that demands the most energy resources is planting, with approximately 84%of total energy use, harvesting contributes 15% and the milling stage only contributes about 1%.

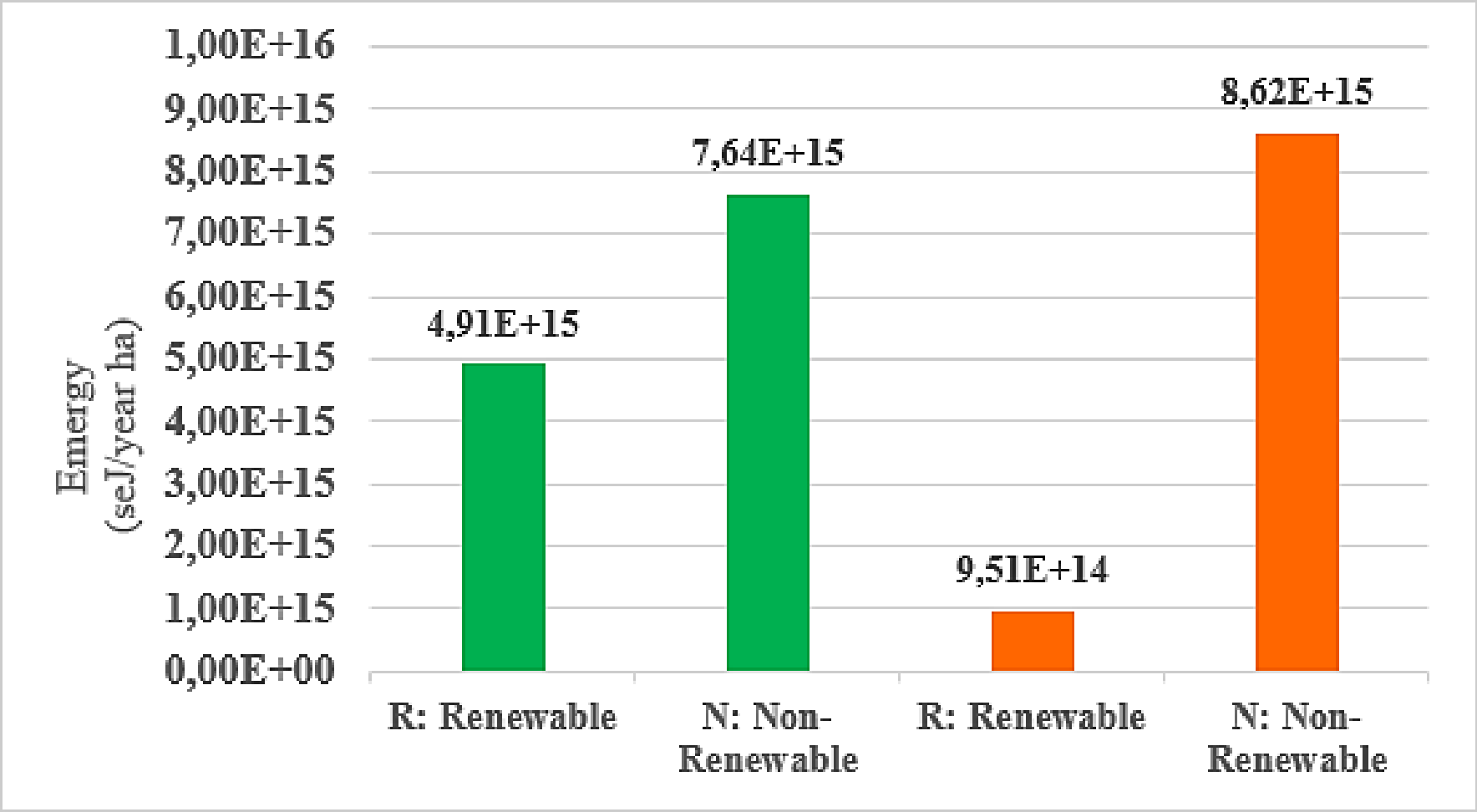

**Fig. 4.** Emergy flows contributed by Nature (green) and the Economy (orange) (Re- newable and non-Renewables), 2018.

**Table 2.** Emergy contributions by stages of the production process, 2018-2020.

| Year | Plantation | Harvesting | Drying | TOTAL Emergy |
|---|---|---|---|---|
| **2018** | 1.86E+16 | 3.38E+15 | 1.75E+14 | 2.21E+16 |
| **2019** | 1,80E+16 | 3,02E+15 | 1,53E+14 | 2,11E+16 |
| **2020** | 1,74E+16 | 3,24E+15 | 1,46E+14 | 2,08E+16 |

The emergy for 2020 decreased by 0.13E+16 (SeJ) with respect to 2018, and 0.3 E+16(SeJ) with reference to 2019, indicating that less energy was required for the operationof the productive system. In addition, it is important to know the significant energy usesin each stage of the productive process (Figure 5). The plantation is the one that con- sumes the most resources with 83.9%. The resources that contribute most energy are the soil, renewable natural resources (sun, wind, rain, etc.), and human labor. In the harvest stage, 15.2% of energy is consumed, and the resource that contributes most is human labor. And only 0.78% of energy is used at the milling stage, with human laborcontributing the most.

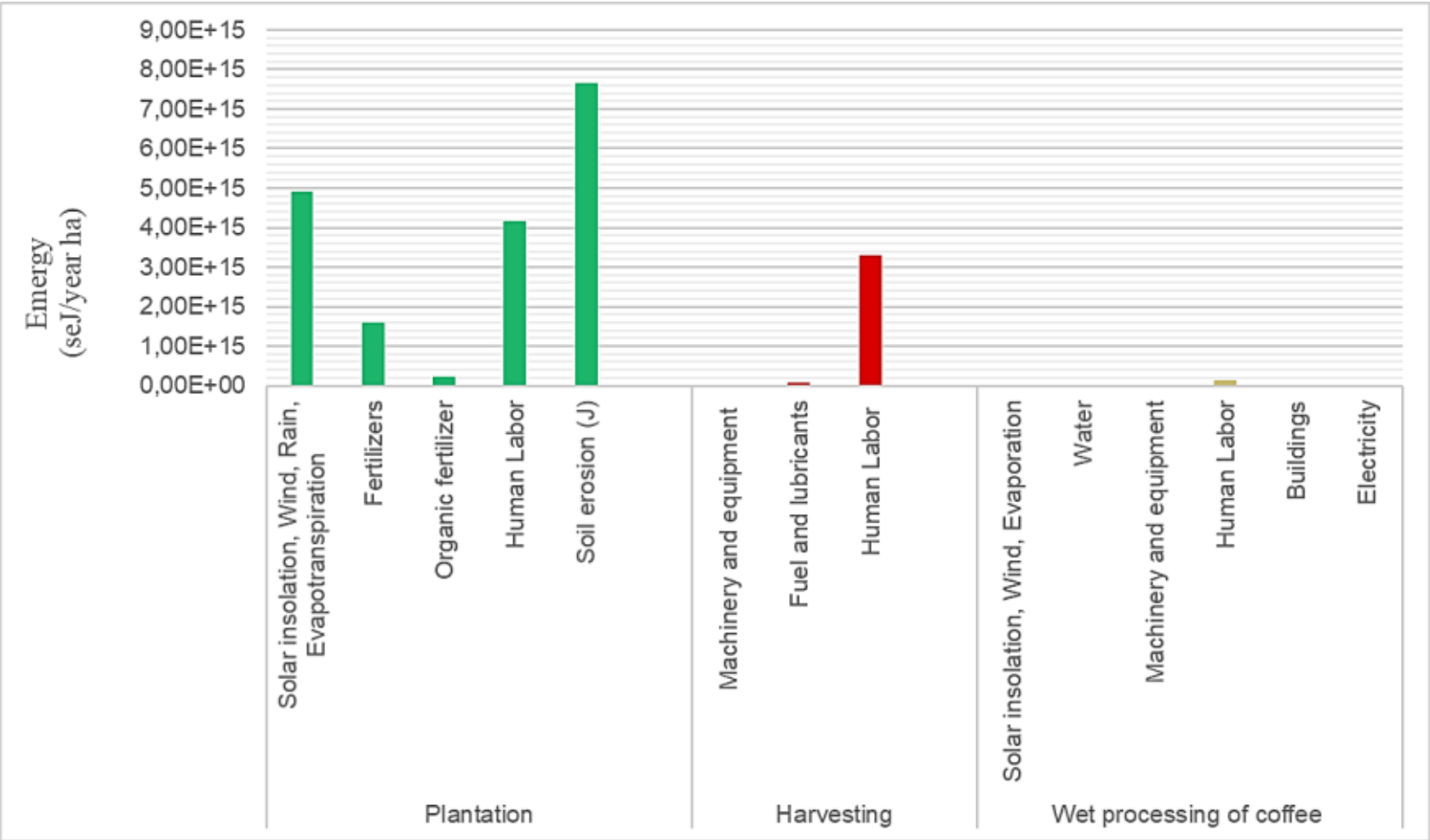

**Fig. 4.** Significant emergy use in each stage of the production process, 2018.

Finally, different indicators were calculated to evaluate the sustainability, efficiency and productivity of the production farm (Table 3).

**Table 3.** Emergy Indices.

| Indices | Formula | 2018 | 2019 | 2020 |
|---|---|---|---|---|
| Total emergy [Sej] | $Y_{coffee} = R + N + F$ | 2.21E+16 | 2.11E+16 | 2.08E+16 |
| Transformity [Sej/J] | $T = \frac{Y_{coffee}}{E} = \frac{Total\ emergy}{Product\ energy}$ | 1.79E+06 | 1.50E+06 | **1.35E+06** |
| Emergy Yield Ratio | $EYR = \frac{Y_{coffee}}{F}$ | 2.31 | 2.46 | 2.52 |
| Environmental Loading Ratio | $ELR = \frac{N + F_N}{R + F_R}$ <br> $F_N = M_N + S_N$ y $F_R = M_R + S_R$ | 2.77 | 2.66 | 2.61 |
| ergy InvestmentRatio | $EIR = \frac{F}{N + R}$ | 0.76 | 0.68 | 0.66 |
| rgy SustainabilityIndex | $ESI = \frac{EYR}{ELR}$ | 0.83 | 0.93 | **0.97** |
| ductivity - cherrycoffee | kg/ha | 7500 | 8200 | **8750** |

Comparing the three years analyzed, the “La Angostura” farm had its best environmen-tal and energy performance in 2020, with a production of 8750 kg/ha of cherry coffee,where the Transformity was the lowest, 1.35E+06 seJ/J, and the Emergy SustainabilityIndex (ESI) increased (0,97). This indicates that the decisions taken in recent years have led to a notable improvement in the efficiency and sustainability of the system. The Emery value decreased each year, demonstrating that the system was more energy ef- ficient in 2020 compared to 2018.

## 4 Conclusions

The sustainable development of the department of Cauca requires the planning of effi-cient agricultural production systems, where the management of resources and respectfor nature are considered. For this purpose, the evaluation of the sustainability of a coffee farm was proposed, considering the energy flows that intervene in the processesof the productive chain. For this purpose, a systemic approach was considered, which would allow for an integral characterization of the components that interact in the sys-tem, and the multiple energy flows that intervene, such as: sun, wind, rain, soil, ferti- lizers, equipment, machinery, labor, etc. This modeling of the production chain was carried out by means of operations and flow modeling, process flow diagrams (PFD), using the ISA 88 standard, thus allowing a broad understanding of the coffee system.

After having the inventory of all the energy flows involved in the system, the Emergy method was used, which helps to establish a model that indicates the energy footprint of a product, in this case coffee. Likewise, the tool provides a series of indicators that help to improve the decision-making processes of the coffee producing communities.

The work was carried out in the traditional farm "La Angostura", located in the villageof Clarete Bajo, in the city of Popayán, data from the years 2018, 2019, and 2020 wereconsidered. For the three years analyzed, the planting stage is the one that consumes the most resources with approximately 84%. The resources that contribute the most energy are: soil, renewable natural resources (sun, wind, rain, etc.), and human labor; the harvest stage contributes approximately 15% of the energy, here the resource that contributes the most is human labor; and only 1% of the energy is used in the milling stage, with human labor contributing the most.

Approximately 59% of the energy used by the farm comes from nature, of which 23%is from renewable sources and 36% from non-renewable sources. The economy con- tributes 41% of the energy, of which only 4% is from renewable sources and 37% fromnon-renewable sources. Energy for 2020 decreased by 0.13E+16 (SeJ) compared to 2018, and 0.3 E+16 (SeJ) compared to 2019, indicating that less energy was required for coffee production on the farm.

This work is an important basis that would help coffee growers in the region to make decisions regarding the management of their farms. Similarly, as a perspective it wouldbe interesting to integrate this type of sustainability evaluation work with environmen-tal management issues and the obtaining of green seals. This last option would be in- teresting to explore, as it would allow the coffee producing communities to have a toolto "demonstrate" that they are carrying out sustainable processes and continuous improvement within their farms, allowing them to have easier access to this type ofseals, also achieving a better positioning in the markets.

## References

1. Federación Nacional de Cafeteros de Colombia, "ESTADISTICAS HISTORICAS," *ESTADISTICAS HISTORICAS - Área cultivada - anual desde2002*, 2017.
2. Federación Nacional de Cafeteros de Colombia, "Organos gremiales de la Fe-deración de Cafeteros," 2017.https://www.federaciondecafeteros.org/clientes/es/que_hacemos/representacion_gremial/organos_gremiales_de_la_federacion_de_cafeteros/ (accessed Nov. 17, 2017).
3. Gobernación del Cauca, "Plan Departamental de Desarrollo 2020-2023." 2020.
4. J. F. Martin, S. A. W. Diemont, E. Powell, M. Stanton, and S. Levy-tacher, "Emergy evaluation of the performance and sustainability of three agriculturalsystems with different scales and management," vol. 115, pp. 128–140, 2006, doi: 10.1016/j.agee.2005.12.016.
5. C. M. Rodríguez, C. F. R. Rodas, J. C. C. Muñoz, and A. F. Casas, "A multi- criteria approach for comparison of environmental assessment methods in the analysis of the energy efficiency in agricultural production systems," *J Clean Prod*, vol. 228, pp. 1464–1471, 2019, doi: https://doi.org/10.1016/j.jcle-pro.2019.04.388.
6. M. Giampietro, G. Pastore, and S. Ulgiati, "Agricultura italiana e conceitos desustentabilidade," 2016.
7. M. Giampietro, "Economic growth, human disturbance to ecological systems, and sustainability," *ECOSYSTEMS OF THE WORLD*, pp. 723–746, 1999.
8. T. Clayton and N. J. Radcliffe, *Sustainability: a systems approach*. Routledge,2015.
9. M. Vegetti and G. P. Henning, "ISA-88 Formalization. A Step Towards its In-tegration with the ISA-95 Standard.," in *FOMI@ FOIS*, 2014.
10. E. Munoz, E. Capon-Garcia, and L. Puigjaner, "ANSI/ISA 88-95 Standards Based-Approach for Improved Integration of Recipes and Operational Tasks Supported by Knowledge Management," in *Computer Aided Chemical Engi- neering*, Elsevier, 2017, pp. 2335–2340.
11. M. T. Brown and S. Ulgiati, "Energy quality, emergy, and transformity : H . T

i. . Odum ' s contributions to quantifying and understanding systems," *Ecological Modelling*, vol. 178, pp. 201–213, 2004, doi: 10.1016/j.ecolmodel.2004.03.002.

12. H. T. Odum, *Environmental accounting: emergy and environmental decisionmaking*. New York: John Wiley and Sons, 1996.
13. G. C. Rótolo and C. Francis, "Los servicios ecosistémicos en el 'corazón' agrí-cola de Argentina," *Ediciones INTA*, vol. 44, 2008.

i. B. F. Giannetti, Y. Ogura, S. H. Bonilla, and C. M. V. B. Almeida, "Accountingemergy flows to determine the best production model of a coffee plantation,"

ii. *Energy Policy*, vol. 39, no. 11, pp. 7399–7407, 2011, doi: 10.1016/j.en- pol.2011.09.005.

14. H. T. Odum, *Environmental accounting*. Wiley, 1996.
15. H. T. Odum and E. P. Odum, "The energetic basis for valuation of ecosystem services," *Ecosystems*, vol. 3, no. 1, pp. 21–23, 2000.

16. M. Cuadra and T. Rydberg, “Emergy evaluation on the production, processingand export of coffee in Nicaragua,” *Ecol Modell*, vol. 6, no. 96, pp. 421–433, 2006, doi: 10.1016/j.ecolmodel.2006.02.010.
17. C. M. V. B. A. B.F. Giannetti, Y. Ogura, S.H. Bonilla, “Accounting emergy flows to determine the best production model of a coffee plantation,” *Energy Policy*, vol. 39, pp. 7399–7407, 2011, doi: 10.1016/j.enpol.2011.09.005.
18. M. Jafari, M. Reza, M. Ramroudi, M. Galavi, and G. Hadarbadi, “Sustainabilityassessment of date and pistachio agricultural systems using energy, emergy andeconomic approaches,” *J Clean Prod*, vol. 193, pp. 642–651, 2018, doi:10.1016/j.jclepro.2018.05.089.
19. J. Björklund, U. Geber, and T. Rydberg, “Emergy analysis of municipalwastewater treatment and generation of electricity by digestion of sewage sludge,” *Resour Conserv Recycl*, vol. 31, no. 4, pp. 293–316, 2001.
20. M. Panzieri, N. Marchettini, and T. G. Hallam, “Importance of the Bradhyrizo- bium japonicum symbiosis for the sustainability of a soybean cultivation,” *EcolModell*, vol. 135, no. 2–3, pp. 301–310, 2000.
21. A. A. Buenfil, *Emergy evaluation of water*. University of Florida, 2001.

**Appendix A.** Emergy "La Angostura" farm, Popayán, Cauca, 2018.

| Item | Description | Class | Annual flow (unit/year ha) | | | Emergy per Unit (seJ/unit) | Emergy (seJ/year ha) | Refs. |
|---|---|---|---|---|---|---|---|---|
| **Plantation*** | | | 2 first years/30 years | Year of 2018 | Total | | | |
| **1** | Solar insolation (J) | *R* | 2,72E+12 | 4,07E+13 | 4,34E+13 | 1,00E+00 | 4,34E+13 | [15] |
| **2** | Wind, kinetic energy (J) | *R* | 3,54E+06 | 5,31E+07 | 5,67E+07 | 2,52E+03 | 1,43E+11 | [15] |
| **3** | Rain, chemical energy (J) | *R* | 4,01E+09 | 6,01E+10 | 6,41E+10 | 3,06E+04 | 1,96E+15 | [15] |
| **4** | Rain, geopotential energy (J) | *R* | 8,65E+07 | 1,30E+09 | 1,38E+09 | 1,76E+04 | 2,44E+13 | [15] |
| **5** | Evapotranspiration (J) | *R* | 4,52E+09 | 6,78E+10 | 7,23E+10 | 3,98E+04 | 2,88E+15 | [16] |
| **6** | Soil erosion (J) | *N* | 6,46E+09 | 9,68E+10 | 1,03E+11 | 7,40E+04 | 7,64E+15 | [15] |
| **7** | Nitrogen (g) | *F* | 3,33E+03 | 5,00E+04 | 5,33E+04 | 6,62E+09 | 3,53E+14 | [17] |
| **8** | Phosphate (g) | *F* | 6,67E+02 | 1,00E+04 | 1,07E+04 | 9,35E+09 | 9,97E+13 | [17] |
| **9** | Potassium (g) | *F* | 4,67E+03 | 7,00E+04 | 7,47E+04 | 9,32E+08 | 6,96E+13 | [17] |
| **10** | Urea (g) | *F* | 1,00E+04 | 1,50E+05 | 1,60E+05 | 6,62E+09 | 1,06E+15 | [17] |
| **11** | Cal (g) | *F* | 6,67E+02 | 1,00E+04 | 1,07E+04 | 1,68E+09 | 1,79E+13 | [18] |
| **12** | Organic fertilizer (J) | 80% *R* | 2,00E+03 | 6,00E+04 | 6,20E+04 | 3,87E+09 | 2,40E+14 | [19] |
| **13** | Seeds (J) | *F* | 3,77E+05 | 0,00E+00 | 3,77E+05 | 5,85E+04 | 2,20E+10 | [17] |
| **14** | Machinery and equipment (g) | *F* | 7,33E+01 | 1,10E+03 | 1,17E+03 | 6,70E+09 | 7,86E+12 | [20] |
| **15** | Human Labor (USD) | *10% FR* | 1,97E+01 | 1,66E+02 | 1,85E+02 | 2,25E+13 | 4,17E+15 | [19] |
| **16** | Pesticides and fungicides (g) | *F* | 6,67E+00 | 1,00E+02 | 1,07E+02 | 1,48E+10 | 1,58E+12 | [21] |
| | **Total for plantation** | | | | | | **1,86E+16** | |
| **Harvesting** | | | **Annual flow (unit/year ha) - 2018** | | | | | |
| **17** | Machinery and equipment (J) | *F* | | | 1,00E+02 | 6,70E+09 | 6,70E+11 | [20] |
| **18** | Fuel and lubricants (J) | *F* | | | 7,90E+08 | 1,11E+05 | 8,78E+13 | [18] |
| **19** | Human Labor (USD) | *10% FR* | | | 1,46E+02 | 2,25E+13 | 3,29E+15 | [19] |
| | **Total for Harvesting** | | | | | | **3,38E+15** | |
| **Drying** | | | | | | | | |
| **20** | Solar insolation (J) | *R* | | | 2,23E+12 | 1,00E+00 | 2,23E+12 | [15] |
| **21** | Wind, kinetic energy (J) | *R* | | | 5,31E+07 | 2,52E+03 | 1,34E+11 | [15] |

| | | | | | | |
|---|---|---|---|---|---|---|
| **22** | Evaporation (g) | *R* | 3,72E+06 | 1,45E+05 | 5,39E+11 | [22] |
| **23** | Water (J) | *R* | 1,78E+07 | 8,60E+04 | 1,53E+12 | [15] |
| **24** | Drying yard or "paseras" (g) | *F* | 7,06E-01 | 2,65E+13 | 1,87E+13 | [17] |
| **25** | Machinery and equip-ment (USD) | *F* | 2,23E+12 | 1,00E+00 | 2,23E+12 | [17] |
| **26** | Human Labor (USD) | *10% FR* | 5,66E+00 | 2,25E+13 | 1,27E+14 | [19] |
| **27** | Buildings (USD) | *F* | 5,08E-01 | 2,65E+13 | 1,35E+13 | [18] |
| **28** | Electricity (USD) | *F* | 3,44E-01 | 2,65E+13 | 9,12E+12 | [17] |
| **29** | Sacos Jute (USD) | *F* | 1,00E+02 | 2,31E+10 | 2,31E+12 | [17] |
| | **Total for Drying** | | | | **1,75E+14** | |
| | | | | | | |
| | **(Y) Total Emergy** | | | | **2,21E+16** | Calculated in this study |

*The planting stage was carried out in 2013, where the energy flows are applied (items 1-16), however, this is an investment that is being made in the long term, in this study, and according to the indications of experts in coffee production, it is estimated that the crop can have a productive life of 30 years (Arcila et al., 2007b). For this reason, it is necessary to divide the energy applied in the first 2 years (where the crop had no production), between the 30 years of useful life. Similarly, each year (year analyzed in the 2018 table), it is necessary to perform the maintenance of the crop where energy flows must be added (items 1-16, except 13 seeds). This energy applied during the first 2 years, which is a long-term investment, is only done for the plantation stage.

# Un sistema de recomendación para la reutilización de fármacos utilizando embeddings sobre Grafos de Conocimiento

**Diego López Yse[1] and Diego Torres[1,2,3]**

[1] Universidad Austral, Buenos Aires, Argentina
[2] LIFIA, CIBPBA-Facultad de Informática, UNLP, La Plata, Argentina
[3] Depto. CyT., Universidad Nacional de Quilmes, Bernal, Argentina

**Abstract.** El proceso de desarrollo tradicional de nuevos fármacos es un camino largo, ineficiente y costoso. Afortunadamente, la reutilizacion de fármacos representa una muy buena alternativa para superar las limitaciones de los métodos de desarrollo tradicionales. En este sentido, los métodos computaciones de Grafos de Conocimiento y su representación vectorial (embeddings) se han utilizado en el campo biomédico para la reutilización de farmacos con gran efectividad, mejorando los resulta- dos producidos por otras técnicas tradicionales de Machine Learning. El presente trabajo propone un esquema para desarrollar un sistema de recomendación para la reutilización de fármacos utilizando embeddings sobre Grafos de Conocimiento, con la flexibilidad suficiente para cen- trarse en múltiples enfermedades e incorporar otros tipos de modelos predictivos para identificar a los mejores.

## 1 Introducción

El proceso de desarrollo tradicional de nuevos fármacos es un camino largo y costoso, que se estima toma en promedio 14 años y cuesta aproximadamente USD 1.8 mil millones para desarrollar un medicamento [7]. Afortunadamente existen otros métodos más aptos para enfrentar la velocidad que nos plantea la biología sobre las enfermedades. La reutilización de fármacos es el proceso de identificación de indicaciones novedosas para fármacos existentes, y representa una muy buena alternativa para superar las limitaciones de los métodos de de- sarrollo tradicionales, reduciendo drásticamente los plazos, las tasas de fracaso y los costos de desarrollo.

Los enfoques principales para la reutilización de fármacos se clasifican en métodos experimentales y computacionales [4], y el éxito de la reutilización está sujeto a cómo se combinan estos dos enfoques en colaboración.

En particular, los Grafo de Conocimiento se han utilizado en el campo biomédico para la reutilización de fármacos y otras actividades como la pri- orización de genes relevantes en enfermedades [5]. La combinación de Grafos de Conocimiento y su representación vectorial (embeddings) propone una visión novedosa para enfrentar el desafío de reutilización de fármaco al establecer nodos y relaciones entre los datos en contraposición a las clásicas bases de información estructuradas, y análisis en espacios vectorizados en lugar de clásicos modelados multidimensionales.

## 2 Trabajos Relacionados

Los enfoques computacionales para la reutilización de fármacos han ganando popularidad debido a la disponibilidad de grandes datos biológicos. Un gran número de trabajos basados en métodos de Machine Learning aprovechan de- scriptores construidos manualmente para realizar predicciones (ej., propiedades moleculares), con el fin de identificar posibles fármacos candidatos para ensayos clínicos posteriores [6].

Pero existen métodos más efectivos para representar información biomédica, y trabajos actuales se han centrado en utilizar técnicas de embeddings sobre Grafos de Conocimiento [2], mapeando la información a un espacio vectorial continuo, y preservando la estructura de proximidad del gráfico de conocimiento para ejecutar algoritmos de Machine Learning.

## 3 Problemática

El proceso de desarrollo de nuevos fármacos es una tarea compleja que requiere el conocimiento de numerosos dominios biológicos y químicos. Posiblemente el mayor obstáculo en este proceso se relaciona con la toxicidad, dando como re- sultado numerosos compuestos que fallan en las últimas etapas de los ensayos clínicos o que se retiran del mercado [1]. De esta forma, la reutilización de fármacos es una posible alternativa a este problema, acelerando el descubrim- iento de nuevas aplicaciones cuando los perfiles de seguridad de los medicamentos que se reutilizan se han evaluado en el contexto del desarrollo de medicamentospara otra enfermedad [3].

La reutilización de fármacos no soloconduce a menores gastos y acelera el desarrollo, sino que también agrega conocimiento sobre los efectos secundarios, la farmacocinética y las interacciones farmacológicas de los fármacos en cuestión [4].

## 4 Contexto

Para enfrentar el desafío de reutilización de fármacos es posible utilizar distintas técnicas y representaciones de datos biológicos. En este sentido, las técnicas de Machine Learning que utilizan descriptores moleculares son populares y suelen alcanzar niveles aceptables de performance. Pero es posible expandir el espaciode exploración de soluciones utilizando técnicas de embeddings sobre Grafos de Conocimiento que vinculen el amplio universo de datos biológicos disponibles (ej., genes o efectos secundarios relacionados a fármacos).

Por esta razón, el presente trabajo propone un esquema para desarrollar un sistema de recomendación para la reutilización de fármacos, integrando datos biológicos y distintos modelos predictivos basados en embeddings sobre Grafos de Conocimiento para lograr resultados con mayores probabilidades de éxito.

# 5 Propuesta

Para desarrollar un sistema de recomendación para la reutilización de fármacos centrado en Grafos de Conocimiento, se proponen las siguientes etapas:

## 5.1 Configuración

La primera fase se basa en definir y estructurar los datos y modelos para analizary construir los modelos predictivos de embeddings.

- Selección de Grafo de Conocimiento con datos significativos para el desafío relacionado. Si la reutilización de fármacos se orienta a una enfermedad determi- nada, se debe seleccionar un set de datos con un volumen adecuado de tripletas para esa enfermedad.

- Visualización de grafo integral y subgrafos seleccionados por el usuario para el analisis de la información y la comprensión de las relaciones establecidas.

- Entrenar y testear multiples modelos de embeddings sobre el Grafo de Conocimiento, asignando la información del grafo a distintas representaciones de baja dimensionalidad, preservando la estructura de proximidad de los datos para explotarla en aplicaciones como la predicción de vértices.

- Realizar predicción de vértices para inferir relaciones entre nodos no conec- tados dentro del Grafos de Conocimiento (e.j., predecir un fármaco dada una query" X, trata, dengue").

## 5.2 Evaluación

Para evaluar los modelos de embeddings sobre la enfermedad objetivo, cada modelo debe medirse: (a) intrínsecamente, utilizando métricas como Adjusted Mean Rank (AMR) y hits@k, y (b) externamente contra una fuente de vali-dación externa para comprender su poder predictivo. sobre la información del mundo real. De esta forma, un modelo de embeddings que acierta todos los com-puestos existentes en la fuente de valiadación para una enfermedad determinada utilizando menos predicciones se considera mejor que otro modelo que necesita más predicciones para alcanzar la misma cantidad de aciertos.

## 5.3 Recomendación

Finalmente, en base a la ponderación que asigne al usuario a los métodos de evaluación y la posibilidad de filtrar resultados por variables tales como tamaño molecular y grupo orgánico, el sistema definirá el ranking final de fármacos sugeridos para reutilizar en la enfermedad target.

# 6 Conclusión

Este trabajo propone un esquema para desarrollar un sistema de recomendación para la reutilización de fármacos utilizando Grafos de Conocimiento, con la flex- ibilidad suficiente para centrarse en múltiples enfermedades. Además, el enfoque propuesto permite incorporar otros tipos de modelos predictivos (e.j., basados ensimilitud química) que pueden competir contra los modelos de embeddings sobre Grafos de Conocimiento para identificar a los mejores. En una etapa posterior a este trabajo, los fármacos más prometedores deben ser testeados in vitro/vivo, retroalimentando los resultados a las bases de datos para continuar mejorando la performance de recomendación posteriores.

## References

1. Barratt, M.J., Frail, D. (eds.): Drug repositioning: bringing new life to shelved assets and existing drugs. John Wiley & Sons, Hoboken, N. J (2012)
2. Florin Ratajczak, et al.: Task-driven knowledge graph filtering im-proves prioritizing drugs for repurposing. BMC Bioinformatics **23**(84) (2022). https://doi.org/https://doi.org/10.1186/s12859-022-04608-y, https://bmcbioinformatics.biomedcentral.com/articles/10.1186/s12859-022-04608-y
3. Li, X., Yu, J., Zhang, Z., Ren, J., Peluffo, A.E., Zhang, W., Zhao, Y., Wu, J., Yan, K., Cohen, D., Wang, W.: Network bioinformatics analysis provides insight into drug repurposing for COVID-19. Medicine in Drug Discovery **10**, 100090 (Jun 2021). elsevier.com/retrieve/pii/S2590098621000117
4. Malik, J.A., Ahmed, S., Jan, B., Bender, O., Al Hagbani, T., Alqarni, A., An-war, S.: Drugs repurposed: An advanced step towards the treatment of breast cancer and associated challenges. Biomedicine & Pharmacotherapy **145**, 112375 (Jan 2022). https://doi.org/10.1016/j.biopha.2021.112375, https://linkinghub.elsevier.com/retrieve/pii/S0753332221011598
5. Nicholson, D.N., Greene, C.S.: Constructing knowledge graphs and their biomedical applications. Computational and Structural Biotechnology Journal **18**, 1414–1428 (2020). https://doi.org/10.1016/j.csbj.2020.05.017, https://linkinghub.elsevier.com/retrieve/pii/S2001037020302804
6. Pan, X., Lin, X., Cao, D., Zeng, X., Yu, P.S., He, L., Nussinov, R., Cheng, F.: Deep learning for drug repurposing: methods, databases, and applications (2022).,
7. Sang, S., Yang, Z., Liu, X., Wang, L., Lin, H., Wang, J., Dumontier, M.: GrEDeL: A Knowledge Graph Embedding Based Method for Drug Discovery From Biomedical Literatures. IEEE Access **7**, 8404–8415 (2019). https://doi.org/10.1109/ACCESS.2018.2886311, https://ieeexplore.ieee.org/document/8574025/

# Collaborative Framework for the Implementation ofDigital Transformation Processes in Higher Education Institutions: Case Study Unicomfacauca

**Alex Armando Torres Bermúdez**[1]

[1] Corporación Universitaria Comfacauca-Unicomfacauca, Popayán, Colombiaatorres@unicomfacauca.edu.co

**Abstract**. The purpose of this Doctoral Research Project is to present a collaborative work frame-work to support the implementation of digital transformation processes in higher education insti-tutions in order to improve the efficiency of mission processes; the case study is the CorporaciónUniversitaria Comfacauca for the implementation of digital transformation processes and the val-idation of the proposed framework in the mission research process. Through a systematic reviewof the literature, it became evident that there are few frameworks validated in higher education scenarios to implement Digital Transformation processes, mediated by a collaborative work en-vironment to contribute to the continuous improvement of the mission processes. The proposal focuses on a projective analytical research methodology, the method of quantitative action re- search type planned, makes it possible to analyze the data of a population in 7 institutions of higher education in the city of Popayan, department of Cauca (Colombia).

## 1 Introduction

Digital transformation is emerging as a topic of interest not only in scientific commu- nities, but also as an increasingly recurrent practice in organizational dynamics, whichare subject to the pressures imposed by digital technologies, new business models that are manifested throughout the value chain, and the personalized demands of users in anenvironment of hyper-connectivity [1]. More than a few authors have tried to explore the concept of digital transformation through three main areas of change: consumer behavior, business processes and business models [2], [3], [4]. From the literature re- view, the concept of digital transformation is defined as an evolutionary process of change in the organizational culture that is governed in an environment of hypercon- nectivity with collaborative principles for the digitization of processes in all activities of the value chain, which is enabled by the adoption of digital technologies and impactson management processes, business models, value creation, efficiency and operationalperformance; feasible to implement from frameworks and can be evaluated through maturity models and key performance indicators. In addition to this, research agrees that the education sector is one of the last industries that has initiated changes in a digitalculture because it is adhering to old methods and practices, the education sector is no exception [5].

## 2 State of the art

Regarding models and frameworks for digital transformation, the systematic literaturereview [6] showed that there are hardly any contributions of systematic methods as a research methodology on Digital Transformation, while it is more frequent to find con-ceptual models, frameworks and strategies; and much more frequent to find case studiesfocused on very specific areas. In addition, as a result of academic interest, numerous digital transformation frameworks have been published to address different sectors andtheir needs [7]. Despite the identified frameworks, however, there are few validated frameworks in higher education scenarios to implement Digital Transformation pro- cesses, mediated by a collaborative work environment to contribute to the improvementof missional processes, as mentioned by [8], there is a shortage of practical, imple- mentable and simple digital transformation models that combine technologies, systems and educational phenomena. This theoretical gap is therefore evidence that it remains unclear what type of sustainable digital transformation model(s) could be adopted [8]. According to the literature, it can be established that there are digital transformation frameworks or models for different sectors or areas of industry from a conceptual pointof view, with very little evidence of practical, implementable frameworks with achieved results, which indicates a knowledge gap as a state of solution in the educationsector.

## 3 Problem statement and contributions

At present, there is little evidence of validated frameworks in higher education scenar-ios to implement digital transformation processes in higher education institutions that contribute to the efficiency of mission processes mediated by a collaborative environ- ment [4] [8]. Similarly, [9] recognize that digital transformation is part of this change faced by all types of organizations. In this context, customer demands have changed [10], demanding customized requirements [10]. For this, organizations will have to re-spond quickly, with the support of collaborative work [11]. This indicates that another factor of Digital Transformation is Collaborative Work. For [1], end-to-end collabora-tion throughout the value chain is one of the key principles of digital transformation. Considering the aforementioned knowledge problems, the research question oriented to the development of this project arises: How to implement a collaborative work framework of digital transformation to improve the efficiency of the missional pro- cesses in a higher education institution?

## 4 Methodology and research approach

In order to carry out the systematic literature review, the procedure defined by Petersenet al. [12] [13] was followed; however, it is complemented by the research proceduresproposed by Hoyos Botero [14]. For the development of the proposal, it focuses on a projective analytical research methodology [15], the quantitative action research method proposed [16], makes it possible to analyze the data of a population in 7 insti- tutions of higher education in the city of Popayán, department of Cauca (Colombia). Asample of one (1) HEI (UNICOMFACAUCA) is extracted where a pre- and post-test will be conducted to know the level of adoption of the digital transformation processesand then the implementation of the process mediated by the technological tool "MCTD"- Collaborative Framework for the Implementation of Digital Transformation Processes, the above conformed from the quantitative research approaches.

## 5 Evaluation plan

The proposed framework will be assessed by validating its efficiency in relation to thecontribution that will be provided in the research mission process at the Corporación Universitaria Comfacauca - UNICOMFACACUA, specifically the research process, allowing it to be a reference framework for higher education institutions in everythingrelated to the current state, maturity level and implementation of digital transformationprocesses. For this, the evaluation plan will carry out: a) evaluation of collaborative work framework components, b) design of metrics and heuristic evaluation instrumentsfor measuring the usability and interaction of the tool, c) validation of framework usa-bility through framework design experts and software developers, d) verification of the impact of the proposed framework design and refinement of elaborated models, e) elab-oration of general document and technical support documents for the project.

## 6 Preliminary or intermediate results

There is a characterization of the different models and/or frameworks of digital trans- formation obtained from the respective literature review, 41 digital transformation frameworks for different industry sectors [7], which supports the knowledge gaps de- scribed in section 2.

## 7 Conclusions and Lessons Learned

There is a lot of literature related to the topic of digital transformation for organizations,facing this topic, there are conceptual, maturity and some exclusive models or frame- works with the name of digital transformation models or frameworks, however, there is no evidence of results of implementation of digital transformation processes in the higher education sector in a specific area. A framework is understood as a standardizedset of concepts, practices, guidelines and criteria to approach a particular type of prob-lem, which serves as a reference in a specific domain and can be used to address or solve new problems of a similar nature, which may serve as a basis for the implemen- tation of transformation processes in the education sector.

## 8 Doctor. Stage: Early.

## References

1. T. Delgado Fernández, "Taxonomía de transformación digital," 2020. [Online]. Available: https://orcid.org/0000-0002-4323-9674
2. K. Schwertner, "Digital transformation of business," *Trakia J. Sci.*, vol. 15, no. Suppl.1, pp. 388–393, 2017, doi: 10.15547/tjs.2017.s.01.065.
3. A. Bockshecker, S. Hackstein, and U. Baumöl, "Systematization of the term digital transformation and its phenomena from a socio-technical perspective – A literature review," 2018.
4. M. H. Ismail, M. Khater, and M. Zaki, "Digital Business Transformation and Strategy: What Do We Know So Far? University of Cambridge," *J. Univ. Cambridge*, no. November 2017, 2018, [Online]. Available: www.cambridgeservicealliance.org
5. K. Kerroum, A. Khiat, A. Bahnasse, E. S. Aoula, and Y. Khiat, "The proposal of an agile model for the digital transformation of the University Hassan II of Casablanca 4.0," *Procedia Comput. Sci.*, vol. 175, pp. 403–410, 2020, doi: 10.1016/j.procs.2020.07.057.
6. C. Gebayew, I. R. Hardini, G. H. A. Panjaitan, N. B. Kurniawan, and Suhardi, "A Systematic Literature Review on Digital Transformation," *2018 Int. Conf. Inf. Technol. Syst. Innov. ICITSI 2018 - Proc.*, pp. 260–265, 2018, doi: 10.1109/ICITSI.2018.8695912.
7. Z. Van Veldhoven and J. Vanthienen, "Digital transformation as an interaction-driven perspective between business, society, and technology," *Electron. Mark.*, 2021, doi: 10.1007/s12525-021-00464-5.
8. M. A. Mohamed Hashim, I. Tlemsani, and R. Duncan Matthews, "A sustainable University: Digital Transformation and Beyond," *Educ. Inf. Technol.*, 2022, doi: 10.1007/s10639-022-10968-y.
9. "Taxonomía de Transformación Digital | Revista Cubana de Transformación Digital." https://rctd.uic.cu/rctd/article/view/62 (accessed Jun. 20, 2021).
10. R. Morakanyane, A. Grace, and P. O'Reilly, "Conceptualizing digital transformation in business organizations: A systematic review of literature," *30th Bled eConference Digit. Transform. - From Connect. Things to Transform. our Lives, BLED 2017*, pp. 427–444, 2017, doi: 10.18690/978-961-286-043-1.30.
11. C. P. Lopez, J. Aguilar, and M. Santorum, "Autonomous VOs management based on industry 4.0: a systematic literature review," *J. Intell. Manuf.*, 2021, doi: 10.1007/s10845-021-01850-8.
12. K. Petersen, R. Feldt, S. Mujtaba, and M. Mattsson, "Systematic mapping studies in software engineering," *12th Int. Conf. Eval. Assess. Softw. Eng. EASE 2008*, pp. 1–10, 2008, doi: 10.14236/ewic/ease2008.8.
13. K. Petersen, S. Vakkalanka, and L. Kuzniarz, "Guidelines for conducting systematic mapping studies in software engineering: An update," *Inf. Softw. Technol.*, vol. 64, pp. 1–18, 2015, doi: 10.1016/j.infsof.2015.03.007.
14. C. H. Botero, *Un modelo para investigacion documental: guia teorico-practica sobre construccion de estados del arte con importantes reflexiones sobre la investigacion*. Señal Editora, 2000. [Online]. Available: https://books.google.com.co/books?id=Wa3PAQAACAAJ
15. M. N. Córdoba and C. Monsalve, "Tipos de investigación, predictiva, interactiva, confirmatoria y evaluativa,"
    i. *Fund. Sypal*, pp. 139–140, 2008.
16. R. Hernández Sampieri, C. Fernández Collado, and M. del P. Baptista Lucio, *Metodología de la Investigación*. Mexico: McGRAW-HILL / INTERAMERICANA EDITORES, S.A. DE C.V, 2005.

FUNDACIÓN
UNIVERSITARIA
DE POPAYÁN

UNIVERSITÉ DE LORRAINE

Lifia

Facultad de Informática - Universidad Nacional de La Plata

Universidad Tecnológica de Pereira

Universidad del Cauca

FUNDACIÓN UNIVERSITARIA DE POPAYÁN

Corporación Universitaria Comfacauca

Unicomfacauca

VIGILADA MINEDUCACIÓN